\documentclass{aa}  

\usepackage{graphicx}
\usepackage{xcolor}
\usepackage{booktabs}
\usepackage{supertabular}
\usepackage[version=4]{mhchem}
\usepackage{ulem}
\usepackage{longtable}
\usepackage{caption}
\usepackage{sidecap}
\usepackage{float}
\usepackage{placeins}
\usepackage{rotating}
\usepackage{txfonts}
\begin{document}

   \title{N-bearing complex organics toward high-mass protostars}

   \subtitle{Constant ratios pointing to formation in similar pre-stellar conditions across a large mass range}

   \author{P. Nazari
          \inst{1}
          \and
          J. D. Meijerhof\inst{1}
          \and 
          M. L. van Gelder\inst{1}
          \and
          A. Ahmadi\inst{1}
          \and
          E. F. van Dishoeck\inst{1, 2}
          \and
          B. Tabone\inst{3}
          \and
          D. Langeroodi\inst{4}
          \and
          N. F. W. Ligterink\inst{5}
          \and 
          J. Jaspers\inst{1}
          \and
          M. T. Beltr\'{a}n\inst{6}
          \and
          G. A. Fuller\inst{7,8}
          \and
          \'{A}. S\'{a}nchez-Monge\inst{8}
          \and
          P. Schilke\inst{8}
          }

    \institute{Leiden Observatory, Leiden University, P.O. Box 9513, 2300 RA Leiden, the Netherlands\\ 
        \email{nazari@strw.leidenuniv.nl}
         \and
        Max Planck Institut f\"{u}r Extraterrestrische Physik (MPE), Giessenbachstrasse 1, 85748 Garching, Germany
        \and 
        Universit\'{e} Paris-Saclay, CNRS, Institut d'Astrophysique Spatiale, 91405 Orsay, France
        \and
        DARK, Niels Bohr Institute, University of Copenhagen, Jagtvej 128, 2200 Copenhagen, Denmark
        \and
        Physics Institute, University of Bern, Sidlerstrasse 5, 3012 Bern, Switzerland
        \and
        INAF-Osservatorio Astrofisico di Arcetri, Largo E. Fermi 5, 50125 Firenze, Italy
        \and
        Jodrell Bank Centre for Astrophysics, Department of Physics and Astronomy, University of Manchester, Oxford Road, Manchester, M13 9PL, UK
        \and
        I. Physikalisches Institut, Universit\"{a}t zu K\"{o}ln, Z\"{u}lpicher Str.77, 50937, K\"{o}ln, Germany}

   \date{Received XXX; accepted YYY}

 
  \abstract
   {Complex organic species are known to be abundant toward low- and high-mass protostars. No statistical study of these species toward a large sample of high-mass protostars with the Atacama Large Millimeter/submillimeter Array (ALMA) has been carried out so far.}
   {We aim to study six N-bearing species: methyl cyanide (CH$_3$CN), isocyanic acid (HNCO), formamide (NH$_2$CHO), ethyl cyanide (C$_2$H$_5$CN), vinyl cyanide (C$_2$H$_3$CN) and methylamine (CH$_3$NH$_2$) in a large sample of line-rich high-mass protostars.}
   {From the ALMA Evolutionary study of High Mass Protocluster Formation in the Galaxy survey, 37 of the most line-rich hot molecular cores with ${\sim} 1\arcsec$ angular resolution are selected. Next, we fit their spectra and find column densities and excitation temperatures of the N-bearing species mentioned above, in addition to methanol (CH$_3$OH) to be used as a reference species. Finally, we compare our column densities with those in other low- and high-mass protostars.}
   {CH$_3$OH, CH$_3$CN and HNCO are detected in all sources in our sample, whereas C$_2$H$_3$CN and CH$_3$NH$_2$ are (tentatively) detected in ${\sim} 78\%$ and ${\sim} 32\%$ of the sources. We find three groups of species when comparing their excitation temperatures: hot (NH$_2$CHO; $T_{\rm ex} \gtrsim 250$\,K), warm (C$_2$H$_3$CN, HN$^{13}$CO and CH$_3^{13}$CN; 100\,K\,$\lesssim T_{\rm ex} \lesssim 250$\,K) and cold species (CH$_3$OH and CH$_3$NH$_2$; $T_{\rm ex} \lesssim 100$\,K). This temperature segregation reflects the trend seen in the sublimation temperature of these molecules and validates the idea that complex organic emission shows an onion-like structure around protostars. Moreover, the molecules studied here show constant column density ratios across low- and high-mass protostars with scatter less than a factor ${\sim} 3$ around the mean.}
   {The constant column density ratios point to a common formation environment of complex organics or their precursors, most likely in the pre-stellar ices. The scatter around the mean of the ratios, although small, varies depending on the species considered. This spread can either have a physical origin (source structure, line or dust optical depth) or a chemical one. Formamide is most prone to the physical effects as it is tracing the closest regions to the protostars, whereas such effects are small for other species. Assuming that all molecules form in the pre-stellar ices, the scatter variations could be explained by differences in lifetimes or physical conditions of the pre-stellar clouds. If the pre-stellar lifetimes are the main factor, they should be similar for low- and high-mass protostars (within factors ${\sim}2-3$).}


   \maketitle
%

\section{Introduction}

Complex organic molecules (COMs) are molecular species containing six or more atoms including carbon and hydrogen with one or more oxygen and/or nitrogen molecules (\citealt{Herbst2009}). The protostellar phase of star formation (for both low- and high-mass stars) is the most rich stage in gaseous complex organics as the temperatures are high (\citealt{Bisschop2007}; \citealt{vantHoff2020c}) and thus, molecules sublimate from the ice grains into the gas. Both Oxygen- and Nitrogen-bearing COMs have been detected in the interstellar medium toward low- and high-mass protostars over the past decades (e.g., \citealt{Blake1987}; \citealt{Turner1991}; \citealt{Ewine1995}; \citealt{Schilke1997}; \citealt{Gibb2000}; \citealt{Cazaux2003}; \citealt{Bottinelli2004}; \citealt{fontani2007}; \citealt{Bisschop2008}; \citealt{Beltran2009}; \citealt{Belloche2013}; \citealt{Jorgensen2016}; \citealt{Rivilla2017}; \citealt{Taniguchi2020}; \citealt{Gorai2021}; \citealt{Zeng2021}; \citealt{Williams2022}, see summary in \citealt{McGuire2021}). The origin of COMs is still debated but several species are likely produced by ice chemistry (methanol (CH$_3$OH); \citealt{Fuchs2009}; ethanol (CH$_3$CH$_2$OH); \citealt{Oberg2009}; aminomethanol (NH$_2$CH$_2$OH); \citealt{Theule2013}; glycerol (HOCH$_2$CH(OH)CH$_2$OH); \citealt{Fedoseev2017}; 1-propanol (CH$_3$CH$_2$CH$_2$OH); \citealt{Qasim2019}; acetaldehyde (CH$_3$CHO); \citealt{Chuang2021}).

If many of the molecular species indeed form as ices in cold dark clouds, ultimately, clues to the origin of such species comes from their direct detection in solid phase through infrared spectroscopy. No COMs except methanol have been securely detected so far in interstellar ices with ground based and space infrared telescopes (\citealt{Grim1991}; \citealt{Taban2003}; \citealt{Boogert2008}). However, OCN$^{-}$ (direct derivative of HNCO; \citealt{Broekhuizen2004}; \citealt{Fedoseev2016}) has been securely detected in interstellar ices (\citealt{Grim1987}; \citealt{vanBroekhuizen2005}; \citealt{Oberg2011}). This points to the fact that other N-bearing species could be residing on grains but we have not been able to detect them due to limitation of observations so far (\citealt{Boogert2015}). The \textit{James Webb} Space Telescope (JWST) will have much better sensitivity and spectral resolution in the critical 3-10\,$\mu$m range than its preceding telescopes such as the \textit{Spitzer} Space Telescope and thus should be able to observe some of the complex organics. In the meantime however, one can use millimeter and radio observations of these species in the gas after they are sublimated from the ices. The variations seen in abundances found in the gas can later be compared with the findings of JWST in both low- and high-mass protostars which will be a direct assessment of the conditions for formation of such species: gas versus ice.   

While single source chemical analyses are useful, more robust and general conclusions on COM formation can be obtained by studying a large sample of objects. For example, \cite{Coletta2020} used the IRAM-30m telescope to observe 39 high-mass star forming regions and find that methyl formate/dimethyl ether (CH$_3$OCHO/CH$_3$OCH$_3$) is remarkably constant across sources hosting different environments such as high-mass star forming regions, low-mass protostars, Galactic Center clouds, outflow shock regions and comets (also see \citealt{Jaber2014}). \cite{Coletta2020} conclude that the chemistry is mainly set in the early stages, stays intact during star formation, and that it is possible that these two molecules are chemically linked. 


The Atacama Large Millimeter/submillimeter Array (ALMA) can achieve higher angular resolution and sensitivity than single-dish telescopes used in previous studies of molecular inventories. It is important to have higher sensitivity to detect optically thin isotopologues of abundant and bright species such as CH$_3$OH and methyl cyanide (CH$_3$CN) and hence, put better constraints on the column densities of the main isotopologues. In addition, high spatial resolution minimizes the beam dilution effect. Moreover, ALMA has recently observed large samples of low- and high-mass protostars with spectral data for chemical analyses. As an example the Perseus ALMA Chemistry Survey (PEACHES) studied COMs towards low-mass protostars in Perseus (\citealt{Yang2021}) and the ALMA Survey of Orion Planck Galactic Cold Clumps (ALMASOP) studied these species towards Class 0/I sources in Orion (\citealt{Hsu2022}). As of yet, chemistry of large samples of high-mass protostars observed by ALMA have not been analyzed. 

In this work we use the data from the ALMA Evolutionary study of High Mass Protocluster Formation in the Galaxy (ALMAGAL) survey (2019.1.00195.L; PI: Sergio Molinari) to do such a chemistry analysis on a large sample of high-mass sources. This study uses 37 of the most line-rich sources observed by the ALMAGAL survey and focuses on six N-bearing species and some of their isotopologues: formamide (NH$_2$CHO), isocyanic acid (HNCO), methyl cyanide, vinyl cyanide (C$_2$H$_3$CN), ethyl cyanide (C$_2$H$_5$CN) and methylamine (CH$_3$NH$_2$). Moreover, methanol is also included in this paper (also see \citealt{vanGelder2022}) for comparison with the N-bearing species. 

The reason for choosing the N-bearing molecules stated above is that some of them are believed to be chemically linked or have ice formation origin. A direct derivative of HNCO, OCN$^{-}$, has been detected in ices (\citealt{Grim1987}). Moreover, there is an extensive literature on the formation of formamide and its potential chemical link to isocyanic acid (e.g., \citealt{Bisschop2007}; \citealt{Lopez2019}). Both ice formation pathways (\citealt{Raunier2004}; \citealt{Jones2011}; \citealt{Rimola2018}; \citealt{Dulieu2019}; \citealt{Martin-Domenech2020}) and gas-phase chemistry routes (\citealt{Barone2015}; \citealt{Skouteris2017}; \citealt{Codella2017}) have been suggested for formamide. In particular, \cite{Haupa2019} found that HNCO and NH$_2$CHO are linked by a dual-cycle of hydrogen addition and abstraction reactions, which convert one species into the other on grains.

Methyl cyanide, vinyl cyanide and ethyl cyanide are part of the cyanide group and they are suggested to be chemically linked through ice chemistry including UV radiation (\citealt{Hudson2004}; \citealt{Bulak2021}). Moreover, chemical models of \cite{Garrod2017} and \cite{Garrod2022} find that vinyl cyanide and ethyl cyanide are chemically linked and mainly form on ices. 

Methylamine is thought to form on grains through recombination of radicals CH$_3$ and NH$_2$ (\citealt{Garrod2008}; \citealt{Kim2011}; \citealt{forstel2017}) in significant abundance based on chemical models (\citealt{Garrod2008}). However, it has only been detected in a handful of high-mass star forming regions due to its intrinsically low Einstein $A_{\rm ij}$ coefficients (\citealt{Kaifu1974}; \citealt{Bogelund2019}; \citealt{Ohishi2019}).

All molecules studied here, except methylamine, have already been observed in previous observations of both low- and high-mass protostars. For example, NH$_2$CHO and HNCO are detected toward low-mass protostars such as the IRAS 16293-2422 binary system (\citealt{Kahane2013}; \citealt{Coutens2016}; \citealt{Manigand2020}), NGC 1333 IRAS 4A2 (\citealt{Lopez2017}), B1-c and S68N (\citealt{Nazari2021}) and high-mass star forming regions such as Sgr B2(N2) (\citealt{Belloche2017}), Orion BN/KL (\citealt{Cernicharo2016}), AFGL 4176 (\citealt{Bogelund2019AFGL}), G10.47+0.03 (\citealt{Gorai2020}), NGC 6334I (\citealt{Ligterink2020}), G10.6-0.4 (\citealt{Law2021}), G31.41+0.31 (\citealt{Colzi2021}). The three cyanides are also studied toward low-mass protostars such as the IRAS 16293-2422 binary (\citealt{Calcutt2018}) 
and high-mass regions such as W43-MM1 (\citealt{Molet2019}) along with many of the sources mentioned above. However, to make robust conclusions on variation of abundances going from low- to high-mass protostars, more sources that are analyzed in a consistent way are needed.

\begin{figure*}
    \centering
    \includegraphics[width=15cm]{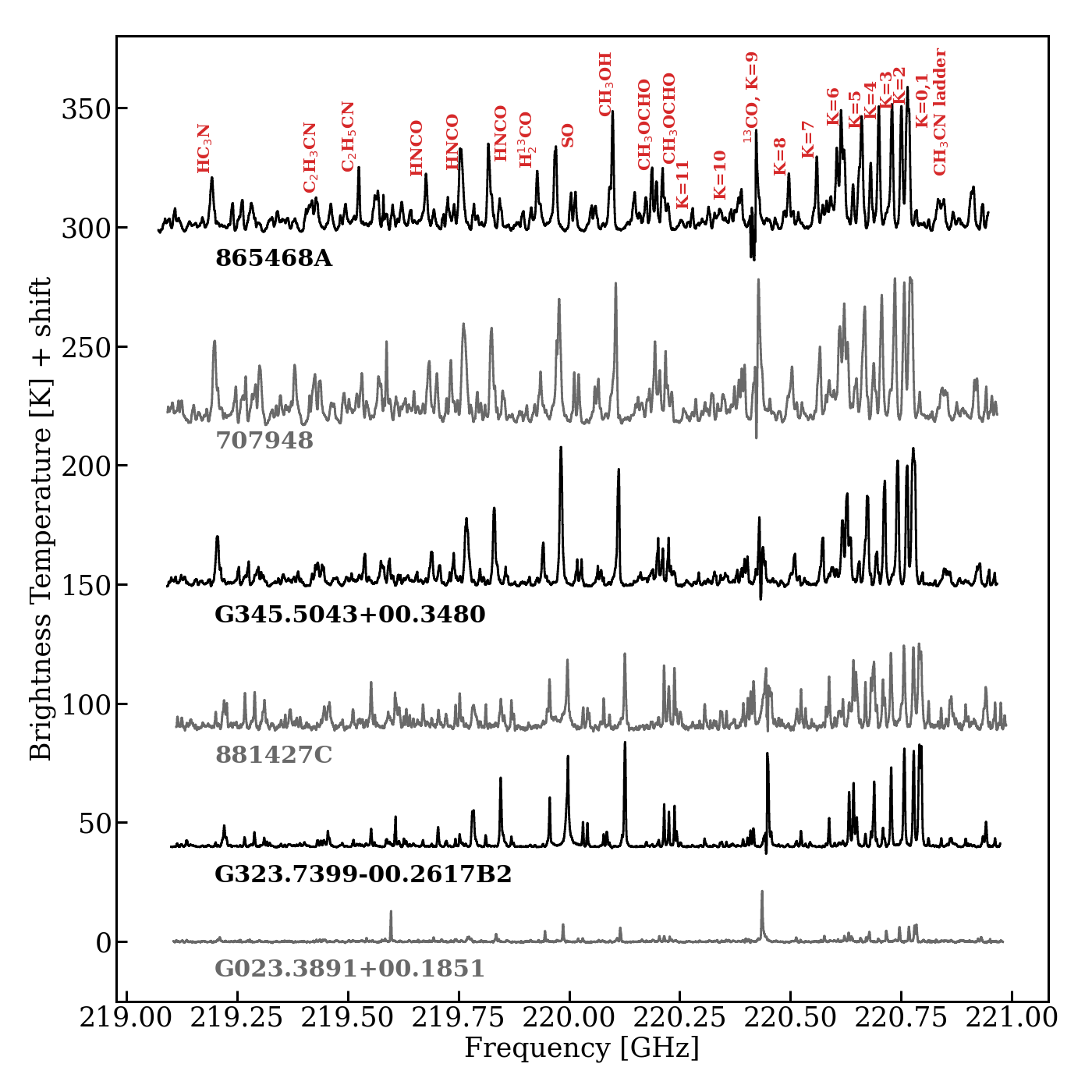}
    \caption{The spectra for six of the sources studied here with the name of each object printed below each spectrum. The molecules associated with several lines are also stated in red next to the lines.}
    \label{fig:spectra}
\end{figure*}

In this paper, a large sample of 37 high-mass sources is investigated. Section \ref{sec:obs_methods} explains the observational parameters in the ALMAGAL sample and methods used to fit the spectra. Section \ref{sec:results} presents the results of the fits and the correlations seen in the data. In Sect. \ref{sec:discussion} we discuss the findings in the context of chemical models and other observations. Finally, Sect. \ref{sec:conclusion} summarizes the main findings and presents our conclusions.

\section{Observations and methods}
\label{sec:obs_methods}
\subsection{The data}

This work uses the archival data observed by ALMA in Band 6 (${\sim 1}$\,mm) as part of the ALMAGAL survey (2019.1.00195.L; PIs: P. Schilke, S. Molinari, C. Battersby, P. Ho). The ALMAGAL survey observed more than 1000 dense clumps with masses larger than 500\,M$_{\odot}$ across the Galaxy with distances less than 7.5\,kpc. The objects observed are in different evolutionary stages, however, many of them are massive young stellar objects (MYSO). The targeted sources are chosen based on the sources observed in the \textit{Herschel} Hi-Gal survey (\citealt{Molinari2010}; \citealt{Elia2017,Elia2021}).

We consider the pipeline calibrated product data of the sources that were publicly available by 19$^{\rm th}$ of June 2021 and had beam sizes between 0.5$\arcsec$ and 1.5$\arcsec$ (${\sim} 1000-5000$\,au for sources at a few kpc). From the ${\sim 200}$ sources available at the time, the most line-rich cores are selected for this work. Therefore, the sample of protostars used here is biased. The criterion for a line rich source in this work is the detection of the CH$_3$CN 12$_K$-11$_K$ $K=7$ line at ${\sim} 2.5-3\sigma$ level. Moreover, a few sources that had very blended lines are eliminated from the sample. This procedure results in 37 line rich sources.

\begin{table*}
\Huge
\renewcommand{\arraystretch}{1.5}
    \caption{Fitted parameters of N-bearing species.}
    \label{tab:results}
    \resizebox{\textwidth}{!}{\begin{tabular}{@{\extracolsep{1mm}}*{11}{c}}
          \toprule
          \toprule      
          & \multicolumn{1}{c}{CH$_3$CN}& \multicolumn{3}{c}{CH$_3^{13}$CN}&\multicolumn{3}{c}{C$_2$H$_3$CN} & \multicolumn{3}{c}{C$_2$H$_5$CN}\\
          \cmidrule{2-2} \cmidrule{3-5} \cmidrule{6-8} \cmidrule{9-11}
        Source & $N (\rm cm^{-2})$ & $N (\rm cm^{-2})$ &  $T_{\rm ex} (\rm K)$ & FWHM (km s$^{-1}$) &  $N (\rm cm^{-2})$ &  $T_{\rm ex} (\rm K)$ & FWHM (km s$^{-1}$) &  $N (\rm cm^{-2})$ &  $T_{\rm ex} (\rm K)$ & FWHM (km s$^{-1}$) \\
        \midrule

101899 & 2.2$^{+0.4}_{-0.4}$ $\times 10^{16}$ & 4.9$^{+0.6}_{-0.6}$ $\times 10^{14}$ & 130$^{+20}_{-30}$ & [9.5] & 1.8$^{+0.2}_{-0.1}$ $\times 10^{15}$ & 150$^{+20}_{-30}$ & [9.5] & 8.2$^{+1.6}_{-1.6}$ $\times 10^{15}$ & [130] & 10.1 \\ 
126348 & 4.2$^{+1.0}_{-1.0}$ $\times 10^{15}$ & 9.0$^{+1.7}_{-1.7}$ $\times 10^{13}$ & 170$^{+60}_{-40}$ & [7.5] & 1.7$^{+0.2}_{-0.3}$ $\times 10^{14}$ & [170] & [7.5] & 1.0$^{+0.2}_{-0.2}$ $\times 10^{15}$ & [170] & 9.8 \\ 
615590 & $^*$1.5$^{+0.1}_{-0.2}$ $\times 10^{16}$ & <\,2.4 $\times 10^{14}$ & [150] & [4.5] & <\,6.2 $\times 10^{14}$ & [150] & [4.5] & 9.7$^{+5.7}_{-4.8}$ $\times 10^{14}$ & [150] & 4.5 \\ 
644284A & $^*$5.0$^{+0.5}_{-0.5}$ $\times 10^{15}$ & <\,1.7 $\times 10^{14}$ & [150] & [6.7] & <\,3.5 $\times 10^{14}$ & [150] & [6.7] & <\,7.6 $\times 10^{14}$ & [150] & [6.7] \\ 
693050 & 1.9$^{+0.3}_{-0.3}$ $\times 10^{16}$ & 3.2$^{+0.3}_{-0.3}$ $\times 10^{14}$ & 150$^{+20}_{-30}$ & [4.7] & 6.4$^{+1.0}_{-0.3}$ $\times 10^{14}$ & 120$^{+20}_{-20}$ & [4.7] & 3.7$^{+0.7}_{-0.7}$ $\times 10^{15}$ & [150] & 5.5 \\ 
705768 & 5.2$^{+1.0}_{-1.6}$ $\times 10^{15}$ & 8.4$^{+1.3}_{-2.5}$ $\times 10^{13}$ & 140$^{+40}_{-30}$ & [6.7] & <\,2.6 $\times 10^{14}$ & [140] & [6.7] & 8.6$^{+1.7}_{-1.7}$ $\times 10^{14}$ & [140] & 6.1 \\ 
707948 & 1.2$^{+0.3}_{-0.3}$ $\times 10^{17}$ & 1.9$^{+0.4}_{-0.4}$ $\times 10^{15}$ & 200$^{+50}_{-50}$ & [8.6] & 8.0$^{+1.6}_{-1.6}$ $\times 10^{15}$ & 130$^{+10}_{-10}$ & [8.6] & 2.9$^{+0.6}_{-0.6}$ $\times 10^{16}$ & [200] & 8.5 \\ 
717461A & 4.5$^{+1.2}_{-0.9}$ $\times 10^{15}$ & 7.8$^{+1.8}_{-1.2}$ $\times 10^{13}$ & [150] & [6.3] & 2.1$^{+0.3}_{-0.3}$ $\times 10^{14}$ & [150] & [6.3] & 3.3$^{+0.9}_{-0.7}$ $\times 10^{14}$ & [150] & 5.5 \\ 
721992 & 1.1$^{+0.2}_{-0.2}$ $\times 10^{16}$ & 2.0$^{+0.1}_{-0.2}$ $\times 10^{14}$ & 90$^{+20}_{-20}$ & [3.5] & 1.3$^{+0.4}_{-0.2}$ $\times 10^{15}$ & 70$^{+20}_{-10}$ & [3.5] & 1.3$^{+0.1}_{-0.2}$ $\times 10^{15}$ & [90] & 3.0 \\ 
724566 & 2.8$^{+0.5}_{-0.5}$ $\times 10^{16}$ & 5.1$^{+0.6}_{-0.7}$ $\times 10^{14}$ & 220$^{+50}_{-40}$ & [6.0] & 6.0$^{+3.5}_{-1.2}$ $\times 10^{14}$ & 150$^{+100}_{-30}$ & [6.0] & 6.4$^{+1.3}_{-1.3}$ $\times 10^{15}$ & [220] & 5.7 \\ 
732038 & $^*$4.1$^{+0.6}_{-0.7}$ $\times 10^{15}$ & <\,1.7 $\times 10^{14}$ & [150] & [6.0] & <\,3.8 $\times 10^{14}$ & [150] & [6.0] & 1.3$^{+0.3}_{-0.3}$ $\times 10^{15}$ & [150] & 6.0 \\ 
744757A & 7.1$^{+1.1}_{-1.1}$ $\times 10^{15}$ & 1.2$^{+0.1}_{-0.1}$ $\times 10^{14}$ & 130$^{+20}_{-30}$ & [5.7] & 7.0$^{+0.6}_{-1.5}$ $\times 10^{14}$ & 190$^{+20}_{-30}$ & [5.7] & 1.2$^{+0.2}_{-0.2}$ $\times 10^{15}$ & [130] & 5.7 \\ 
767784 & 6.1$^{+0.9}_{-0.9}$ $\times 10^{15}$ & 1.2$^{+0.1}_{-0.1}$ $\times 10^{14}$ & 110$^{+20}_{-20}$ & [4.0] & <\,4.1 $\times 10^{14}$ & [120] & [4.0] & 9.2$^{+1.8}_{-1.9}$ $\times 10^{14}$ & [110] & 3.3 \\ 
778802 & 3.3$^{+0.7}_{-0.7}$ $\times 10^{15}$ & 5.9$^{+0.9}_{-1.1}$ $\times 10^{13}$ & [150] & [7.0] & 1.3$^{+0.3}_{-0.3}$ $\times 10^{14}$ & [150] & [7.0] & 5.1$^{+0.4}_{-0.4}$ $\times 10^{14}$ & [150] & 7.0 \\ 
779523 & 3.1$^{+0.8}_{-0.9}$ $\times 10^{15}$ & 5.6$^{+1.2}_{-1.5}$ $\times 10^{13}$ & [150] & [8.4] & <\,1.7 $\times 10^{14}$ & [150] & [8.4] & 3.4$^{+0.9}_{-0.7}$ $\times 10^{14}$ & [150] & 5.2 \\ 
779984 & 2.5$^{+0.4}_{-0.7}$ $\times 10^{15}$ & 4.9$^{+0.6}_{-1.2}$ $\times 10^{13}$ & [150] & [7.0] & 3.1$^{+0.8}_{-0.6}$ $\times 10^{14}$ & 250$^{+50}_{-50}$ & [7.0] & 5.4$^{+1.1}_{-1.1}$ $\times 10^{14}$ & [150] & 9.0 \\ 
783350 & 5.3$^{+1.4}_{-1.2}$ $\times 10^{15}$ & 1.1$^{+0.2}_{-0.2}$ $\times 10^{14}$ & [150] & [6.5] & 1.5$^{+0.2}_{-0.2}$ $\times 10^{14}$ & [150] & [6.5] & 1.1$^{+0.2}_{-0.2}$ $\times 10^{15}$ & [150] & 7.2 \\ 
787212 & $^*$4.3$^{+0.4}_{-0.3}$ $\times 10^{16}$ & 5.1$^{+0.4}_{-0.6}$ $\times 10^{14}$ & 170$^{+20}_{-30}$ & [7.5] & 2.3$^{+0.4}_{-0.4}$ $\times 10^{15}$ & 190$^{+40}_{-30}$ & [7.5] & 8.0$^{+0.9}_{-0.6}$ $\times 10^{15}$ & [170] & 8.0 \\ 
792355 & 2.8$^{+1.1}_{-1.2}$ $\times 10^{15}$ & 5.8$^{+2.1}_{-2.4}$ $\times 10^{13}$ & [150] & [6.7] & <\,1.2 $\times 10^{14}$ & [150] & [6.7] & 4.2$^{+1.1}_{-0.8}$ $\times 10^{14}$ & [150] & 6.0 \\ 
800287 & 2.8$^{+0.5}_{-0.5}$ $\times 10^{16}$ & 6.1$^{+0.5}_{-0.7}$ $\times 10^{14}$ & [140] & [7.5] & 2.9$^{+0.6}_{-0.6}$ $\times 10^{15}$ & 130$^{+20}_{-10}$ & [7.5] & 9.6$^{+1.9}_{-1.9}$ $\times 10^{15}$ & [140] & 8.9 \\ 
800751 & 8.3$^{+1.5}_{-1.5}$ $\times 10^{15}$ & 1.7$^{+0.2}_{-0.2}$ $\times 10^{14}$ & 140$^{+30}_{-20}$ & [5.5] & 3.1$^{+0.3}_{-0.3}$ $\times 10^{14}$ & 120$^{+20}_{-30}$ & [5.5] & 1.6$^{+0.3}_{-0.3}$ $\times 10^{15}$ & [140] & 6.2 \\ 
865468A & 1.1$^{+0.2}_{-0.2}$ $\times 10^{17}$ & 2.2$^{+0.2}_{-0.1}$ $\times 10^{15}$ & 170$^{+50}_{-20}$ & [8.0] & 5.6$^{+1.1}_{-1.1}$ $\times 10^{15}$ & 180$^{+10}_{-10}$ & [8.0] & 2.3$^{+0.1}_{-0.2}$ $\times 10^{16}$ & [170] & 6.9 \\ 
876288 & 3.7$^{+1.5}_{-1.5}$ $\times 10^{15}$ & 1.1$^{+0.2}_{-0.2}$ $\times 10^{14}$ & [150] & [4.5] & <\,2.2 $\times 10^{14}$ & [150] & [4.5] & 7.0$^{+0.6}_{-0.5}$ $\times 10^{14}$ & [150] & 3.0 \\ 
881427C & 9.3$^{+1.5}_{-1.3}$ $\times 10^{16}$ & 1.6$^{+0.2}_{-0.1}$ $\times 10^{15}$ & 170$^{+30}_{-20}$ & [5.5] & 2.2$^{+0.6}_{-0.4}$ $\times 10^{15}$ & 120$^{+50}_{-30}$ & [5.5] & 1.3$^{+0.3}_{-0.3}$ $\times 10^{16}$ & [170] & 6.1 \\ 
G023.3891+00.1851 & 5.4$^{+1.8}_{-1.4}$ $\times 10^{15}$ & 1.1$^{+0.3}_{-0.2}$ $\times 10^{14}$ & 200$^{+70}_{-60}$ & [4.0] & 2.0$^{+0.4}_{-0.4}$ $\times 10^{14}$ & 170$^{+50}_{-40}$ & [4.0] & 7.2$^{+1.4}_{-1.4}$ $\times 10^{14}$ & [200] & 3.0 \\ 
G025.6498+01.0491 & 1.9$^{+0.3}_{-0.7}$ $\times 10^{16}$ & 3.3$^{+0.2}_{-1.1}$ $\times 10^{14}$ & [150] & [8.3] & 1.9$^{+0.5}_{-0.6}$ $\times 10^{15}$ & 250$^{+50}_{-70}$ & [8.3] & 1.7$^{+0.2}_{-0.2}$ $\times 10^{15}$ & [150] & 7.5 \\ 
G305.2017+00.2072A1 & 1.1$^{+0.3}_{-0.2}$ $\times 10^{16}$ & 1.9$^{+0.4}_{-0.2}$ $\times 10^{14}$ & 130$^{+40}_{-20}$ & [5.5] & 9.6$^{+1.9}_{-2.0}$ $\times 10^{14}$ & 190$^{+10}_{-40}$ & [5.5] & 1.6$^{+0.3}_{-0.3}$ $\times 10^{15}$ & [130] & 5.1 \\ 
G314.3197+00.1125 & 9.7$^{+1.7}_{-1.7}$ $\times 10^{15}$ & 1.7$^{+0.2}_{-0.2}$ $\times 10^{14}$ & [150] & [8.5] & 3.2$^{+0.4}_{-0.5}$ $\times 10^{14}$ & [150] & [8.5] & 1.7$^{+0.3}_{-0.3}$ $\times 10^{15}$ & [150] & 10.1 \\ 
G316.6412-00.0867 & 3.1$^{+0.4}_{-0.5}$ $\times 10^{16}$ & 5.4$^{+0.3}_{-0.6}$ $\times 10^{14}$ & 160$^{+20}_{-20}$ & [6.0] & 1.3$^{+0.1}_{-0.1}$ $\times 10^{15}$ & 130$^{+30}_{-10}$ & [6.0] & 6.7$^{+1.3}_{-1.3}$ $\times 10^{15}$ & [160] & 5.8 \\ 
G318.0489+00.0854B & 8.8$^{+1.6}_{-1.9}$ $\times 10^{15}$ & 1.6$^{+0.2}_{-0.3}$ $\times 10^{14}$ & [150] & [7.3] & 2.2$^{+0.2}_{-0.2}$ $\times 10^{14}$ & [150] & [7.3] & 1.0$^{+0.2}_{-0.2}$ $\times 10^{15}$ & [150] & 6.6 \\ 
G318.9480-00.1969A1 & $^*$5.4$^{+0.8}_{-0.6}$ $\times 10^{16}$ & 7.4$^{+0.6}_{-1.1}$ $\times 10^{14}$ & 150$^{+30}_{-10}$ & [6.3] & 1.2$^{+0.2}_{-0.1}$ $\times 10^{15}$ & 140$^{+20}_{-20}$ & [6.3] & 6.2$^{+1.2}_{-1.2}$ $\times 10^{15}$ & [150] & 5.4 \\ 
G323.7399-00.2617B2 & 1.7$^{+0.2}_{-0.2}$ $\times 10^{16}$ & 3.2$^{+0.3}_{-0.3}$ $\times 10^{14}$ & 120$^{+40}_{-20}$ & [5.5] & 6.2$^{+1.1}_{-0.6}$ $\times 10^{14}$ & 160$^{+30}_{-20}$ & [5.5] & 3.0$^{+0.4}_{-0.4}$ $\times 10^{15}$ & [120] & 5.2 \\ 
G326.4755+00.6947 & 2.1$^{+1.0}_{-0.7}$ $\times 10^{15}$ & 3.7$^{+1.5}_{-1.1}$ $\times 10^{13}$ & [150] & [6.7] & 1.5$^{+0.1}_{-0.1}$ $\times 10^{14}$ & [150] & [6.7] & 3.4$^{+0.7}_{-0.7}$ $\times 10^{14}$ & [150] & 7.7 \\ 
G326.6618+00.5207 & 5.2$^{+0.9}_{-0.9}$ $\times 10^{15}$ & 9.2$^{+1.2}_{-1.2}$ $\times 10^{13}$ & 150$^{+60}_{-40}$ & [5.5] & 2.0$^{+0.7}_{-0.1}$ $\times 10^{14}$ & 150$^{+80}_{-20}$ & [5.5] & 3.4$^{+0.9}_{-0.7}$ $\times 10^{14}$ & [150] & 5.2 \\ 
G327.1192+00.5103 & 2.6$^{+0.4}_{-0.4}$ $\times 10^{16}$ & 5.4$^{+0.3}_{-0.2}$ $\times 10^{14}$ & [150] & [8.0] & 1.3$^{+0.1}_{-0.2}$ $\times 10^{15}$ & 200$^{+20}_{-30}$ & [8.0] & 4.2$^{+0.8}_{-0.8}$ $\times 10^{15}$ & [150] & 8.7 \\ 
G343.1261-00.0623 & 3.0$^{+0.8}_{-0.5}$ $\times 10^{16}$ & 5.3$^{+1.2}_{-0.6}$ $\times 10^{14}$ & 170$^{+40}_{-60}$ & [8.5] & 3.7$^{+1.0}_{-0.7}$ $\times 10^{15}$ & 170$^{+50}_{-10}$ & [8.5] & 6.1$^{+0.7}_{-0.4}$ $\times 10^{15}$ & [170] & 10.0 \\ 
G345.5043+00.3480 & $^*$1.3$^{+0.2}_{-0.2}$ $\times 10^{17}$ & 1.1$^{+0.1}_{-0.1}$ $\times 10^{15}$ & 160$^{+60}_{-20}$ & [7.8] & 4.5$^{+1.1}_{-0.4}$ $\times 10^{15}$ & 200$^{+50}_{-30}$ & [7.8] & 1.1$^{+0.2}_{-0.2}$ $\times 10^{16}$ & [160] & 7.2 \\

\bottomrule
        \end{tabular}}
        \tablefoot{CH$_3$CN and HNCO column densities are found from column densities of CH$_3^{13}$CN and HN$^{13}$CO using the isotopologue ratios presented in Table \ref{tab:source_features}. Stars indicate the column densities found from fitting to the weakest clearly detected line ($S/N > 3$) of HNCO and CH$_3$CN by fixing the $T_{\rm ex}$ to the excitation temperature of CH$_3^{13}$CN. For HNCO, this is done when no measurement is possible for the $^{13}$C isotopologue due to line blending. For CH$_3$CN this is done either when an upper limit is found on the isotopologue of this species or the detected $^{13}$C isotopologue lines become more optically thick than the CH$_3$CN 12$_{10}$-11$_{10}$ line. In the latter case, the CH$_3$CN 12$_{10}$-11$_{10}$ line is used for the column density measurement. Where molecules are fitted by eye, the uncertainties on column densities come from the variation seen when the $T_{\rm ex}$ is changed from its lower limit to its upper limit. If the species are fitted using grid fitting the errors show the $2\sigma$ found from the reduced-$\chi^2$ error calculation. Square brackets around excitation temperatures show where the temperature is fixed in fitting. They are generally fixed to the excitation temperature of CH$_3^{13}$CN except for CH$_3$NH$_2$ where the temperature is fixed to 90\,K (temperature of CH$_3$NH$_2$ in G345.5043+00.3480 is fixed to 60\,K as this is the upper limit on $T_{\rm ex}$ for this source). Square brackets around FWHM show that it is fixed to the FWHM of CH$_3^{13}$CN. The FWHM values of HNC$^{13}$CO are fixed to the FWHM of single HNCO lines. The uncertainties on FWHM are ${\sim} 0.5$\,km\,s$^{-1}$.}
\end{table*}

\begin{table*}
\ContinuedFloat 
\Huge
\renewcommand{\arraystretch}{1.5}
    \caption{Continued.}
    \resizebox{\textwidth}{!}{\begin{tabular}{@{\extracolsep{1mm}}*{11}{c}}
          \toprule
          \toprule      
          & \multicolumn{1}{c}{HNCO} & \multicolumn{3}{c}{HN$^{13}$CO}&\multicolumn{3}{c}{NH$_2$CHO} &\multicolumn{3}{c}{CH$_3$NH$_2$}\\
          \cmidrule{2-2} \cmidrule{3-5} \cmidrule{6-8} \cmidrule{9-11}
        Source & $N (\rm cm^{-2})$ & $N (\rm cm^{-2})$ &  $T_{\rm ex} (\rm K)$ & FWHM (km s$^{-1}$) &  $N (\rm cm^{-2})$ &  $T_{\rm ex} (\rm K)$ & FWHM (km s$^{-1}$) &  $N (\rm cm^{-2})$ &  $T_{\rm ex} (\rm K)$ & FWHM (km s$^{-1}$) \\
\midrule   


101899 & 1.1$^{+0.2}_{-0.2}$ $\times 10^{17}$ & 2.5$^{+0.4}_{-0.3}$ $\times 10^{15}$ & 150$^{+30}_{-30}$ & 10.3 & 7.0$^{+1.4}_{-1.7}$ $\times 10^{15}$ & 370$^{+130}_{-100}$ & [9.5] & 4.3$^{+0.9}_{-1.0}$ $\times 10^{15}$ & 70$^{+30}_{-40}$ & [9.5] \\ 
126348 & 2.2$^{+0.3}_{-0.4}$ $\times 10^{16}$ & 4.6$^{+0.3}_{-0.5}$ $\times 10^{14}$ & 240$^{+30}_{-40}$ & 8.5 & 4.2$^{+1.1}_{-1.7}$ $\times 10^{14}$ & [170] & [7.5] & <\,2.8 $\times 10^{15}$ & [90] & [7.5] \\ 
615590 & 9.8$^{+2.3}_{-2.3}$ $\times 10^{16}$ & 1.4$^{+0.3}_{-0.3}$ $\times 10^{15}$ & [150] & 4.1 & 2.3$^{+0.3}_{-0.4}$ $\times 10^{15}$ & [150] & [4.5] & <\,4.2 $\times 10^{15}$ & [90] & [4.5] \\ 
644284A & <\,6.4 $\times 10^{16}$ & <\,9.3 $\times 10^{14}$ & [150] & 7.3 & 6.4$^{+1.9}_{-0.9}$ $\times 10^{14}$ & [150] & [6.7] & <\,4.7 $\times 10^{15}$ & [90] & [6.7] \\ 
693050 & 5.0$^{+0.7}_{-1.0}$ $\times 10^{16}$ & 8.1$^{+0.7}_{-1.4}$ $\times 10^{14}$ & [150] & 3.9 & 1.6$^{+0.3}_{-0.4}$ $\times 10^{15}$ & [160] & [4.7] & <\,2.0 $\times 10^{15}$ & [90] & [4.7] \\ 
705768 & <\,1.2 $\times 10^{16}$ & <\,2.0 $\times 10^{14}$ & [140] & 6.8 & 3.0$^{+0.7}_{-0.5}$ $\times 10^{14}$ & [140] & [6.7] & <\,3.9 $\times 10^{15}$ & [90] & [6.7] \\ 
707948 & 6.6$^{+1.1}_{-0.9}$ $\times 10^{17}$ & 1.1$^{+0.1}_{-0.1}$ $\times 10^{16}$ & 250$^{+30}_{-30}$ & 9.5 & 2.9$^{+1.4}_{-0.8}$ $\times 10^{16}$ & 350$^{+150}_{-100}$ & [8.6] & 1.5$^{+0.1}_{-0.2}$ $\times 10^{16}$ & 60$^{+20}_{-10}$ & [8.6] \\ 
717461A & 3.2$^{+1.0}_{-0.5}$ $\times 10^{16}$ & 5.7$^{+1.6}_{-0.4}$ $\times 10^{14}$ & 260$^{+40}_{-40}$ & 6.9 & 7.7$^{+2.3}_{-2.0}$ $\times 10^{14}$ & [150] & [6.3] & <\,2.8 $\times 10^{15}$ & [90] & [6.3] \\ 
721992 & <\,9.4 $\times 10^{15}$ & <\,1.7 $\times 10^{14}$ & [90] & 3.2 & <\,1.6 $\times 10^{14}$ & [90] & [3.5] & <\,2.7 $\times 10^{15}$ & [90] & [3.5] \\ 
724566 & 4.1$^{+1.5}_{-1.0}$ $\times 10^{16}$ & 7.3$^{+2.5}_{-1.4}$ $\times 10^{14}$ & 200$^{+100}_{-40}$ & 7.8 & 7.3$^{+1.7}_{-1.1}$ $\times 10^{14}$ & [220] & [6.0] & <\,5.3 $\times 10^{15}$ & [90] & [6.0] \\ 
732038 & <\,2.3 $\times 10^{16}$ & <\,4.3 $\times 10^{14}$ & [150] & 6.5 & 7.3$^{+1.3}_{-1.1}$ $\times 10^{14}$ & [150] & [6.0] & <\,5.5 $\times 10^{15}$ & [90] & [6.0] \\ 
744757A & 2.9$^{+0.6}_{-0.6}$ $\times 10^{16}$ & 5.0$^{+0.9}_{-0.9}$ $\times 10^{14}$ & [130] & 7.5 & 8.1$^{+5.1}_{-1.1}$ $\times 10^{14}$ & [130] & [5.7] & 2.9$^{+0.6}_{-0.9}$ $\times 10^{15}$ & 80$^{+40}_{-20}$ & [5.7] \\ 
767784 & 2.1$^{+0.4}_{-0.4}$ $\times 10^{16}$ & 4.1$^{+0.5}_{-0.5}$ $\times 10^{14}$ & 120$^{+20}_{-20}$ & 4.3 & 3.5$^{+0.7}_{-0.4}$ $\times 10^{15}$ & 320$^{+80}_{-80}$ & [4.0] & 5.7$^{+1.3}_{-0.5}$ $\times 10^{15}$ & 100$^{+20}_{-10}$ & [4.0] \\ 
778802 & 1.6$^{+0.4}_{-0.4}$ $\times 10^{16}$ & 2.8$^{+0.6}_{-0.6}$ $\times 10^{14}$ & 240$^{+60}_{-50}$ & 6.4 & 1.8$^{+0.4}_{-0.5}$ $\times 10^{14}$ & [150] & [7.0] & <\,3.1 $\times 10^{15}$ & [90] & [7.0] \\ 
779523 & 1.2$^{+0.3}_{-0.4}$ $\times 10^{16}$ & 2.2$^{+0.5}_{-0.7}$ $\times 10^{14}$ & 200$^{+50}_{-90}$ & 8.0 & 2.6$^{+0.5}_{-0.5}$ $\times 10^{14}$ & [150] & [8.4] & <\,3.2 $\times 10^{15}$ & [90] & [8.4] \\ 
779984 & <\,6.3 $\times 10^{15}$ & <\,1.2 $\times 10^{14}$ & [150] & 6.8 & 1.8$^{+0.4}_{-0.4}$ $\times 10^{14}$ & [150] & [7.0] & <\,2.6 $\times 10^{15}$ & [90] & [7.0] \\ 
783350 & 9.8$^{+2.5}_{-3.0}$ $\times 10^{15}$ & 2.0$^{+0.4}_{-0.6}$ $\times 10^{14}$ & 150$^{+60}_{-60}$ & 6.9 & 1.1$^{+0.4}_{-0.2}$ $\times 10^{15}$ & 340$^{+120}_{-110}$ & [6.5] & <\,2.6 $\times 10^{15}$ & [90] & [6.5] \\ 
787212 & 1.3$^{+0.2}_{-0.2}$ $\times 10^{17}$ & 2.5$^{+0.3}_{-0.3}$ $\times 10^{15}$ & [170] & 8.0 & 7.2$^{+1.2}_{-1.0}$ $\times 10^{15}$ & 300$^{+120}_{-70}$ & [7.5] & 3.7$^{+0.9}_{-0.6}$ $\times 10^{15}$ & [90] & [7.5] \\ 
792355 & <\,1.1 $\times 10^{16}$ & <\,2.4 $\times 10^{14}$ & [150] & 8.0 & 2.6$^{+0.5}_{-0.5}$ $\times 10^{14}$ & [150] & [6.7] & <\,2.4 $\times 10^{15}$ & [90] & [6.7] \\ 
800287 & 3.2$^{+1.3}_{-0.6}$ $\times 10^{16}$ & 6.9$^{+2.6}_{-0.7}$ $\times 10^{14}$ & [140] & 8.2 & 2.2$^{+0.2}_{-0.3}$ $\times 10^{15}$ & [140] & [7.5] & <\,8.0 $\times 10^{15}$ & [90] & [7.5] \\ 
800751 & $^*$5.9$^{+0.7}_{-0.5}$ $\times 10^{15}$ & -- & -- & -- & 2.1$^{+0.6}_{-0.5}$ $\times 10^{15}$ & 400$^{+100}_{-150}$ & [5.5] & <\,3.0 $\times 10^{15}$ & [90] & [5.5] \\ 
865468A & 3.1$^{+0.5}_{-0.6}$ $\times 10^{17}$ & 6.1$^{+0.6}_{-0.7}$ $\times 10^{15}$ & 170$^{+30}_{-20}$ & 8.5 & 4.3$^{+1.2}_{-0.9}$ $\times 10^{15}$ & [170] & [8.0] & 3.0$^{+0.3}_{-0.2}$ $\times 10^{16}$ & 100$^{+20}_{-10}$ & [8.0] \\ 
876288 & <\,1.3 $\times 10^{16}$ & <\,3.7 $\times 10^{14}$ & [150] & 5.8 & 3.2$^{+1.4}_{-1.6}$ $\times 10^{14}$ & [150] & [4.5] & 2.7$^{+0.6}_{-0.7}$ $\times 10^{15}$ & 70$^{+30}_{-30}$ & [4.5] \\ 
881427C & 2.0$^{+0.3}_{-0.3}$ $\times 10^{17}$ & 3.4$^{+0.3}_{-0.4}$ $\times 10^{15}$ & 250$^{+50}_{-40}$ & 8.5 & 6.3$^{+1.9}_{-1.2}$ $\times 10^{15}$ & 400$^{+100}_{-90}$ & [5.5] & 4.8$^{+2.5}_{-2.0}$ $\times 10^{15}$ & 70$^{+30}_{-30}$ & [5.5] \\ 
G023.3891+00.1851 & 1.1$^{+0.3}_{-0.3}$ $\times 10^{16}$ & 2.4$^{+0.4}_{-0.5}$ $\times 10^{14}$ & [200] & 3.9 & 1.3$^{+0.4}_{-0.2}$ $\times 10^{15}$ & 420$^{+80}_{-60}$ & [4.0] & <\,2.3 $\times 10^{15}$ & [90] & [4.0] \\ 
G025.6498+01.0491 & 2.8$^{+0.8}_{-0.6}$ $\times 10^{16}$ & 5.0$^{+1.1}_{-0.6}$ $\times 10^{14}$ & [150] & 8.5 & 4.2$^{+0.7}_{-0.3}$ $\times 10^{15}$ & 250$^{+70}_{-50}$ & [8.3] & <\,3.6 $\times 10^{15}$ & [90] & [8.3] \\ 
G305.2017+00.2072A1 & 1.5$^{+0.3}_{-0.3}$ $\times 10^{16}$ & 2.6$^{+0.4}_{-0.4}$ $\times 10^{14}$ & [130] & 5.8 & 2.1$^{+0.3}_{-0.5}$ $\times 10^{15}$ & 230$^{+70}_{-50}$ & [5.5] & <\,3.0 $\times 10^{15}$ & [90] & [5.5] \\ 
G314.3197+00.1125 & <\,2.1 $\times 10^{16}$ & <\,3.6 $\times 10^{14}$ & [150] & 10.0 & 7.2$^{+1.8}_{-1.4}$ $\times 10^{14}$ & [150] & [8.5] & <\,4.8 $\times 10^{15}$ & [90] & [8.5] \\ 
G316.6412-00.0867 & 4.3$^{+0.9}_{-0.6}$ $\times 10^{16}$ & 7.5$^{+1.2}_{-0.6}$ $\times 10^{14}$ & [160] & 6.3 & 3.5$^{+1.1}_{-0.8}$ $\times 10^{15}$ & 400$^{+100}_{-140}$ & [6.0] & 3.5$^{+0.3}_{-0.3}$ $\times 10^{15}$ & 60$^{+10}_{-10}$ & [6.0] \\ 
G318.0489+00.0854B & 3.9$^{+1.1}_{-1.1}$ $\times 10^{16}$ & 7.0$^{+1.8}_{-1.8}$ $\times 10^{14}$ & 170$^{+60}_{-60}$ & 7.7 & 7.0$^{+1.8}_{-1.1}$ $\times 10^{14}$ & [150] & [7.3] & <\,2.7 $\times 10^{15}$ & [90] & [7.3] \\ 
G318.9480-00.1969A1 & 1.4$^{+0.3}_{-0.3}$ $\times 10^{17}$ & 2.4$^{+0.3}_{-0.4}$ $\times 10^{15}$ & 170$^{+20}_{-40}$ & 7.0 & 9.3$^{+2.5}_{-1.0}$ $\times 10^{15}$ & 450$^{+50}_{-70}$ & [6.3] & 3.4$^{+0.3}_{-0.3}$ $\times 10^{15}$ & [90] & [6.3] \\ 
G323.7399-00.2617B2 & 5.5$^{+1.1}_{-1.1}$ $\times 10^{16}$ & 1.0$^{+0.2}_{-0.2}$ $\times 10^{15}$ & 130$^{+40}_{-30}$ & 5.7 & 4.4$^{+0.2}_{-0.2}$ $\times 10^{15}$ & 280$^{+50}_{-60}$ & [5.5] & <\,3.0 $\times 10^{15}$ & [90] & [5.5] \\ 
G326.4755+00.6947 & <\,7.9 $\times 10^{15}$ & <\,1.4 $\times 10^{14}$ & [150] & 8.0 & 2.0$^{+0.7}_{-0.7}$ $\times 10^{14}$ & [150] & [6.7] & <\,2.5 $\times 10^{15}$ & [90] & [6.7] \\ 
G326.6618+00.5207 & 2.1$^{+0.7}_{-0.4}$ $\times 10^{16}$ & 3.7$^{+1.2}_{-0.6}$ $\times 10^{14}$ & 180$^{+50}_{-50}$ & 5.0 & 3.5$^{+0.3}_{-0.3}$ $\times 10^{15}$ & 420$^{+80}_{-70}$ & [5.5] & 2.6$^{+0.3}_{-0.3}$ $\times 10^{15}$ & 90$^{+20}_{-10}$ & [5.5] \\ 
G327.1192+00.5103 & 1.1$^{+0.2}_{-0.3}$ $\times 10^{17}$ & 2.4$^{+0.4}_{-0.5}$ $\times 10^{15}$ & 190$^{+50}_{-70}$ & 9.0 & 1.1$^{+0.2}_{-0.1}$ $\times 10^{16}$ & 310$^{+70}_{-70}$ & [8.0] & <\,7.5 $\times 10^{15}$ & [90] & [8.0] \\ 
G343.1261-00.0623 & 1.6$^{+0.3}_{-0.4}$ $\times 10^{17}$ & 2.8$^{+0.5}_{-0.5}$ $\times 10^{15}$ & 200$^{+50}_{-60}$ & 7.5 & 8.8$^{+1.8}_{-0.4}$ $\times 10^{15}$ & 240$^{+50}_{-40}$ & [8.5] & <\,5.3 $\times 10^{15}$ & [90] & [8.5] \\ 
G345.5043+00.3480 & 2.1$^{+0.3}_{-0.5}$ $\times 10^{17}$ & 3.8$^{+0.3}_{-0.8}$ $\times 10^{15}$ & 190$^{+20}_{-60}$ & 8.3 & 1.7$^{+0.2}_{-0.2}$ $\times 10^{16}$ & 350$^{+60}_{-70}$ & [7.8] & 7.0$^{+1.2}_{-1.2}$ $\times 10^{15}$ & [60] & [7.8] \\

\bottomrule
        \end{tabular}}
\end{table*}

Most of the objects analyzed here have been studied by \cite{vanGelder2022} for CH$_3$OH and its isotopologues, see also that paper for description of the data and analysis. Two spectral windows are used for each source with frequencies of ${\sim} 217.00-218.87$\,GHz and ${\sim} 219.07-220.95$\,GHz. The frequency setting covers transitions of many N-bearing species such as CH$_{3}$CN 12$_K$-11$_K$ for $K$ going from 0 to 11. The spectral resolution of the data used here is ${\sim 0.7}$\,km\,s$^{-1}$. All sources have line widths of ${\gtrsim 3}$\,km\,s$^{-1}$, thus, the lines are spectrally resolved. The data were pipeline calibrated and imaged with the Common Astronomy Software Applications (CASA) package version 5.6.1 (\citealt{McMullin2007}). The observational parameters for the line rich sources used in this paper are presented in Table \ref{tab:obs_params}, whereas Table \ref{tab:source_features} lists source properties.

The spectrum for each source is extracted from the peak pixel of the CH$_3$CN 12$_4$-11$_4$ integrated intensity map. The reason for this choice is that here we focus on the N-bearing complex organics emission and CH$_3$CN is used as a proxy for overall COM emission. An exception to this is source 787212, where the spectrum is extracted ${\sim} 1$ beam off-source to decrease line blending and infall-related signatures (see Table \ref{tab:obs_params} for the coordinates of the pixel at which spectra are extracted). It is important to note that the spectrum extracted from a pixel contains the information from the beam and thus, the column densities reported in this work are the column densities within one beam. 

This is different from what is done in \cite{vanGelder2022} where the spectra are extracted from the peak pixel of the continuum. The choice in \cite{vanGelder2022} is based on the fact that continuum fluxes of many line-poor sources are an important part of the discussion in that study. Moreover, some of the high mass objects considered here and in \cite{vanGelder2022} exist within clusters. \cite{vanGelder2022} studied all cores within a cluster whereas here we only look at the brightest source in COMs. Therefore, we only extract the spectrum from the brightest pixel in the moment zero map of CH$_3$CN 12$_4$-11$_4$ line. Most sources have similar peak positions for CH$_3$CN and continuum within half a beam size (${\sim} 0.5\arcsec$). Only source 693050 has more than half a beam difference between these two peaks. This source was particularly discussed in \cite{vanGelder2022} as one having highly optically thick dust continuum blocking the methanol emission on-source with only a ring of emission observed. The names used for the various sources in this paper match the names in \cite{vanGelder2022} for the sources that overlap between the two works. Figure \ref{fig:spectra} shows the spectra of six line rich sources studied here in one of the spectral windows used ($\nu {\sim} 219-221$\,GHz). Various O- and N-bearing COMs and simple molecules have been labeled to indicate the major features that are present in these spectra.

\subsection{Spectral modeling}
\label{sec:spectral_modelling}

This work considers six N-bearing species in 37 MYSOs covered in the observed spectral windows. These molecules are CH$_3$CN, HNCO, NH$_2$CHO, C$_2$H$_5$CN, C$_2$H$_3$CN and CH$_3$NH$_2$. Moreover, we include methanol in this work as a benchmark for comparison between N- and O-bearing species. The isotopologues of the most abundant and bright species such as HNCO, CH$_3$CN and CH$_3$OH are also considered for a more accurate derivation of the column densities of the main isotopologues. Table \ref{tab:lines} shows the transitions of these molecules available in the data.

The column density, excitation temperature and full width at half maximum (FWHM) of each molecule are measured by fitting the spectrum for each source as a whole using the CASSIS\footnote{\url{http://cassis.irap.omp.eu/}} spectral analysis tool (\citealt{Vastel2015}) assuming local thermodynamics equilibrium (LTE). The procedure here is similar to the grid fitting and fit by eye method of \cite{Nazari2021} (also see \citealt{vanGelder2020}). The spectroscopic data for each molecule are obtained either from the Jet Propulsion Laboratory database (JPL; \citealt{Pickett1998}) or the Cologne Database for Molecular Spectroscopy (CDMS; \citealt{Muller2001}; \citealt{Muller2005}). For more information on the spectroscopy see Appendix \ref{sec:spec_data} and Table \ref{tab:lines}.

The grid fitting method works if there are enough relatively unblended lines of a species present in the data. Therefore, this method is used for C$_2$H$_3$CN, C$_2$H$_5$CN and CH$_3^{18}$OH in most sources. The fits are then carefully inspected by eye and if the grid fitting fails to provide a good fit they are fitted using the by-eye method (CH$_3^{18}$OH for most sources was done by eye). The rest of the species (HN$^{13}$CO, CH$_3^{13}$CN, NH$_2$CHO, CH$_3$NH$_2$ and $^{13}$CH$_3$OH) are fitted by eye. This is because these species mainly have blended lines and hence it is not reliable to blindly use a grid of column densities and excitation temperatures and assign the model with the lowest $\chi^{2}$ as the best fit model. Instead it is more beneficial to look at each spectrum individually and find the best-fit model. 

In the grid fitting method $N$ is generally varied from 10$^{13}$\,cm$^{-2}$ to 10$^{17}$\,cm$^{-2}$ with spacing of 0.1 in logarithmic space, $T_{\rm ex}$ (if enough lines with a large range of E$_{\rm up}$ are covered) is varied from 10\,K to 300\,K with spacing of 10\,K and FWHM (if fitted) is varied from 3\,km s$^{-1}$ to 11\,km s$^{-1}$ with spacing of 0.1\,km s$^{-1}$. The 2\,$\sigma$ uncertainties on these values are calculated using the reduced-$\chi^{2}$ error calculation of the grid which in turn depends on the number of free parameters (usually two) in the fitting process. If this calculation gives lower or upper uncertainties smaller than 20$\%$, an uncertainty of 20$\%$ is assumed to avoid any error underestimation. This 20$\%$ is to account for some systematic errors and is chosen based on the typical errors found from the fit-by-eye method (see below for more discussion on the error calculation).

When fitting by eye for molecules with very blended lines, the FWHM is fixed to the FWHM of single CH$_3$CN lines, except for HN$^{13}$CO where they are fixed to the FWHM of single HNCO lines. Then the temperature is fixed to the highest temperature possible based on the data available for partition functions (usually 300\,K, except for NH$_2$CHO where the partition function is available up to 500\,K) and the column density is varied until a good model is found. If no good models can be fitted for that temperature the temperature is decreased until a suitable model is fitted to the spectrum. This gives the upper limit on the temperature. The same is done to find the lower limit on the temperature but now starting from a low temperature (usually 50\,K unless models with lower temperatures fit) and increasing it. The best-fit model parameters fall between the upper and lower limits found for $T$ and $N$. Typically the uncertainties on the temperatures are ${\sim} 50$\,K (see Table \ref{tab:results}).  

\begin{figure*}
    \centering
    \includegraphics[width=17cm]{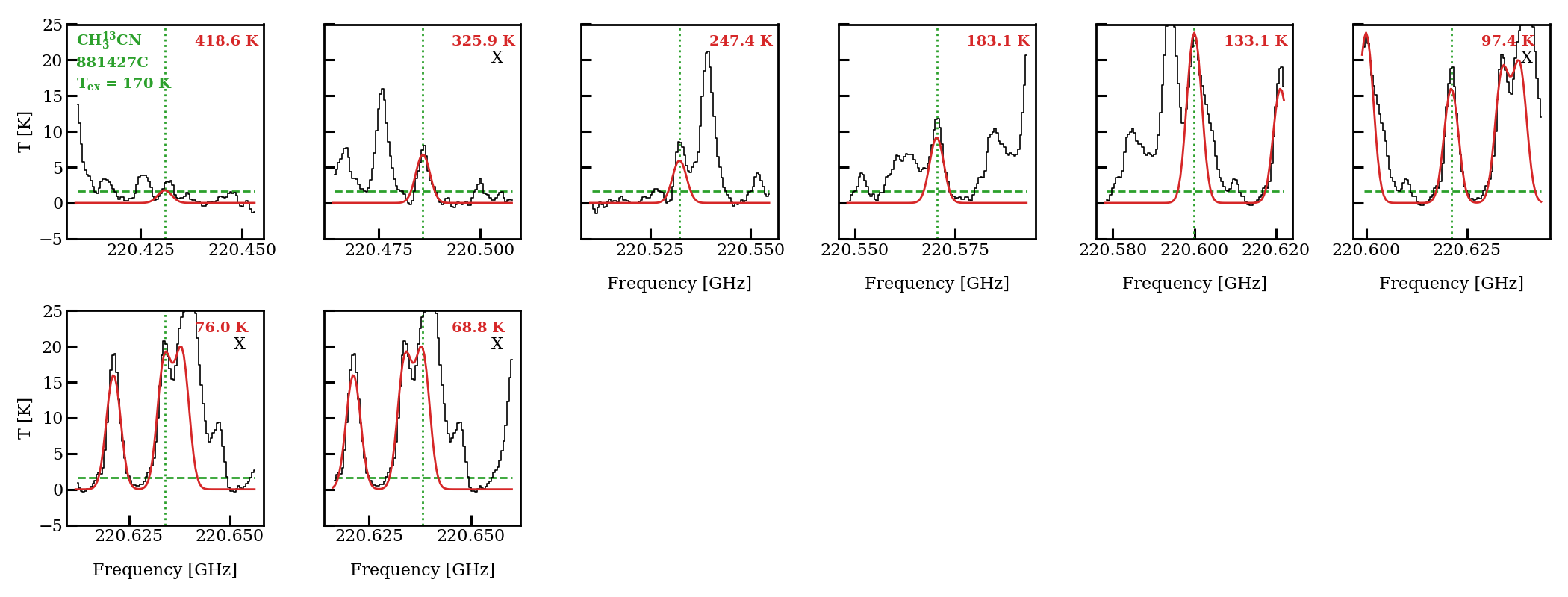}
    \caption{The best model for CH$_3^{13}$CN of the source 881427C (red), which has narrow lines with FWHM = 5.5\,km\,s$^{-1}$ on top of the data (black). In this figure only lines with $A_{\rm ij} > 10^{-6}$\,s$^{-1}$ and $E_{\rm up}< 500$\,K are shown. The detected lines that are mainly considered for the fits are marked with an `X' on the top right. The dashed line shows the 3$\sigma$ level.}
    \label{fig:881427_CH3C-13-N}
\end{figure*}

\begin{figure*}
    \centering
    \includegraphics[width=17cm]{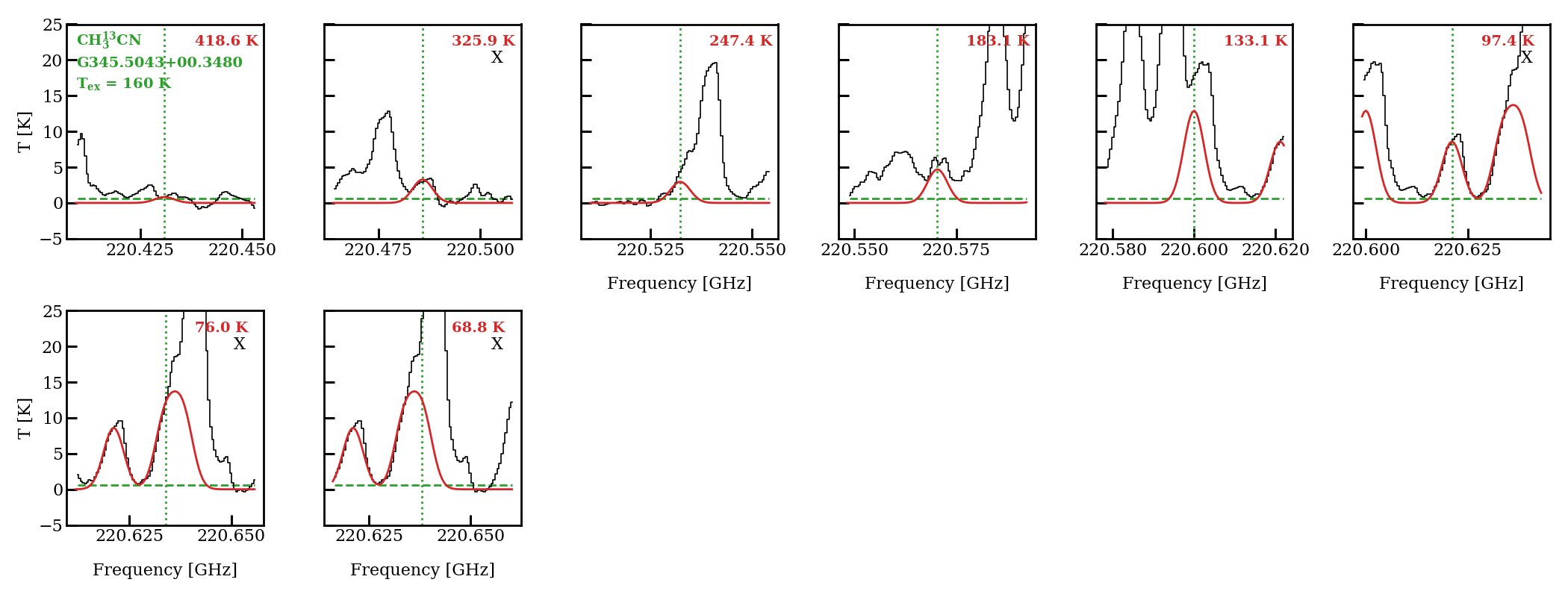}
    \caption{The same as Fig. \ref{fig:881427_CH3C-13-N} but for source G345.5043+00.3480 with broad lines (FWHM = 7.8\,km\,s$^{-1}$).}
    \label{fig:G345_CH3C-13-N}
\end{figure*}

Figures \ref{fig:881427_CH3C-13-N} and \ref{fig:G345_CH3C-13-N} show the best-fit models of CH$_3^{13}$CN for two sources, 881427C and G345.5043+00.3480. The best-fit models for the other molecules are shown in Appendix \ref{sec:add_plots}. G345.5043+00.3480 is one of the most line-rich sources in our sample with large FWHM (${\sim} 8$\,km\,s$^{-1}$) and many blended lines whereas source 881427C is a less line-rich source with narrow (${\sim}5.5$\,km\,s$^{-1}$) and less blended lines. These two are chosen as representative of a difficult source for line identification, fitting and continuum subtraction (G345.5043+00.3480) and a simpler one (881427C) where the source has fewer blended lines. Although the lines are blended for many sources, the different velocity components are usually not clearly separated for the N-bearing COMs or $^{18}$O isotopologue of methanol. Hence here we fit the data with a single component, while \cite{vanGelder2022D_H} fit a two-component profile in their work on deuteration of methanol for three of the sources overlapping with this work.  

The column densities of HNCO and CH$_3$CN are found from their $^{13}$C isotopologues (HN$^{13}$CO and CH$_3^{13}$CN). There are no transitions of $^{13}$CH$_3$CN and CH$_3$C$^{15}$N covered by the available frequency range. There is only one transition of H$^{15}$NCO in the frequency range but it is not detected for any of the sources. There are no transitions of HNC$^{18}$O in the data. Moreover, the column density of methanol is found from CH$_3^{18}$OH when possible (the lines are detected and not too blended) and otherwise the $^{13}$CH$_3$OH isotopologue is used (the same approach as \citealt{vanGelder2022}). These sources are indicated by a star in Table \ref{tab:results_methanol}. The isotopologue ratios of  $^{16}$O/$^{18}$O and $^{12}$C/$^{13}$C are calculated using the equations in \cite{Wilson1994} and \cite{Milam2005} and the distances to the galactic center presented in Table \ref{tab:source_features}. The typical uncertainties on the distances are ${\sim}0.5$\,kpc. This uncertainty on distances together with the uncertainty on the slope in the relations of the isotopologue ratios with distance to the galactic center given in \cite{Wilson1994} and \cite{Milam2005} have been used to calculate the final uncertainties on the isotopologue ratios. The reason that we only include the uncertainty on the slope and not that on the y-intercept is that the trend seen in the isotopologue ratio as a function of distance to the galactic center is of interest here. The uncertainty on the y-intercept depends on the source sample used in those particular studies. Next, the uncertainties on the isotopologue ratios are propagated so that they are included in the uncertainties of the column densities of CH$_3$CN, CH$_3$OH and HNCO. If the uncertainty on the y-intercept were included as well, the uncertainties on column densities would increase by ${\sim}10-30\%$. For the rest of the species the main isotopologue is used.

The temperature is fitted (both when running a grid and when fitting by eye) only when there are at least two (and no anti-coincidences) transitions of a molecule detected with a range of upper energy levels. For example, CH$_3^{13}$CN has several lines with an $E_{\rm up}$ range of ${\sim} 70-300$\,K (Figures \ref{fig:881427_CH3C-13-N} and \ref{fig:G345_CH3C-13-N}) and hence, it is possible to derive reliable excitation temperatures. Moreover, HN$^{13}$CO also has a few transitions with a range of upper energy levels (${\sim} 60-400$\,K) and it is possible to constrain the temperature using the fit-by-eye method (Figures \ref{fig:881427_HNC-13-O} and \ref{fig:G345_HNC-13-O}). CH$_3$NH$_2$, C$_2$H$_3$CN and CH$_3^{18}$OH also have enough transitions for an excitation temperature measurement (see Figures \ref{fig:plots_CH3O18H}-\ref{fig:bad_CH3NH2_upper} for examples of how the excitation temperatures of CH$_3^{18}$OH and CH$_3$NH$_2$ are measured in source 767784). On the other hand, C$_2$H$_5$CN has three strong lines but all have upper energy levels of ${\sim} 140$\,K and therefore, the temperature is fixed for this molecule in all sources to the temperature of CH$_3^{13}$CN. 

\begin{figure*}
    \centering
    \includegraphics[width=0.7\textwidth]{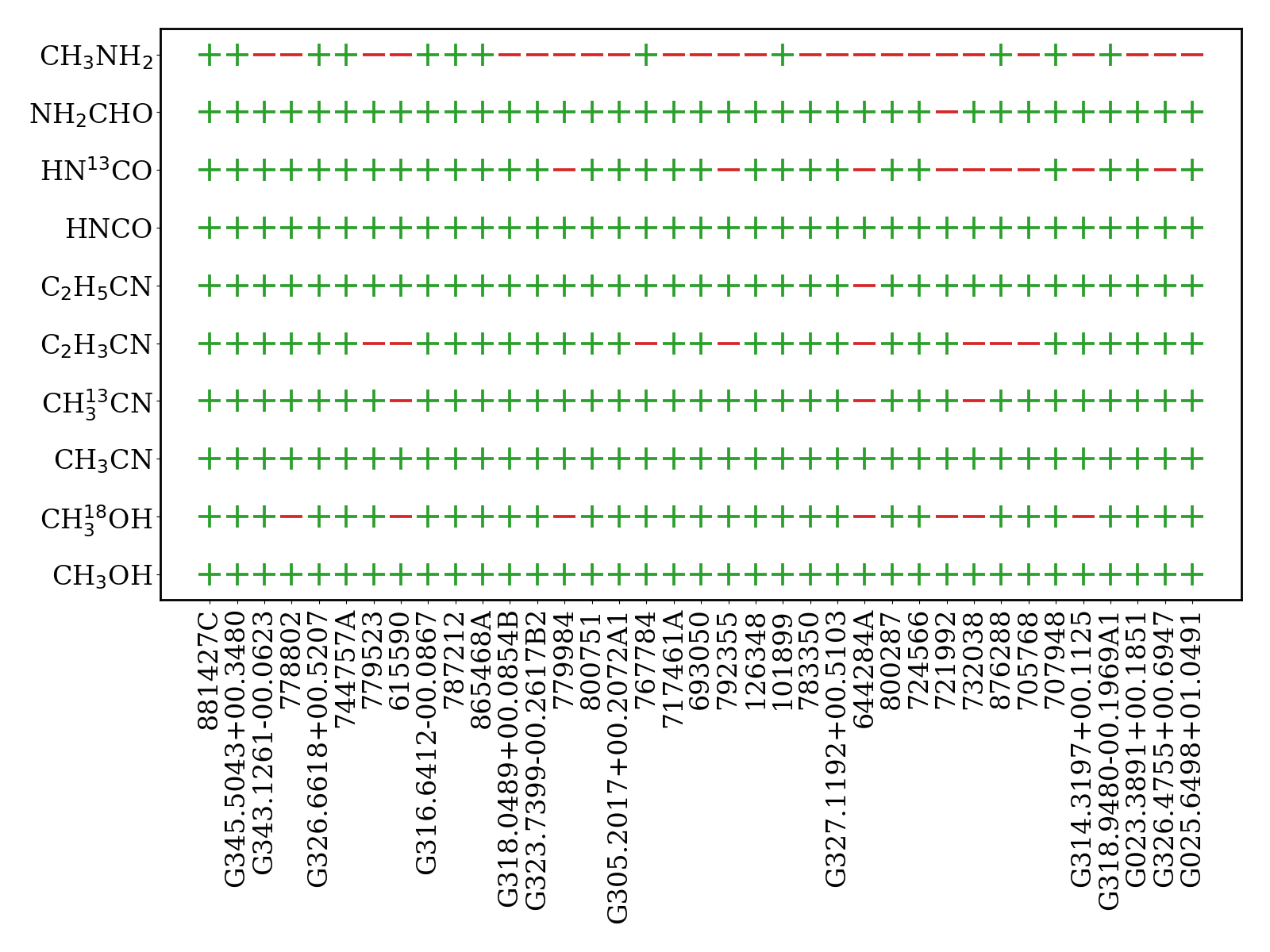}
    \caption{Summary of the detected species toward the sources. Plus signs show the detected and tentatively detected molecules and minus signs show the non-detections. The sources are ordered from left to right from the lowest to highest distance. The tentative detection of CH$_3$NH$_2$ in sources 707948 and 787212 is uncertain due to line blending and more data is needed for a more robust confirmation.} 
    \label{fig:detections}
\end{figure*}

Formamide has one rotational line originating from the $v=0$ ground state vibrational level with a low upper energy level ($E_{\rm up}=60.8$\,K) but some sources show detection of another line of this molecule in the vibrationally excited state $v_{12}=1$ with $E_{\rm up}=476.5$\,K. It is assumed that this line is dominated by formamide. Other plausible molecules with a transition close to this line are C$_2$H$_3$CN and c-C$_3$HD. Therefore, first a fit to C$_2$H$_3$CN is found and the fit for NH$_2$CHO is done on top of that. We found that there is no contribution from c-C$_3$HD to this line. This is because the lines with transitions from c-C$_3$H$_2$ either are not detected in our sources or if they are, the fit to c-C$_3$H$_2$ results in low column densities of this molecule. Thus, the expected column density of c-C$_3$HD using a high D/H ratio of 0.01 will be negligible and there will be no contribution from c-C$_3$HD to the line. Hence, we attribute the line at frequency of ${\sim 218.180}$\,GHz mainly to the vibrationally excited state of formamide.

When the vibrationally excited line is not detected, the temperature is fixed to the temperature of CH$_3^{13}$CN for that source and only column density is fitted. However, when both lines are detected the excitation temperature of formamide is constrained using the fit-by-eye method (see Fig. \ref{fig:plots_NH2CHO} for an example showing how the temperature of formamide was measured for source 767784). The vibrationally excited line of formamide could be radiatively excited instead of collisionally excited. Hence, fitting the temperature for formamide could result in a mix of kinetic and radiative temperature. To deal with this potential issue one can find the column density of formamide by fixing the temperature to those of CH$_3^{13}$CN and fitting each line separately. However, the resulting column densities with this method will change by an approximately constant factor for all ALMAGAL sources which would not significantly change the scatter in column density ratios including formamide discussed in Sect. \ref{sec:low-to-high}. Therefore, the main conclusions of this work are not affected by the method used for finding column densities of formamide. In this work we choose to fit for the temperature using both formamide lines when possible to get approximate measurements of the temperatures that formamide traces. 

A final complication for NH$_2$CHO is that the line with the low $E_{\rm up}$ is also blended with NH$_2$CN, C$_2$H$_5$OH and C$_2$H$_3$CN. Therefore, first a best fit model is found for these molecules and then the best fit for NH$_2$CHO is found by fitting on top of the already existing models of NH$_2$CN, C$_2$H$_5$OH and C$_2$H$_3$CN. Given that NH$_2$CN and C$_2$H$_5$OH are not targets of this paper, only approximate column densities are found for these species by fixing the temperature to the temperature of CH$_3^{13}$CN in each source. It is notable that fixing the temperature does not change the derived column densities significantly and hence reliable column densities can be found.

In the procedure of fitting the spectra for various species the source velocities are fixed. These values are found for each source using the CH$_3$CN ladder mainly $K=0$ to $K=5$ lines and are presented in Table \ref{tab:source_features}. For some sources and molecules the sources velocities are slightly shifted and hence adjusted during the fitting process. A beam dilution factor of one is assumed given that the exact source sizes are not known. In reality the beam dilution factor can be different from unity. However, since our results focus only on column density ratios, the precise source size does not matter as long as lines are optically thin (\citealt{vanGelder2020}; \citealt{Nazari2021}).

\section{Results}
\label{sec:results}
\subsection{Detection statistics}
\label{sec:detections}

Figure \ref{fig:detections} presents the molecules detected in each of the sources. Plus signs show the (tentatively) detected molecules and minus signs show the non-detected molecules. The sources in this figure are ordered by their distances from the Hi-Gal or RMS surveys (see Table \ref{tab:source_features}; \citealt{Lumsden2013}; \citealt{Mege2021}) where 881427C is the source with the closest distance (1.5\,kpc) and G025.6498+01.0491 is the source with the farthest distance (12.20\,kpc).

CH$_3$OH, CH$_3$CN and HNCO are detected in all sources because the sample is chosen so that the sources are all line-rich. The isotopologues of these species are detected in ${\gtrsim} 76\%$ of the sources. C$_2$H$_5$CN and NH$_2$CHO are (tentatively) detected in all sources except one and C$_2$H$_3$CN is (tentatively) detected in ${\sim} 78\%$ of all sources.

From the molecules shown in Fig. \ref{fig:detections}, CH$_3$NH$_{2}$ is the least (tentatively) detected molecule, in only 12 sources (in ${\sim} 32\%$ of the sources). This molecule has intrinsically weak lines ($A_{\rm ij} \lesssim 5 \times 10^{-5}$\,s$^{-1}$; see Table \ref{tab:lines}) making it difficult to detect. In fact, it has only been detected toward a handful of high-mass sources in the ISM such as Sgr B2 (\citealt{Kaifu1974}; \citealt{Belloche2013}; \citealt{Neill2014}), Orion KL B (\citealt{Cernicharo2016}), three cores in NGC 6334I (\citealt{Bogelund2019}), the Galactic Center cloud G+0.693 (\citealt{Zeng2018}) and the hot core G10.47+0.03 (\citealt{Ohishi2019}). It is also detected in a spiral galaxy at redshift 0.89 (\citealt{Muller2011}). But it was not detected towards the low-mass protostar IRAS 16293-2422B (\citealt{Ligterink2018}).  Therefore, (tentative) detection of this molecule in 12 additional sources is increasing this sample size by more than a factor of 2. 


\begin{figure*}
    \centering
    \includegraphics[width=\textwidth]{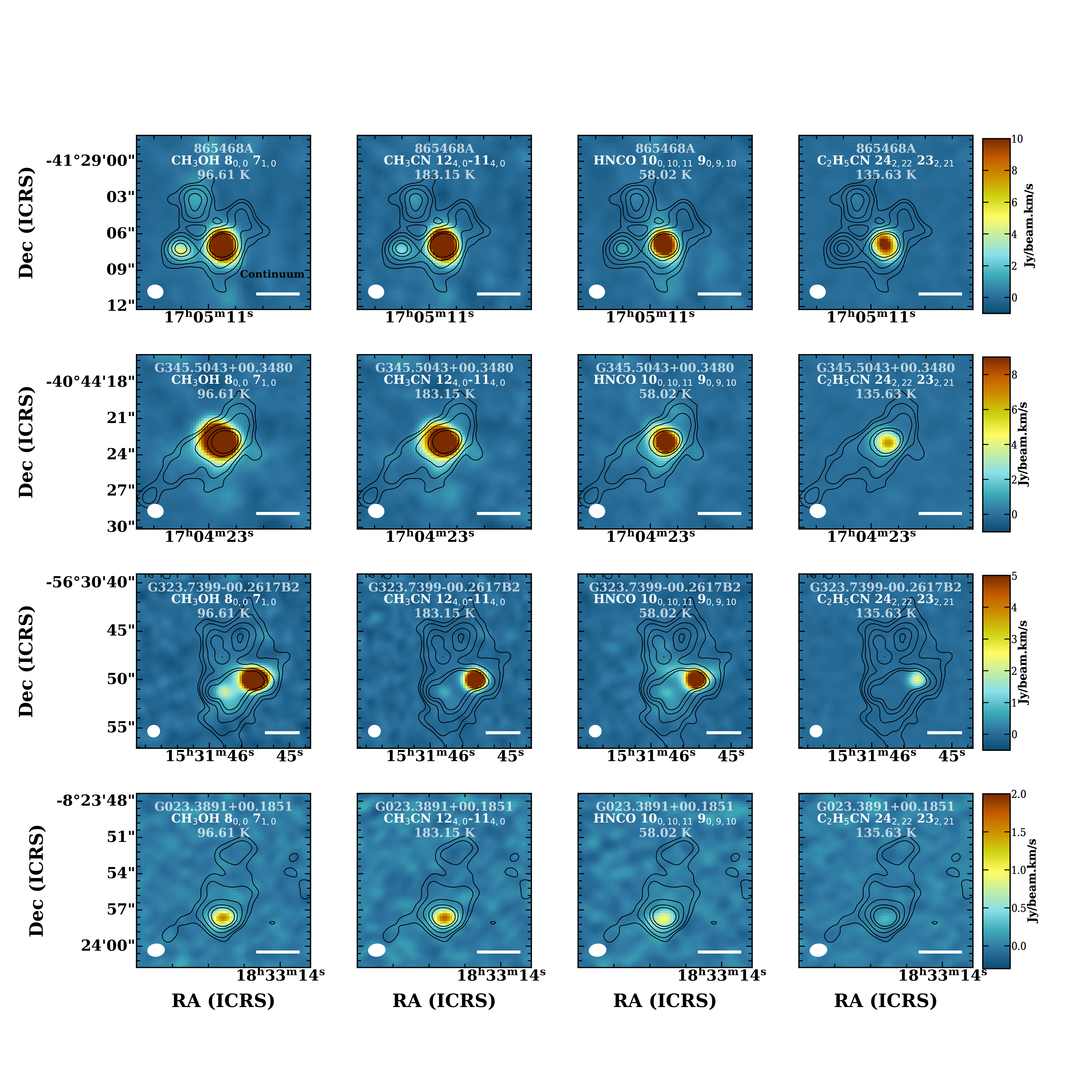}
    \caption{Moment zero maps of four selected lines for four sources included in Fig. \ref{fig:spectra}. The line, name of the source and $E_{\rm up}$ of the line are printed in white at the middle top of each panel. The beam is also shown on the bottom left of each panel. The continuum contours are shown in black and are at $[3, 5, 10, 20, 30]\sigma_{\rm cont}$ with $\sigma_{\rm cont}$ for the sources from top to bottom being 7.6, 4.1, 1.9 and 0.8\,mJy\,beam$^{-1}$, respectively. The scale bars on the bottom right of each panel from top to bottom show lengths of 10942\,au, 7199\,au, 11518\,au, 39089\,au at the distances of the sources.} 
    \label{fig:mom0}
\end{figure*}

Moreover, as seen from Fig. \ref{fig:detections}, there are more minus signs toward the right hand side of this figure, i.e., for sources located at larger distances. Therefore, the fact that some molecules are not detected in some sources could be the sensitivity limit. However, there is not a clear correlation between luminosity and detection as shown in \ref{fig:detections_L}.

\subsection{Spatial extent}
\label{sec:spatial_extent}

Figure \ref{fig:mom0} presents the moment zero (integrated intensity) maps of four lines for four of the sources whose spectra are included in Fig. \ref{fig:spectra}. The lines and sources selected are a few examples of the types of sources and line spatial distributions seen in the sample. The transitions with their upper energy levels are printed on each panel of Fig. \ref{fig:mom0} in white. Moreover, the continuum contours are shown for each source in black.

\begin{SCfigure*}
    \centering
    \includegraphics[width=12cm]{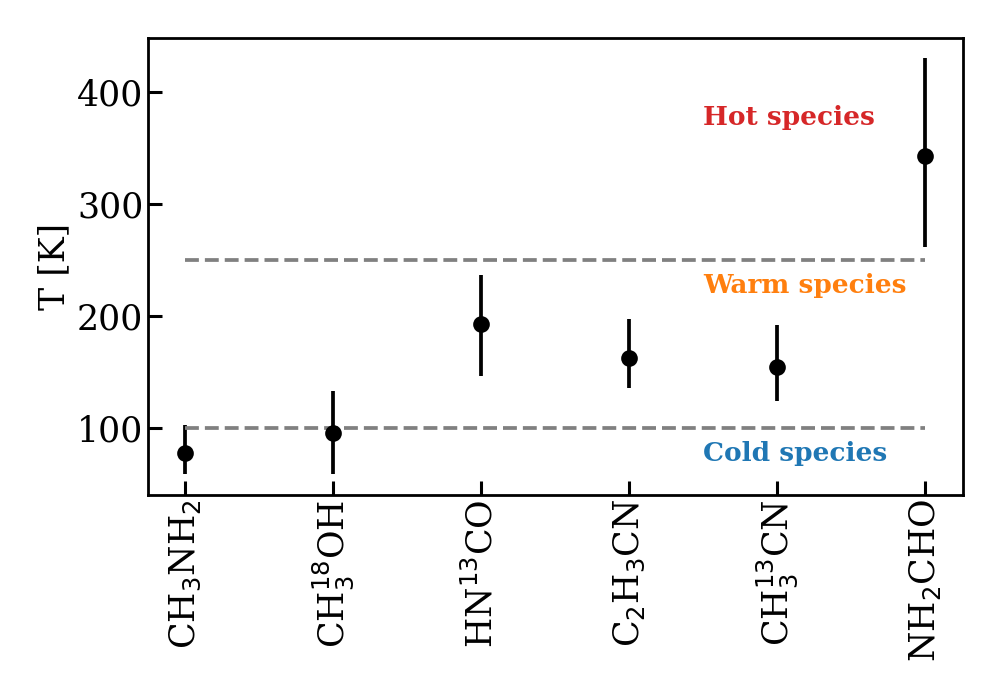}
    \caption{Mean of the excitation temperatures of the species in sources where the temperature is fitted. The error bars are calculated as explained in the text. The species are ordered by their binding energies from low (left) to high (right). The dashed lines are to separate the three groups of species seen.} 
    \label{fig:Tex}
\end{SCfigure*}


The high mass clumps considered in this work show complex structures in their dust emission. They often present multiple continuum sources in a single field of view. However, not all the continuum bright regions show emission from the species considered here. Often, only a single source in each region is bright in complex organic emission which is the source that we extract the spectrum from.

The species considered here (a few presented in Fig. \ref{fig:mom0}) usually originate from a single compact region with radii of ${\sim} 1000-7500$\,au. Therefore, assuming a single $T_{\rm ex}$ and FWHM for one molecule is valid as long as lines are relatively optically thin.

The emission is mostly unresolved, although it is clear that some lines show more extended emission. For example, the HNCO and C$_2$H$_5$CN transitions presented here show more compact emission compared to CH$_3$OH and CH$_3$CN lines. However, these speculations are line dependent as the emission region depends on the upper energy level of the line and its optical depth. Higher spatial resolution data are needed to measure the column densities and excitation temperatures as a function of distance from the central source. Although the emission is usually unresolved, some spatial information can be obtained by looking into the excitation temperatures of various species as discussed in Sect. \ref{sec:implication_Tex}.

\subsection{Excitation temperatures}
\label{sec:Tex}


It is possible to measure the excitation temperatures of CH$_3^{18}$OH, $^{13}$CH$_3$CN, C$_2$H$_3$CN, HN$^{13}$CO, NH$_2$CHO and CH$_3$NH$_2$ for at least nine sources. Table \ref{tab:results} presents these values. The excitation temperature of CH$_3^{18}$OH is likely biased toward lower temperatures given that it is mainly measured from two lines with $E_{\rm up} < 100$\,K and one transition with $E_{\rm up} = 238.9$\,K. The latter usually provides a good upper limit on the excitation temperature. Figure \ref{fig:Tex} shows the average excitation temperatures for these species where measurement of $T_{\rm ex}$ was possible (see Figures \ref{fig:plots_NH2CHO}- \ref{fig:bad_CH3NH2_upper} for examples of how excitation temperatures were measured). The error bars in this figure are calculated as follows. For each molecule the average of excitation temperatures, average of their upper (i.e. $T_{\rm ex} + T_{\rm err,up}$) and lower values (i.e. $T_{\rm ex} - T_{\rm err,low}$) for different sources are calculated. Then the error bars show the difference between the upper and lower averages with the mean of the excitation temperatures. The molecules in this figure are ordered by increasing binding energy from left to right.

This figure shows that the species studied here fall in three categories in terms of their excitation temperatures. Cold species ($T_{\rm ex} \lesssim 100$\,K) are found to be CH$_3^{18}$OH and CH$_3$NH$_2$. Warm species (100\,K\,$\lesssim T_{\rm {ex}} \lesssim 250$\,K) are found to be $^{13}$CH$_3$CN, HN$^{13}$CO and C$_2$H$_3$CN. NH$_2$CHO is found to be a hot molecule ($T_{\rm {ex}} \gtrsim 250$\,K).


\subsection{Column densities}
\label{sec:col_NH2CHO_HNCO}
The fitted column densities of the species investigated here are presented in Tables \ref{tab:results} and \ref{tab:results_methanol}. It is important to note that methanol column densities here are found for spectra extracted from the peak of moment zero maps of CH$_3$CN 12$_4$-11$_4$ while the values in \cite{vanGelder2022} are found from spectra extracted centered at the continuum peak. Moreover, here the temperatures are fitted for $^{18}$O methanol while in \cite{vanGelder2022} the temperatures were fixed to 150\,K. Therefore, variations in column densities of methanol between these two works are expected. These variations are always within a factor ${\sim}4$ but for most sources they are only within a factor ${\sim}2$.


Given the large number of sources considered in this work, one can investigate if there is any correlation between the column densities of various species. We do this by using the Pearson's $r$ coefficient. It is calculated by the following equation for our sample in logarithmic space

\begin{equation}
    r_{\rm P} = \frac{\sum_i (N_{1,i} - \Bar{N_{1}}) (N_{2,i} - \Bar{N_{2}})}{\left(\sum_i (N_{1,i} - \Bar{N_{1}})^{2} \sum_i (N_{2,i} - \Bar{N_{2}})^{2}\right)^{0.5}},
    \label{eq:Pearson}
\end{equation}

\noindent where $N_{1}$ and $N_{2}$ are the column densities of the two species that $r_{\rm P}$ is calculated for, $\Bar{N_{1}}$ and $\Bar{N_{2}}$ are the mean of the column densities of species 1 and 2 and the sums are over the sources in our sample indicated by $i$.  

\begin{figure}
  \resizebox{\columnwidth}{!}{\includegraphics{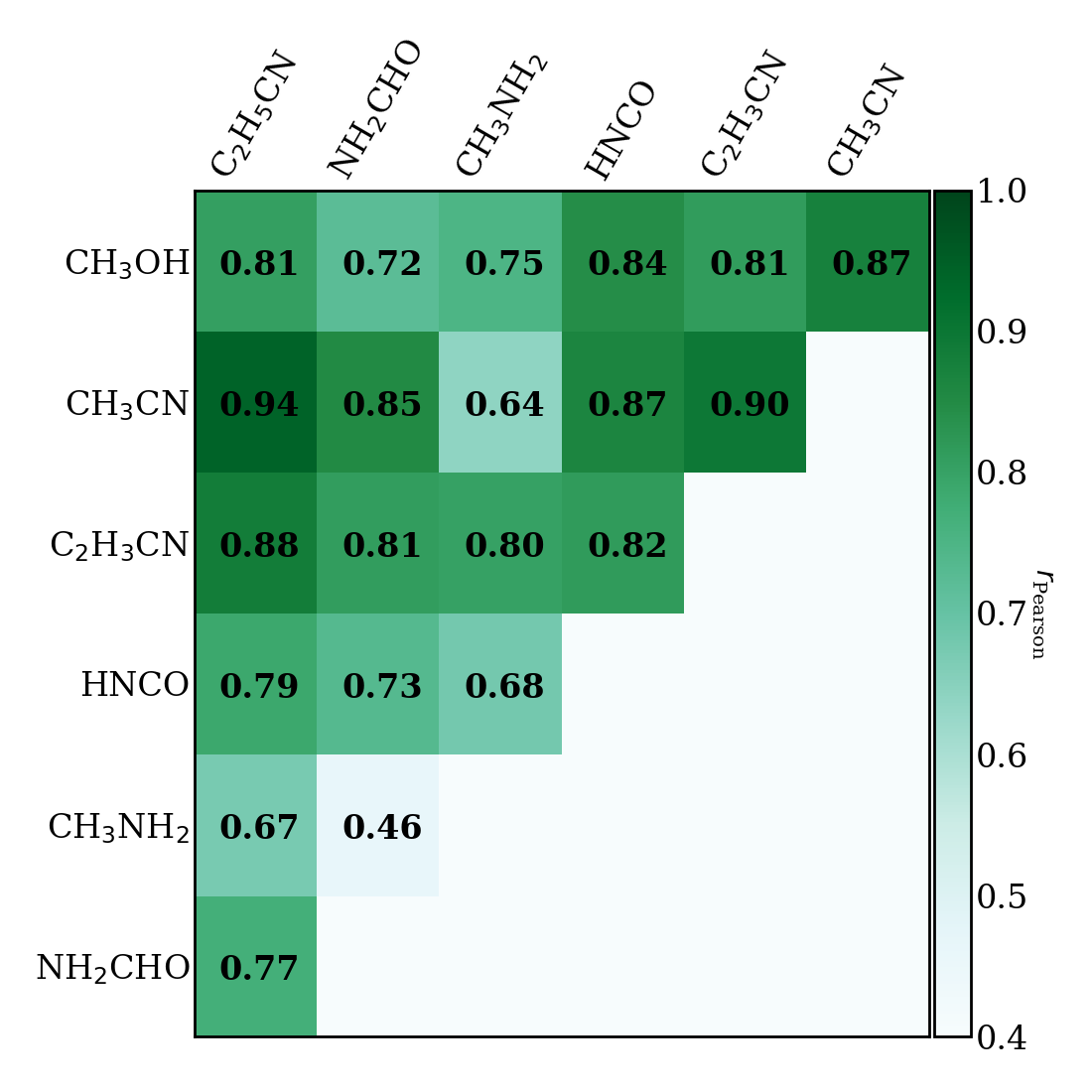}}
  \caption{Correlation matrix for Pearson coefficient of the column densities. The darkest green shows the highest correlation. In this figure sources with their methanol column density found from $^{13}$CH$_3$OH are excluded due to optical depth effects.}
  \label{fig:matrix_cor}
\end{figure}

Figure \ref{fig:matrix_cor} presents the correlation matrix for the Pearson's $r$ coefficients of the column densities of the studied species. Additionally, Fig. \ref{fig:column_column} presents column densities of the species plotted versus each other. It is important to note that sources whose methanol column density is derived from $^{13}$CH$_3$OH (instead of CH$_3^{18}$OH) are excluded in Fig. \ref{fig:matrix_cor}. This is to eliminate the potential optical depth issues with $^{13}$CH$_3$OH. For these sources CH$_3^{18}$OH was either not detected or the lines were too blended to derive reliable column densities of this molecule (see Sect. \ref{sec:spectral_modelling}). Figure \ref{fig:matrix_cor_thick} presents the same figure but including the sources where methanol column density is found from $^{13}$CH$_3$OH and all species except CH$_3$NH$_2$ show weaker correlation with methanol. This is likely due to the optical depth effects of $^{13}$CH$_3$OH. Figure \ref{fig:matrix_cor} shows that most species have strong ($r_{\rm P}>0.7$) positive correlations with each other except for CH$_3$NH$_2$ that does not show such strong correlation with most other species. 

\begin{figure}
  \resizebox{\hsize}{!}{\includegraphics{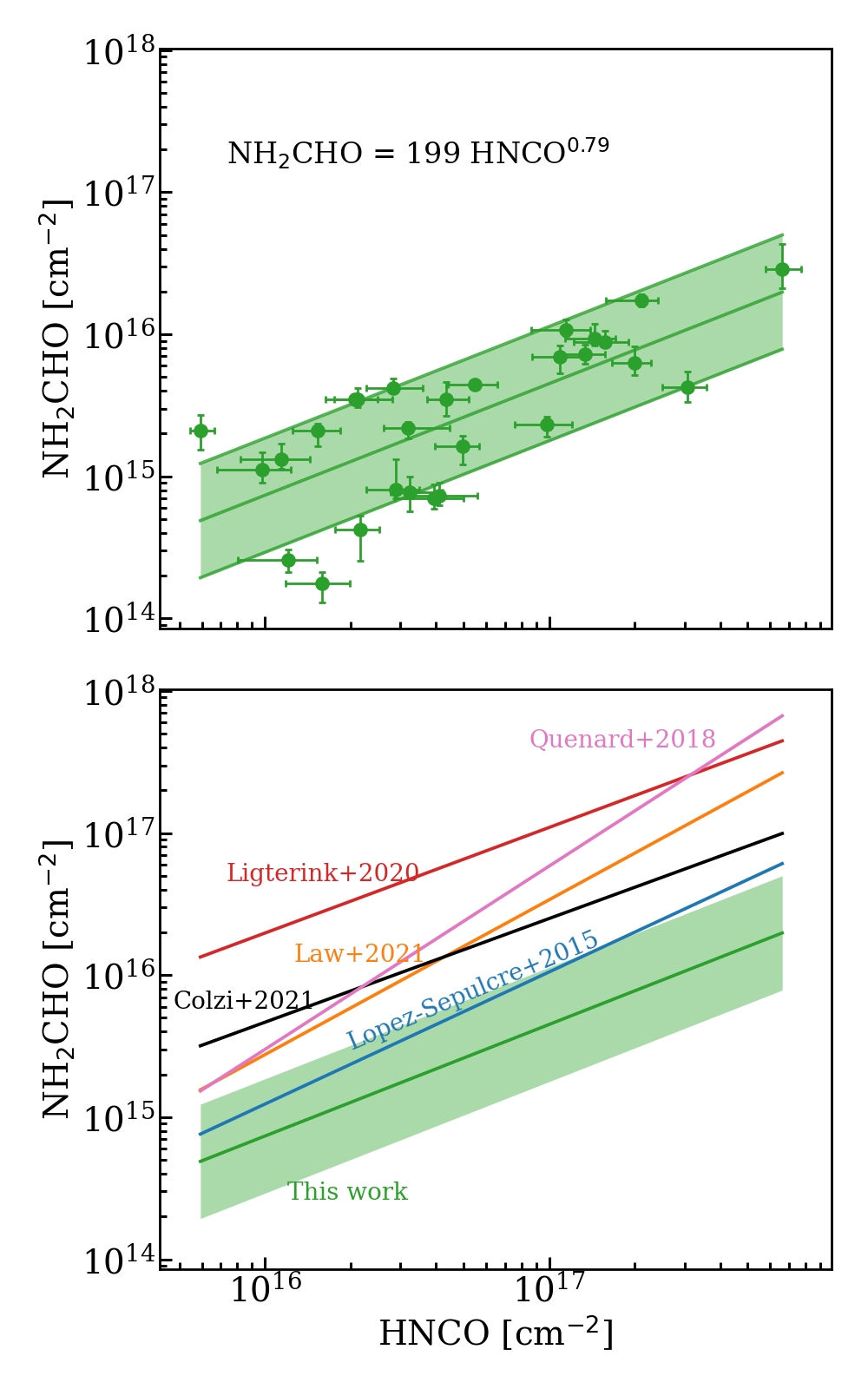}}
  \caption{Column density of NH$_2$CHO versus column density of HNCO. The top panel shows data points from the ALMAGAL survey found in this work with the best fit plotted as the middle solid line and the shaded green area showing the 68 percentile scatter in the data. Bottom panel compares the fitted data in our work by studies from \cite{Lopez2015}, \cite{Quenard2018}, \cite{Ligterink2020}, \cite{Law2021} and \cite{Colzi2021}. The curves for \cite{Lopez2015}, \cite{Quenard2018} and \cite{Colzi2021} are taken from the fit to abundances of NH$_2$CHO and HNCO and are normalized here as explained in the text: only slopes can be compared, not the absolute values.}
  \label{fig:NH2CHO_HNCO_T}
\end{figure}

C$_{2}$H$_5$CN and CH$_3$CN are found to have the tightest correlation with $r_{\rm P}$ of 0.94. In fact, there is a tight correlation ($r_{\rm P} \geq 0.88$) between all three molecules in the cyanide group: CH$_3$CN, C$_{2}$H$_3$CN and C$_{2}$H$_5$CN. Other observational studies have also seen the strong correlation between CH$_3$CN and C$_{2}$H$_5$CN. \cite{Yang2021} find $r_{\rm P} = 0.85$ for the column density normalized by the continuum brightness of these molecules in the PEACHES sample for low-mass protostars. Moreover, \cite{Law2021} find a correlation coefficient of 0.91 for the column densities of these two molecules in their spatially resolved data of the high-mass star forming region, G10.6-0.4. 

It is notable that absolute values of column densities depend on the source size assumed and the amount of warm gas mass in the beam of the observations. Therefore, if a source has more warm mass in the beam all the species are enhanced. Hence, the most interesting fact in Fig. \ref{fig:matrix_cor} is the weaker correlation of CH$_3$NH$_2$ with most species (see Sect. \ref{sec:chemistry}).

\section{Discussion}
\label{sec:discussion}


\begin{figure*}
    \centering
    \includegraphics[width=13.6cm]{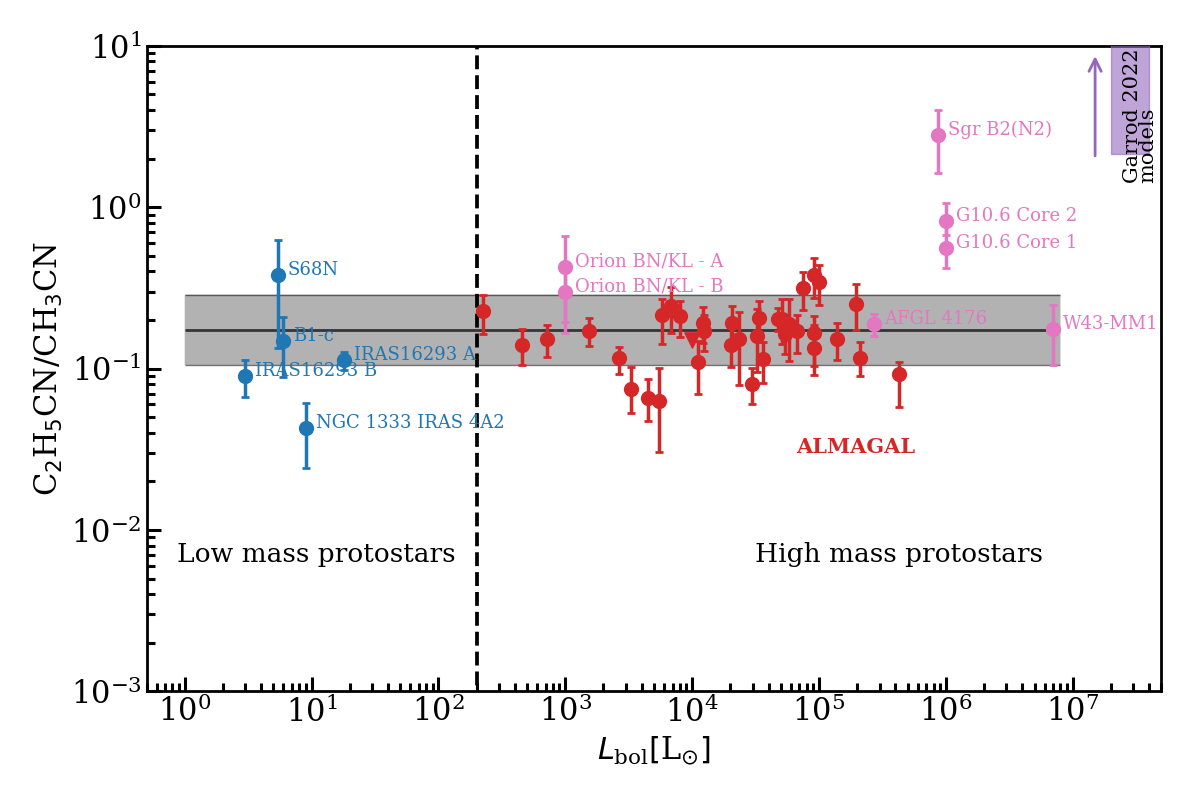}
    \caption{Column density ratio of C$_2$H$_5$CN to CH$_3$CN versus bolometric luminosity for low- and high-mass sources. The red points show the values for ALMAGAL sources from this work. Blue and pink show values for low- and high-mass protostars taken from the literature (see Table \ref{tab:refs} for the references). Moreover, the purple bar shows the range of C$_2$H$_5$CN/CH$_3$CN peak gas-phase values from \cite{Garrod2022} in the warm-up stages with slow, medium and fast pace (from their Table 17, final model setup). The purple arrow shows that the range of models are higher than the range in this plot. Upward triangles show lower limits and downward triangles show upper limits. Grey solid line and shaded gray area show the mean and standard deviation weighted by the uncertainty on the $\log_{10}$ of each column density ratio after eliminating upper and lower limits. Moreover, Sgr B2(N2) is eliminated in the derivation of mean and standard deviation, as it often seems to be an outlier.} 
    \label{fig:C2H5CN_CH3CN}
\end{figure*}

\subsection{Implications of the measured excitation temperatures}
\label{sec:implication_Tex}

Figure \ref{fig:Tex} shows three categories of species based on their excitation temperatures (see Sect. \ref{sec:Tex}). A similar temperature segregation was found toward IRAS 16293–2422B (\citealt{Jorgensen2018}) and seven high-mass young stellar objects (\citealt{Bisschop2007}). The differences seen in the excitation temperatures of various molecules can be due to the differences in their desorption temperatures which depends on the binding energies of these species in their respective ice matrix. Taking the binding energies of these species in amorphous solid water, formamide has the highest binding energy (\citealt{Chaabouni2018}; \citealt{Ferrero2020}; \citealt{Minissale2022}) and CH$_3$NH$_2$ has the lowest one (\citealt{Chaabouni2018}) compared with the other species. The rest of the molecules (CH$_3^{18}$OH; HN$^{13}$CO; C$_2$H$_3$CN; and $^{13}$CH$_3$CN) have similar binding energies in between (\citealt{Collings2004}; \citealt{Das2018}; \citealt{Song2016}; \citealt{Wakelam2017}; \citealt{Bertin2017}; \citealt{Penteado2017}; \citealt{Ferrero2020}; \citealt{Minissale2022}). Therefore, assuming that the excitation temperature represents the temperature of the environment to some extent, the pattern seen in Fig. \ref{fig:Tex} is consistent with the pattern seen in binding energies and desorption temperatures of these species.

One can assume that the excitation temperatures of these species give information on the temperature of the environment that these species are tracing. If this is the case NH$_2$CHO is tracing the hotter regions closer to the protostar, CH$_3^{18}$OH and CH$_3$NH$_2$ trace the cold temperatures farther from the protostar and the rest of the molecules trace regions in between. In other words the pattern seen in the excitation temperatures point to existing of an `onion-like' structure around the protostar. A similar onion-like structure was also observed around the hot corinos of IRAS 16293-2422 and SVS13-A binary systems (\citealt{Manigand2020}; \citealt{Bianchi2022}). This is particularly interesting as it can be an indication of different emitting areas of these species while this information cannot be obtained with the angular resolution of the data used here. In particular the low excitation temperature of CH$_3^{18}$OH indicates a more extended emission of this molecule compared with some of the other species such as formamide. Although the low $E_{\rm up}$ lines of $^{12}$CH$_3$OH also trace emission from outflows (\citealt{Tychoniec2021}), in this work we do not use methanol itself but rather its $^{18}$O isotopologue to find the column density of $^{12}$C-methanol for most sources. Therefore, the $^{12}$C-methanol column densities should mainly trace the gas in the envelope or disk around the protostar. The implication of different emitting areas on column density measurements and their ratios is discussed in Sect. \ref{sec:source_structure}.

\subsection{NH$_2$CHO versus HNCO}
\label{sec:HNCO_NH2CHO}

This section is focused on two molecules whose relationship has been discussed extensively in the literature: NH$_2$CHO and HNCO. Figure \ref{fig:NH2CHO_HNCO_T} presents the column densities of the two plotted against each other. In the top panel the data points are presented and a simple curve of the form $N_{\rm{NH_{2}CHO}} = 199 \times N_{\rm{HNCO}}^{0.79}$ is fitted to the column densities. In the bottom panel, we compare the result of our fit with other studies. Red shows the fit to the column densities of these two molecules towards the high-mass star forming region NGC 6334I (\citealt{Ligterink2020}). Orange shows the same from \cite{Law2021} for the massive star forming region G10.6-0.4. Blue, pink and black are the fits to the abundances of NH$_2$CHO and HNCO with respect to $N_{\rm H_{2}}$ from \cite{Lopez2015}, \cite{Quenard2018} and \cite{Colzi2021} normalized by a factor $10^{23(1-y)}$, where $y$ is the exponent of the fit to the abundances and $N_{\rm H_{2}}$ is assumed to be $10^{23}$ for all (see Appendix \ref{app:norm} for the explanation of the normalization). Therefore, one can only compare the slopes of the curves shown in the bottom panel of Fig. \ref{fig:NH2CHO_HNCO_T} rather than the absolute values of the column densities. In general, the slopes of the relations between NH$_2$CHO and HNCO column densities (i.e., the exponents) agree well between the different data sets with the slopes being slightly steeper in the observations of \cite{Quenard2018} and \cite{Law2021}.

\begin{figure*}
    \centering
    \includegraphics[width=13.6cm]{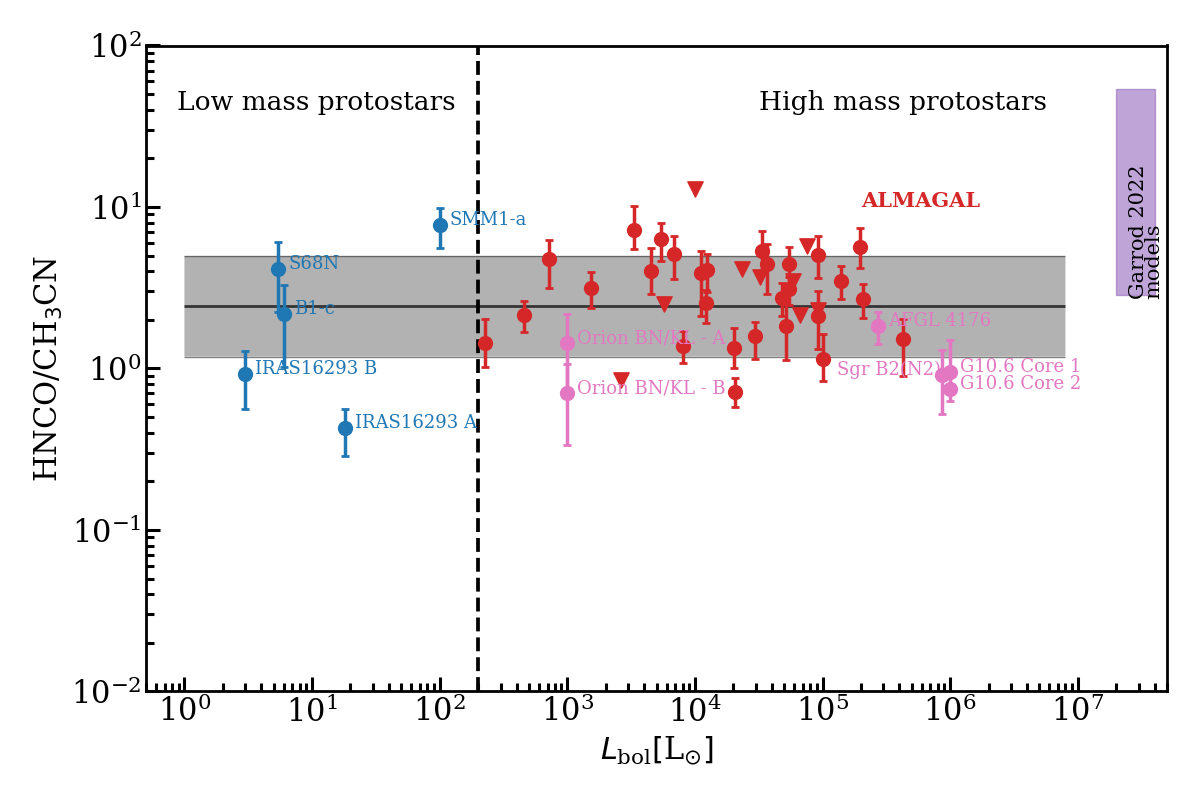}
    \caption{Same as Fig. \ref{fig:C2H5CN_CH3CN} but for HNCO/CH$_3$CN. The literature values are taken from studies presented in Table \ref{tab:refs}. In this plot SMM1-a is colored blue although it is an intermediate mass protostar.} 
    \label{fig:HNCO_CH3CN}
\end{figure*}

\subsection{Going from low- to high-mass protostars}
\label{sec:low-to-high}

The absolute values of column densities discussed in Sections \ref{sec:col_NH2CHO_HNCO} and \ref{sec:HNCO_NH2CHO} are not as informative as column density ratios because the former are dependent on the beam size which is not the same in all sources. The column density ratios are not dependent on the beam size, but they are dependent on the emitting areas of the two species in the ratio. This effect needs to be considered and is explained in depth in Sect. \ref{subsec:source_structure}. However, in this Section we consider the column density ratios and the general trends seen for them.

\begin{figure*}
    \centering
    \includegraphics[width=13.6cm]{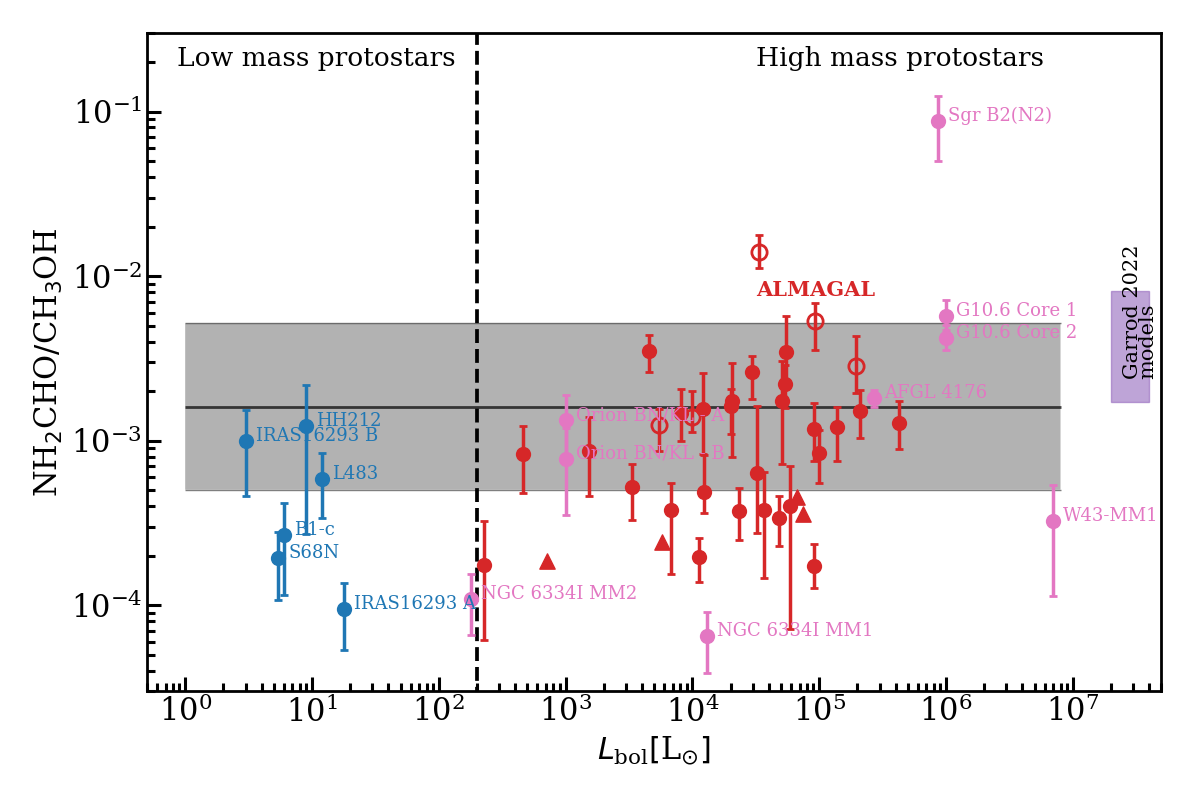}
    \caption{Same as Fig. \ref{fig:C2H5CN_CH3CN} but for NH$_2$CHO/CH$_3$OH. The red hollow circles indicate the sources for which $^{13}$CH$_3$OH was used to find the column density of CH$_3$OH.} 
    \label{fig:NH2CHO_CH3OH}
\end{figure*}

\begin{figure}
  \resizebox{\columnwidth}{!}{\includegraphics{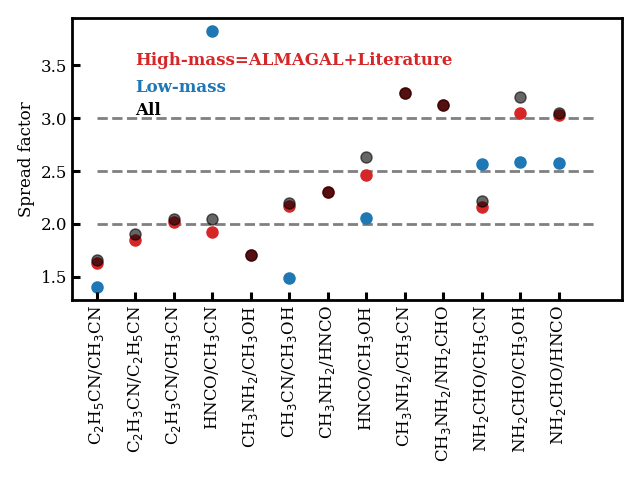}}
  \caption{1$\sigma$ scatter around the mean of $\log_{10}$ of the column density ratios. Both mean and standard deviation are weighted by the errors on each data point. Blue shows the values for low-mass protostars where there were more than two sources available for this calculation. Red shows the scatter for high-mass protostars and black presents those for all sources available (i.e. low- and high-mass protostars).}
  \label{fig:spread_factor}
\end{figure}

Figures \ref{fig:C2H5CN_CH3CN}, \ref{fig:HNCO_CH3CN} and \ref{fig:NH2CHO_CH3OH} present the column density ratios of C$_2$H$_5$CN to CH$_3$CN, HNCO to CH$_3$CN and NH$_2$CHO to CH$_3$OH for the sources studied in this work as well as other low- and high-mass protostars from the literature. Appendix \ref{sec:add_plots} presents the same figures for some additional combination of species. Each plot covers four orders of magnitude in ratio. The solid black line in each figure shows the mean of the ($\log_{10}$) column density ratios. The shaded-gray regions indicate the 1$\sigma$ scatter of the ($\log_{10}$) column density ratios around this mean value. The means and standard deviations include the data from this work and the literature, for low- and high-mass protostars, and are weighted by the errors on each data point. Moreover, the upper limits, lower limits and the Sgr B2(N2) source are excluded from this calculation. The $1\sigma$ scatter values in $\log$ space (i.e. dex) can be converted to a `factor of spread' around the mean value as $10^{\rm dex}$. These factors are summarized in Table \ref{tab:spread_factor} and are shown in Fig. \ref{fig:spread_factor}.

It is important to note that the calculated scatter for the low-mass protostars are mainly based on a handful (four or five) of measurements available from the literature. Therefore, one cannot make any conclusions solely based on the calculated scatter in low-mass protostars. The PEACHES survey that studies ${\sim} 50$ low-mass protostars does not include HNCO, and with their relatively short integration times $^{18}$O methanol isotopologue lines are unfortunately not detected for most of their sources. Moreover, the column density of CH$_3$CN is found from the main isotopologue lines which are likely optically thick. Hence, their column density values are not included in the quantification of the scatter.     

\begin{table}
    \caption{Factor of spread around the mean defined as 10$^{\rm dex}$}
    \label{tab:spread_factor}
    \centering
    \begin{tabular}{l l l l l} 
    \toprule
    \toprule    
Ratio & ALMAGAL & Low- & High- & All\\
&&mass&mass&\\
\midrule     

C$_2$H$_3$CN/CH$_3$CN    &1.66   &--   &2.02   &2.04 \\
C$_2$H$_3$CN/C$_2$H$_5$CN        &1.74   &--   &1.85   &1.91 \\
C$_2$H$_5$CN/CH$_3$CN    &1.45   &1.40   &1.63   &1.66 \\
CH$_3$NH$_2$/CH$_3$CN    &2.65   &--   &3.23   &3.23 \\
CH$_3$NH$_2$/CH$_3$OH    &1.66   &--   &1.71   &1.71 \\
CH$_3$NH$_2$/HNCO        &2.37   &--   &2.30   &2.30 \\
CH$_3$NH$_2$/NH$_2$CHO   &2.45   &--   &3.13   &3.13 \\
NH$_2$CHO/CH$_3$CN       &2.13   &2.57   &2.16   &2.21 \\
NH$_2$CHO/CH$_3$OH       &3.15   &2.58   &3.05   &3.20 \\
NH$_2$CHO/HNCO   &2.31   &2.57   &3.03   &3.05 \\
HNCO/CH$_3$CN    &1.84   &3.83   &1.93   &2.04 \\
HNCO/CH$_3$OH    &2.49   &2.06   &2.46   &2.63 \\
CH$_3$CN/CH$_3$OH        &2.05   &1.49   &2.17   &2.19 \\

\bottomrule
\end{tabular}
\tablefoot{The first column presents the column density ratio considered. The second column shows the factor of spread in the ALMAGAL column density ratios. Whereas the third column shows that for the low-mass protostars (from literature, see Table \ref{tab:refs} for references), the fourth column shows the same for high-mass protostars (ALMAGAL and literature) and the last column shows the spread factor for all low- and high-mass protostars. The averages are given in Table \ref{tab:means}}.
\end{table}

The scatter in column density ratios of all sources, as shown in Fig. \ref{fig:spread_factor}, is less than a factor of about three. In other words they are remarkably constant. Given the sensitivity of COM abundances to temperature and time (\citealt{Garrod2013}; \citealt{Garrod2022}; \citealt{Aikawa2020}), these constant ratios imply that the physical conditions in which the complex organics are formed are similar (see e.g. \citealt{Quenard2018}; \citealt{Coletta2020}; \citealt{Belloche2020}). Hence, our data indicate that COMs are formed during the phase of star formation in which the physical conditions are mainly constant and similar for all species. The studied hot cores, span a range of orders of magnitude in luminosities (see Table \ref{tab:source_features}). This means that the hot core stage of the various ALMAGAL sources should have large differences in gas temperature and radiation field. Therefore, the most likely phase in which the physical conditions are mainly constant is that of the cold cloud prior to collapse, i.e. the pre-stellar phase during which most molecules are frozen out and formation of complex molecules in ices can take place. Another possibility is that the precursors of these species are formed in the pre-stellar ices while COMs form later in the gas in a manner that the column density ratios stay constant. Given that this latter argument adds an extra level of complexity (i.e., needing constant physical conditions in the pre-stellar phase and the protostellar phase), the former argument, where COMs are likely forming in the pre-stellar ices, could be a more probable interpretation.

In Fig. \ref{fig:spread_factor} the C$_2$H$_5$CN/CH$_3$CN column density ratio has the lowest scatter compared to other species combinations. In fact there is only a factor of 1.66 spread around the mean for C$_2$H$_5$CN/CH$_3$CN in all sources. The spread factor is in general small ($\lesssim 2$) for all ratios that relate the species in the cyanide group with each other. The scatter increases for other ratios presented in Fig. \ref{fig:spread_factor}. For instance, all ratios that include NH$_2$CHO have a large spread factor, with most having a scatter larger than a factor 2.5 around the mean. Moreover, the ratios that include CH$_3$NH$_2$, on average, show higher spread factors compared to the ratios that do not include this molecule and/or NH$_2$CHO. These trends are also seen in Fig. \ref{fig:matrix_cor}. It should be noted that the number of sources with CH$_3$NH$_2$ detection is lower than that of the other species studied here, and therefore, the conclusions made here should be confirmed with more data points in the future.

The remaining species show spread factors ${\lesssim}2.5$ (Table \ref{tab:spread_factor}). In general, this is a small scatter given the range of envelope masses (${\sim}1- 5000$\,M$_{\odot}$) and luminosities (${\sim}2- 10^{7}$\,L$_{\odot}$) of the sources studied here, all scattered across various regions in the sky. However, there are some differences in the spread factors of various column density ratios in Fig. \ref{fig:spread_factor}. The reason for such differences in the spread factor can be either physical or chemical. The physical and chemical effects are discussed in Sections \ref{sec:source_structure} and \ref{sec:chemistry}.

\subsection{Physical effects}
\label{sec:source_structure}

\subsubsection{Optical depth}

First, optical depth of the lines can be an issue when calculating the column density ratios and their scatter. This is especially an important problem for abundant species and those with bright lines such as HNCO, CH$_3$CN, CH$_3$OH and often NH$_2$CHO. Therefore, the column densities of these species could be underestimated in some sources because all their lines are potentially optically thick.

In this work the values for the column densities of HNCO, CH$_3$CN and CH$_3$OH are found from the column densities of their less abundant isotopologues as an attempt to solve the problem of optical depth. However, lines originating from the isotopologues can also become optically thick for very line-rich sources. The modeled optical depths of the HN$^{13}$CO lines toward the most line-rich source, 707948, are less than 0.1. Moreover, H$^{15}$NCO is not detected toward any of our sources. For 707948, HN$^{13}$CO/H$^{15}$NCO is ${\gtrsim} 0.4$ compared with the $^{13}$C/$^{15}$N isotopologue ratio of ${\sim}7$. Although these limits are not enough to prove that HN$^{13}$CO is optically thin but it can be an indication that optical depth might not be a major problem for HN$^{13}$CO. It is also less likely that CH$_3^{18}$OH lines are optically thick but for the sources where $^{13}$CH$_3$OH is used to obtain the column densities of CH$_3$OH (indicated by a star in Table \ref{tab:results_methanol}), optical depth can be an issue. These sources are shown as hollow red circles in Fig. \ref{fig:NH2CHO_CH3OH} for the ratio of NH$_2$CHO/CH$_3$OH. The column density ratios of the three ALMAGAL sources with luminosities $>3 \times 10^4$\,L$_{\odot}$ are on the higher end suggesting that the methanol column density is likely underestimated due to optical thickness of CH$_3^{13}$OH lines. These are indeed the three line rich sources (101899, 707948 and G343.1261-00.0623) for which due to extreme line blending it was not possible to find the column density of CH$_3^{18}$OH and hence $^{13}$CH$_3$OH was used. Moreover, it is possible that for the most line-rich sources some of CH$_3^{13}$CN lines start becoming optically thick but for most such sources this problem is mitigated by finding the column density from the CH$_3$CN 12$_{10}$-11$_{10}$ line if it is less optically thick than the CH$_3^{13}$CN lines. Finally, the column densities found for NH$_2$CHO can suffer from line optical depth issues as was found in NGC 6334I (\citealt{Ligterink2020}). However, NH$_2^{13}$CHO was not detected for some of the most line rich sources in this work. The $3\sigma$ upper limits of NH$_2^{13}$CHO can be converted to NH$_2$CHO upper limits using $^{12}$C/$^{13}$C ratios in Table \ref{tab:source_features}. These values obtained for NH$_2$CHO are indeed larger than the measured column density of formamide as expected and result in more optically thick lines of NH$_2$CHO (while over producing the lines). For some of the most line-rich sources (e.g. 707948) the low $E_{\rm up}$ line of formamide becomes marginally optically thick ($\tau \simeq 0.1$) if a beam dilution factor of 2 is assumed. Therefore, the column densities of NH$_2$CHO for most line-rich sources should be taken with care.

Considering other values presented in Figures \ref{fig:C2H5CN_CH3CN}, \ref{fig:HNCO_CH3CN} and \ref{fig:NH2CHO_CH3OH} from the literature, similar issues hold when calculating the column densities. The column densities from the literature are chosen carefully so this problem is minimized. However, as an example S68N in Fig. \ref{fig:C2H5CN_CH3CN} is a data point with large error bars and falls above the main spread. This is because the lines are weak in S68N and none of the isotopologues of CH$_3$CN except its deuterated versions are detected (\citealt{Nazari2021}). Moreover, the values of HNCO, CH$_3$CN and CH$_3$OH in G10.6 cores may be suffering from optical depth problems as explained carefully in \cite{Law2021}. The HNCO column densities for IRAS 16293 A and B can also be suffering from line optical thickness issues. This is because HNCO value for IRAS 16293 A was found from the main isotopologue of this molecule (\citealt{Manigand2020}) and in IRAS 16293 B except HN$^{13}$CO other isotopologues of HNCO such as DNCO are also detected (\citealt{Coutens2016}) indicating that even HN$^{13}$CO might be optically thick. However, in general it is safe to assume that for most sources shown here (especially the ALMAGAL data) line optical depth is unlikely to have a large effect.

Dust optical depth can also have an effect on the column density ratios. It is true that the dust will affect all species coming from the same region to the same extent. However, if the species considered have different emitting areas, dust optical depth can affect the species closer to the protostar more and alter the column density ratios. This is especially evident in source 693050 as discussed by \cite{vanGelder2022} where the methanol emission is seen in a ring shaped structure around the continuum emission. Given Fig. \ref{fig:Tex} and as discussed in Sect. \ref{sec:implication_Tex}, NH$_2$CHO should be tracing the closest regions to the protostar. Therefore, the effect of dust opacity should be the largest for the ratios including formamide. From these ratios NH$_2$CHO/HNCO is notable as many studies have found a correlation between these two species (see Sect. \ref{sec:HNCO_NH2CHO}). But Fig. \ref{fig:spread_factor} shows that the spread factor calculated for this ratio is one of the largest when compared with the other ratios considered here. One of the main reasons of this high spread is potentially the larger effect of dust opacity on NH$_2$CHO.   

\subsubsection{Source structure}

\label{subsec:source_structure}

The second physical effect on column density ratios is the difference in emitting areas of the two species in a ratio. The column density found from an emitting region is directly proportional to the area of the emission. In this study the area of the emission is assumed to be the same for all species in a source. However, due to variations in sublimation temperatures some species can trace hotter regions closer to the protostar as seen for NH$_2$CHO in Fig. \ref{fig:Tex} and some species can trace colder regions further from the protostar as seen for CH$_3^{18}$OH in the same figure. Therefore, the assumption that the emitting area of these two species are the same is not correct. \cite{Nazari2021} show that for a toy model of a spherically symmetric envelope with power law structure in density and temperature (their Sect. 4.3 and Appendix B) the ratio of column densities for two molecules is related to the ratio of their abundances by

\begin{equation}
     \frac{N_{1}}{N_{2}} = \frac{X_{1}}{X_{2}} \left(\frac{T_{\rm 1,sub}}{T_{\rm 2,sub}}\right)^{-3.75},
    \label{eq:sub_T}
\end{equation}

\noindent where $X$ is the abundance with respect to hydrogen and $T_{\rm sub}$ is the sublimation temperature of a molecule. This means that in order to determine the true abundance ratio for NH$_2$CHO and CH$_3$OH which have different sublimation temperatures, a constant factor needs to be multiplied to all the data points in Fig. \ref{fig:NH2CHO_CH3OH}. 

This constant factor can be the ratio of their average $T_{\rm ex}$ found in Fig. \ref{fig:Tex} or the ratio of their sublimation temperatures from the lab experiments as proposed by \cite{Nazari2021}. This will not change the spread seen in this figure, it only moves all the data points by a factor $\left(\frac{T_{\rm CH_3OH,sub}}{T_{\rm NH_2CHO,sub}}\right)^{-3.75}$ if sublimation temperatures are assumed as in Eq. \eqref{eq:sub_T}. However, is the assumption that these sources are spherically symmetric (i.e. with no source structure such as disks) valid for all sources considered here? 


\cite{vanGelder2022} find a scatter of ${\sim} 4$ orders of magnitude in warm methanol mass if both line-rich and line-poor sources are included. It is important to note that methanol is known to be a molecule that forms only in the ice and then sublimates into the gas. They explain the reason for the ${\sim} 4$ orders of magnitude scatter based on non-chemical effects including a discussion on the possible existence of disks. When considering the line-rich sources the scatter in \citealt{vanGelder2022} becomes only ${\sim} 2$ orders of magnitude which points to the fact that for line-rich sources the effect of a disk is not as considerable. However, ${\sim} 2$ orders of magnitude scatter in methanol mass is still large enough that existence of (smaller) disks are needed to explain all of the spread in methanol mass. Therefore, in the line-rich sample analyzed here not all sources can be assumed as spherically symmetric and may host disks with various (perhaps small) sizes.

\begin{figure}
  \resizebox{\columnwidth}{!}{\includegraphics{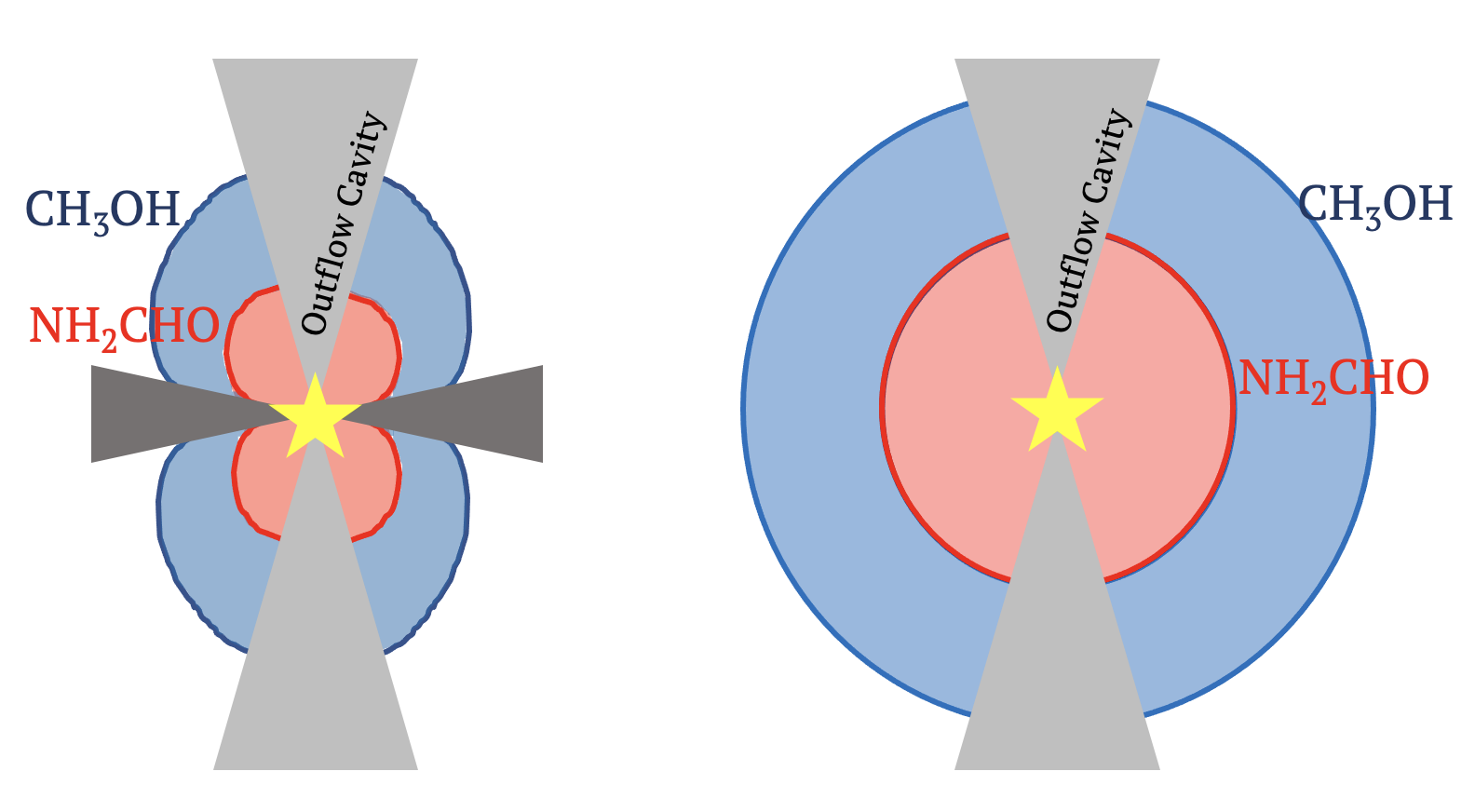}}
  \caption{A cartoon of two sources: one with disk (left) and one without (right). The red and blue regions show the emitting regions of NH$_2$CHO and CH$_3$OH, respectively. There is not a priory reason as to why the ratio of the two sublimation regions between the two sources should remain constant.}
  \label{fig:cartoon}
\end{figure}

Now the question becomes: If a source has a disk, will the ratio of emitting areas of NH$_2$CHO and CH$_3$OH stay constant compared with when there is no disk present? It is true that the ratio of sublimation temperatures remains constant but the temperature structure depends on the source structure as seen in the work of \cite{Nazari2022} on low-mass protostars. Figure \ref{fig:cartoon} shows a cartoon of possible emitting areas of NH$_2$CHO and CH$_3$OH for a case where a disk is present and a case where it is not. An analogue to Eq. \eqref{eq:sub_T} for a case where a disk is present is

\begin{equation}
     \frac{N_{1}}{N_{2}} = \frac{X_{1}}{X_{2}} \frac{f(R_{\rm 1,sub},z_{\rm 1,sub})}{f(R_{\rm 2,sub},z_{\rm 2,sub})},
    \label{eq:sub_T_disk}
\end{equation}

\noindent where $f(R,z)$ is a function for the emitting area of a species which is dependent on both radius ($R$) and height ($z$). The exact shape of this function is not easy to find analytically as the dependence of $f$ on the disk radius cannot be simply written as a power-law relation (see \citealt{Nazari2022}). Therefore, the ratio of the two functions across various sources is not necessarily constant.

A more quantitative understanding can be achieved by taking the mid-plane temperature in the fiducial envelope-only and envelope plus disk models of \cite{Nazari2022} for an 8\,L$_{\odot}$ source, and calculate the emitting radius of NH$_2$CHO and CH$_3$OH in the mid-plane by assuming the mean excitation temperatures found in Fig. \ref{fig:Tex} (i.e. ${\sim} 100$\,K for methanol and ${\sim} 300$\,K for formamide). These are given by 5.1\,au and 28.6\,au for NH$_2$CHO and CH$_3$OH in the envelope-only model and by 1.6\,au and 5.7\,au in the envelope plus disk model. Given that area is proportional to radius squared one can calculate the factor (ratio of the two area functions) in Eq. \eqref{eq:sub_T_disk} assuming NH$_2$CHO being species one and CH$_3$OH species two as 0.032 for the envelope-only model and 0.079 for the envelope plus disk model. There is a factor 2.5 difference between the two multiplication factors calculated that can alter the ratio of NH$_2$CHO/CH$_3$OH for the less line-rich sources. However, this is only a rough approximation by taking the mid-plane temperature. Radiative transfer models are needed for more quantitative understanding of the effect of source structure on the ratios in both low- and high-mass protostars and its effect on chemical implication of the spread. This is beyond the scope of this paper and is considered in other works (see Nazari et al. in preparation).  


Finally, the discussion above shows that the effect of dust optical depth and a first order approximation of the disk effect are the highest for ratios including NH$_2$CHO given its largest excitation temperature among all the species considered here. Therefore, to explain the larger scatter seen in ratios including NH$_2$CHO in Fig. \ref{fig:spread_factor}, first one needs to correct for the physical/non-chemical effects.

\subsection{Chemical implications}
\label{sec:chemistry}

\subsubsection{Comparison with chemical models}
\label{sec:chem_models}

In this section we compare our findings with the existing \cite{Garrod2022} chemical models for protostars. The models discussed in \cite{Garrod2022} build on the models in \cite{Garrod2013} (also see \citealt{Garrod2008}) altering and updating various chemical processes in different steps. Their models go through two physical stages. Collapse, where the gas temperature is kept at 10\,K but the density increases and warm-up, where the density is kept constant to the final density in the previous stage (2 $\times 10^8$\,cm$^{-3}$) and temperature increases to 400\,K with three different speeds.      

The purple bar in figures that present column density ratios (e.g. Figures \ref{fig:C2H5CN_CH3CN}-\ref{fig:NH2CHO_CH3OH}) show the peak gas-phase abundances from \cite{Garrod2022} models in the second stage with variations in the warm-up speed. Starting with C$_2$H$_5$CN/CH$_3$CN, Fig. \ref{fig:C2H5CN_CH3CN} shows that the `final' model setup of \cite{Garrod2022} does not agree well with the observations. To dive deeper into these models, Fig. \ref{fig:garrod_C2H5CN_CH3CN} shows our observational ratios overplotted with their values for different models during the warm-up stage with medium speed (the same plot for C$_2$H$_3$CN/CH$_3$CN has a similar trend). It is seen that the models agree best with the observations where gas-phase chemistry routes (the model on the left of the final model) are not added. They find that for their `final' model CH$_3$CN first forms to some extent in the ice but at later stages gas-phase chemistry takes over and most of its production occurs at later stages. Moreover, they find that C$_2$H$_5$CN and C$_2$H$_3$CN mainly form in ices. However, given that the `final' model does not agree well with the observational data (more than one order of magnitude difference) could mean that either not enough CH$_3$CN is produced in the cold stages, too much C$_2$H$_5$CN is produced, it is not as effectively destroyed at later hot stages or gas-phase production of CH$_3$CN is not observed. 

The observed column density ratios including CH$_3$NH$_2$ mostly differ with the \cite{Garrod2022} models (see purple bar in Figures \ref{fig:CH3NH2_CH3OH}-\ref{fig:CH3NH2_CH3CN}). Figure \ref{fig:garrod_CH3NH2_CH3OH} shows the comparison of various models in \cite{Garrod2022} with the spread in the data. It is more clear from this figure that most models in \cite{Garrod2022} overproduce CH$_3$NH$_2$ assuming that their methanol abundances are more accurate.

Figure \ref{fig:garrod_HNCO_CH3OH} shows the comparison of models with our results for HNCO/CH$_3$OH. In general the models do a good job (except the two models where they turn off diffusion in the ice mantles) at explaining the data. If the net rate of production of these two species in their final model are compared (their Figures 13 and 14) much of the HNCO and CH$_3$OH form at temperatures below 10\,K in ices supporting the argument that HNCO and CH$_3$OH could be forming in the dense core phase.

\subsubsection{Observed scatter}
\label{sec:ratios_obs}

One of the key findings of this paper is the relatively small scatter in column density ratios. However, a second finding is that this scatter, while usually small, shows some variations from species to species. This in turn may hold clues on the chemistry and the physical conditions where these molecules are formed. In Appendix \ref{sec:toy}, we develop a simple toy model to explain the scatter of abundance ratios around the mean value (i.e., spread factor, discussed in Sect. \ref{sec:low-to-high} and presented in Fig. \ref{fig:spread_factor}) from a chemical point of view. To minimize the contribution from the physical structure and dust optical depth on the scatter of abundance ratios around the mean, we only consider the molecules CH$_3$OH, CH$_3$CN, C$_2$H$_5$CN, C$_2$H$_3$CN, HNCO and CH$_3$NH$_2$ (i.e., all except formamide; see Sect. \ref{sec:source_structure}). 

The abundance of a molecule is primarily determined by the environmental conditions (temperature, density, and UV radiation) as well as the amount of time a source has spent in those conditions (\citealt{Lee2003}; \citealt{Sakai2013}). A cloud experiences various stages before forming a main sequence star, each of which are characterized by a distinct set of environmental variables. The bulk of most complex organic species is thought to form in the pre-stellar stage (\citealt{Herbst2009}). Moreover, as discussed in Sect. \ref{sec:low-to-high}, the almost constant column density ratios (spread $\lesssim 3$) point to formation of these species in similar physical conditions, likely the pre-stellar ices. Hence, we focus on pre-stellar stage in this section.

We determine the abundances in the pre-stellar ices with a toy model, inspired by the available chemical models. Most importantly, we assume that the abundance of each molecule is a function of time (\citealt{Lee2003}; \citealt{Aikawa2020}; \citealt{Garrod2022}). The details of the toy model are given in Appendix \ref{sec:toy}. Equations \eqref{eq:abund_ratio} and \eqref{eq:abund_ratio_mean} can be used to find 

\begin{equation}
    \frac{X_{1,2}}{\overline{X_{1,2}}} = \left(t/\Bar{t}\right)^{\alpha_{1} - \alpha_{2}}.
    \label{eq:time_scatter}
\end{equation}

\noindent This is a particularly interesting relation because it clearly shows that the scatter in column density ratios (assumed to be the same as abundance ratios; $X_{1,2}/\overline{X_{1,2}}$) can be explained by the scatter in pre-stellar timescales ($t/\Bar{t}$) without a direct dependence on the exact values of parameters $A_1$ and $A_2$ (the ice abundance of each species at 10$^6$\,yr). In other words, Eq. \eqref{eq:time_scatter} shows that the scatter in the pre-stellar timescales is the only parameter driving the scatter in abundance ratios (for a given temperature and density condition). Therefore, the distribution of the pre-stellar timescales can be inferred from the distributions of abundance ratios. 

\begin{figure}
  \resizebox{\columnwidth}{!}{\includegraphics{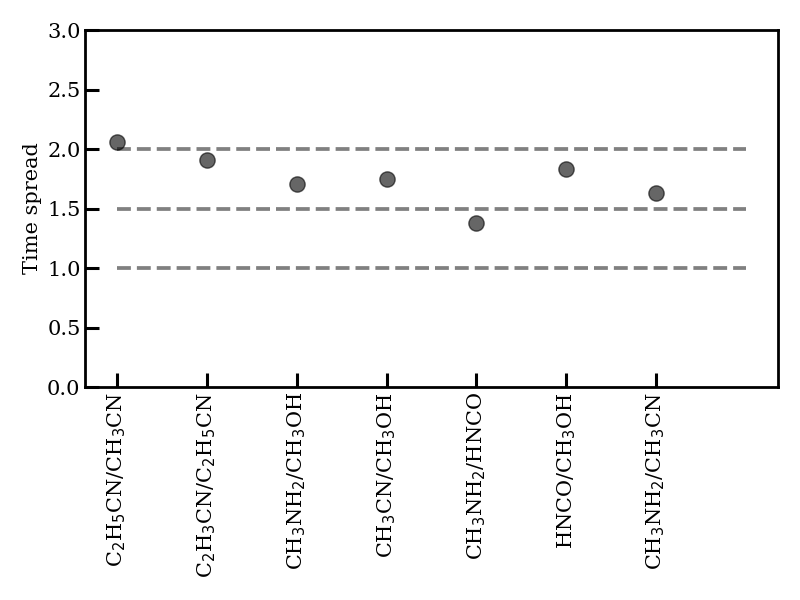}}
  \caption{The factor of spread around the mean of the pre-stellar timescales for various low- and high-mass protostars using the spread factor found for low-mass, ALMAGAL and high-mass literature sources (see black points in Fig. \ref{fig:spread_factor}).}
  \label{fig:time_spread}
\end{figure}

Figure \ref{fig:time_spread} presents the scatter in the pre-stellar timescales ($t/\Bar{t}$) for low- and high-mass protostars. This figure shows that the scatter in column density ratios of some species can be explained by a factor ${\sim} 1.5-2$ spread in the pre-stellar timescales of the low- and high-mass protostars. It is reassuring that the spread in the pre-stellar timescales is approximately constant and independent from the ratio used to find this spread. The small variation in the spreads seen in Fig. \ref{fig:time_spread} is expected. This is because all the scatter seen in the column density ratios (Fig. \ref{fig:spread_factor}) is not purely due to the variations in pre-stellar timescales but also physical effects explained in Sect. \ref{sec:source_structure}. It is important to add that these are all based on a simple toy model. Therefore, the exact values for the spread in pre-stellar times or the column density ratios that have a strong correlation with time might change. However, the general idea that the scatter in column density ratios can be (at least partially) explained by the scatter in pre-stellar timescales stays valid regardless of the toy model used. For example if the ice abundances in the `static phase' models of \cite{Aikawa2020} are taken (see their Figures 3 and 5), a similar toy model can be constructed with a similar trend seen in Fig. \ref{fig:abundances}. The result would again be that the scatter in abundance ratios can be related to the scatter in pre-stellar lifetimes. As \cite{Aikawa2020} conclude, the variations in COM abundances (mainly O-bearing species in their work) are less than one order of magnitude when the pre-stellar lifetimes are changed. This is consistent with our observations and toy model as we only observe spread factors $\lesssim 3$.       

The implication of such parametrization could be important. For all species, the mean column density ratio inferred from the low-mass sample is roughly similar to the value inferred from the high-mass sample (see Table \ref{tab:means}). Knowing $\overline{X_{1,2}}$, one can estimate $\Bar{t}$ in Eq. \eqref{eq:abund_ratio_mean} for all low- and high-mass protostars. Therefore, given that $\overline{X_{1,2}}$ is the same for low- and high-mass protostars the average duration of pre-stellar phase ($\Bar{t}$) is similar for the low- and high-mass protostars. This is only valid if the other parameters determining the abundances other than time (i.e., $\alpha_{1}$, $\alpha_{2}$ and $A_{1,2}$) are constant. Therefore, the general consensus that high-mass stars have a more rapid pre-main sequence evolution compared to the low-mass stars (e.g., see \citealt{Davies2011}; \citealt{Mottram2011}) implies that this faster evolution only occurs after the pre-stellar phase (i.e warm-up and protostellar phase onward). Hence, pre-stellar phases of low- and high-mass protostars should be similar (pre-stellar lifetimes within factor ${\sim}2-3$): cold low-mass clouds that can interact to form the high-mass protostars. This implication is in line with the non-detection of high-mass pre-stellar cores suggesting that such high-mass pre-stellar cores simply do not exist (\citealt{Motte2018}). Moreover, this conclusion agrees with the findings of \cite{Pitts2022} who show that low- and high-mass protostars are statistically indistinguishable on larger envelope scales. Because of the sparsity of the sample of low-mass protostars with such detailed chemistry analyses with ALMA, robust conclusions based on the low-mass sample remain pending increased sample size.

It is also important to mention that the argument above assumes that the temperatures of the pre-stellar environments are constant at ${\sim} 10$\,K for both low- and high-mass protostars. In addition, it assumes that the densities increase from ${\sim}10^3$\,cm$^{-3}$ to ${\sim}10^8$\,cm$^{-3}$ during the collapse phase. However, if these temperature and densities change (also see \citealt{Aikawa2020}), the point at which the abundances of these species stop their rapid growth (which is ${\sim} 10^5$\,yr with the assumptions of \citealt{Garrod2022} models) can also change to shorter or longer timescales. In other words, the other parameters that the abundance depends on (i.e., $A$ and $\alpha$) can be dependent on the temperature and density of the pre-stellar phase of each source. Therefore, changes in temperatures and densities can also affect the scatter seen in column density ratios. Finally, we emphasize that these conclusions are based on a simple toy model and for more robust conclusions further study is needed.

\section{Conclusions}
\label{sec:conclusion}
We analyzed the spectra of 37 line-rich MYSOs from the ALMAGAL survey. This work focuses on the study of six N-bearing species: CH$_3$CN, HNCO, NH$_2$CHO, C$_2$H$_5$CN, C$_2$H$_3$CN and CH$_3$NH$_2$. Our main conclusions are as follows:

\begin{itemize}
    \item CH$_3$CN and HNCO are detected in all the sources selected for this study. CH$_3^{13}$CN and HN$^{13}$CO are (tentatively) detected in $> 76\%$ of the sources. C$_2$H$_5$CN and NH$_2$CHO are (tentatively) detected in all sources except one and C$_2$H$_3$CN in ${\sim} 78\%$ of all sources. CH$_3$NH$_2$ is (tentatively) detected in 12 of the objects in our sample which more than doubles the number of sources with CH$_3$NH$_2$ detection in the ISM.
    \item Three groups of species are observed based on their excitation temperatures (Fig. \ref{fig:Tex}). NH$_2$CHO falls under the hot category ($T_{\rm ex} \gtrsim 250$\,K), HN$^{13}$CO, CH$_3^{13}$CN and C$_2$H$_3$CN fall under the warm category (100\,K\,$\lesssim T_{\rm ex} \lesssim 250$\,K) and CH$_3^{18}$OH and CH$_3$NH$_2$ fall under the cold category ($T_{\rm ex} \lesssim 100$\,K). The reason for these different groups can mainly be associated with the sublimation temperatures of these molecules.
    \item Assuming excitation temperatures indicate the temperature of the environment in which the species reside, the hot species trace regions closer to the protostar, the cold species trace regions far from the protostar and the warm species trace the regions in between.
    \item Column density ratios of the species studied here are remarkably constant across low- and high-mass protostars (with scatter less than ${\sim} 3$) indicating that these species and/or their precursors form in similar physical environments, most plausibly in pre-stellar ices.
    \item The scatter in column density ratios varies with C$_2$H$_5$CN/CH$_3$CN showing the smallest scatter (spread factor = 1.66) and ratios including formamide and methylamine showing the largest scatter (Fig. \ref{fig:spread_factor}). Given that formamide is most probably tracing the closest regions to the protostars compared with the other species studied here, the column density ratios including NH$_2$CHO are most affected by source structure and dust optical depth. Therefore, the scatter in its column density ratios mostly have a physical origin. Source structure and dust optical depth effects are smaller for the ratios including the other molecules and hence are likely not the main origin of the observed scatter. 
    \item Assuming that the bulk of complex organics form in the pre-stellar ices, the scatter in the column density ratios could be readily explained by the differences in the time that various clouds spend in the pre-stellar phase in addition to the variations in temperature and density of this phase. If the scatter mainly depends on the time spent in the pre-stellar phase, the data show similar pre-stellar lifetimes for low- and high-mass protostars (within factors of ${\sim} 2-3$).     
\end{itemize}

More robust conclusions can be made by increasing the number of low-mass protostars with such chemistry analysis observed by ALMA. However, with the current data available the abundances are similar toward low- and high-mass protostars indicating that these species form in similar pre-stellar physical conditions.


%
%

\begin{acknowledgements}
We would like to thank the referee for very constructive comments. We also thank R. T. Garrod for providing us with some of their model outputs. Astrochemistry in Leiden is supported by the Netherlands Research School for Astronomy (NOVA), by funding from the European Research Council (ERC) under the European Union’s Horizon 2020 research and innovation programme (grant agreement No. 101019751 MOLDISK), and by the Dutch Research Council (NWO) grants 648.000.022 and 618.000.001. Support by the Danish National Research Foundation through the Center of Excellence “InterCat” (Grant agreement no.: DNRF150) is also acknowledged. This paper makes use of the following ALMA data: ADS/JAO.ALMA\#2019.1.00195.L. ALMA is a partnership of ESO (representing its member states), NSF (USA) and NINS (Japan), together with NRC (Canada), MOST and ASIAA (Taiwan), and KASI (Republic of Korea), in cooperation with the Republic of Chile. The Joint ALMA Observatory is operated by ESO, AUI/NRAO and NAOJ. The National Radio Astronomy Observatory is a facility of the National Science Foundation operated under cooperative agreement by Associated Universities, Inc. D.L. was supported by a grant from VILLUM FONDEN (project number 16599). N.F.W.L. acknowledges funding by the Swiss National Science Foundation (SNSF) Ambizione grant 193453. 

\end{acknowledgements}

\bibliographystyle{aa}
\bibliography{ALMAGAL-N-COMs}

\begin{appendix} 

\section{Spectroscopic data}
\label{sec:spec_data}

The line list and the respective spectroscopic data for CH$_3$CN and CH$_3^{13}$CN are taken from the JPL database (\citealt{Kukolich1973}; \citealt{Boucher1977}; \citealt{Demaison1979}; \citealt{Kukolich1982}; \citealt{Anttila1993}; \citealt{Pearson1996}; \citealt{Cazzoli2006}; \citealt{Muller2009}). The vibration factor can be assumed to be similar for CH$_3$CN and CH$_3^{13}$CN at the excitation temperatures found in this work. The vibration factor for CH$_3$CN is less than 1.1 for temperatures below 180\,K (\citealt{Muller2015}) and hence are ignored in this paper.  

The line list for C$_2$H$_5$CN is taken from the CDMS (\citealt{Mader1974}; \citealt{Johnson1977}; \citealt{Boucher1980}; \citealt{Pearson1994}; \citealt{Fukuyama1996}). The vibrational correction factor for this molecule is between ${\sim} 1.1$ and ${\sim} 2.1$ for the temperatures assumed here (\citealt{Heise1981}). For temperatures below 120\,K we have assumed no vibrational correction factor as it is negligible. For sources with assumed excitation temperature 120\,K $\leq T_{\rm ex} \leq$ 140\,K we use a vibrational correction factor of 1.3174, if $T_{\rm ex}=150$\,K a correction factor of 1.3848 is used, if $T_{\rm ex}$ is 160\,K or 170\,K we use 1.4591, if $T_{\rm ex}=200$\,K a correction factor of 1.8316 is used and finally when $T_{\rm ex}=220$\,K we use 2.1318.

For C$_2$H$_3$CN the spectroscopic data are taken from the CDMS (\citealt{Gerry1973}; \citealt{Stolze1985}; \citealt{Cazzoli1988}; \citealt{Demaison1994}; \citealt{Baskakov1996}; \citealt{Colmont1997}; \citealt{Muller2008}). Low lying vibrational modes are included in its partition function. The line lists for HNCO and its isotopologues are taken from the JPL database (\citealt{Hocking1975}; \citealt{Pickett1998}). 

NH$_2$CHO v=0 and NH$_2$CHO v$_12$=1 entries are taken from the CDMS (\citealt{Kukolich1971}; \citealt{Hirota1974}; \citealt{Moskienko1991}; \citealt{Vorobeva1994}; \citealt{Kryvda2009}; \citealt{Motiyenko2012}). For all the sources a vibrational correction factor of 1.5 for 300\,K is used. CH$_3$NH$_2$ spectroscopic data were taken from the JPL database (\citealt{Ilyushin2005}; \citealt{Kreglewski1992}; \citealt{Kreglewski1992b}; \citealt{Ohashi1987}; \citealt{Takagi1971}; \citealt{Nishikawa1957}; \citealt{Shimoda1954}). Methanol line list is taken from the CDMS database (\citealt{Lees1968}; \citealt{Pickett1981}; \citealt{Xu2008}).

\section{Normalization in Figure \ref{fig:NH2CHO_HNCO_T}}
\label{app:norm}

Figure \ref{fig:NH2CHO_HNCO_T} shows the absolute values of column densities of different works in its bottom panel. However, not all the works from the literature present their fitted lines to the absolute values of column densities but they show the fit to the abundances of the two molecules NH$_2$CHO and HNCO. Therefore, to use their fits in Fig. \ref{fig:NH2CHO_HNCO_T} one needs to use a method to convert a fit to the abundances to a fit to column densities. Here this method is explained. 

The fit to the column densities can be written as:

\begin{equation}
    N_{\rm{NH}_{2}\rm{CHO}} = x^{\prime} \times N_{\rm HNCO}^{y^{\prime}},
    \label{eq:col_fit}
\end{equation}
\noindent and the fit to the abundances can be written as:

\begin{equation}
    N_{\rm{NH}_{2}\rm{CHO}}/N_{\rm H} = x \times \left(\frac{N_{\rm HNCO}}{N_{\rm H}}\right)^{y}.
    \label{eq:abund_fit}
\end{equation}

\noindent By rearranging Eq. \eqref{eq:abund_fit}

\begin{equation}
    N_{\rm{NH}_{2}\rm{CHO}} = x N_{\rm H}/N_{\rm H}^{y} \times N_{\rm HNCO}^{y}.
    \label{eq:abund_re}
\end{equation}

Therefore, by comparing Equations \eqref{eq:col_fit} and \eqref{eq:abund_re} one can see that $y = y^{\prime}$ and hence the slopes of the two fits can be compared. However, $x^{\prime} = x N_{\rm H}^{1-y}$ and thus a value for the column density of Hydrogen is needed. Here we assume an $N_{\rm H} = 10^{23}$ for simplicity to normalize all the values from different works.  

\section{Chemical toy model}
\label{sec:toy}

We define the pre-stellar phase to be as when the cloud is becoming more centrally concentrated and the densities increase while the temperatures remain low (${\sim} 10$\,K). This is the phase during which the extinction, $A_{\rm v}$, increases from ${\sim}1-2$ (i.e. translucent cloud; \citealt{Snow2006}) to ${\gtrsim} 9$, the CO freeze-out takes place (\citealt{Boogert2015}), and is similar to the `collapse stage' in chemical models of \cite{Garrod2022}. Based on the abundances calculated in the collapse stage of \cite{Garrod2022} models (their Fig. 9), they increase rapidly in the first $< 10^5$\,yr of the collapse and then increase with a shallower slope from $10^5$\,yr to $10^6$\,yr. Given that the pre-stellar phase of star formation usually lasts around $10^5$\,yr and $10^6$\,yr (supported by the values found from water and CO observations; \citealt{Schmalzl2014}; \citealt{Jorgensen2005}), we particularly focus on molecular abundances between ${\sim} 10^5$ and ${\sim} 10^6$ years in our toy model. Therefore, we define the abundance, $X_{\rm i}$, of species $\rm i$ with respect to hydrogen as 

\begin{equation}
    X_{\rm i} = A_{\rm i} \left(\frac{t}{10^6 \rm{yr}}\right)^{\alpha_{\rm i}},
    \label{eq:dens_abund}
\end{equation}

\noindent where $A_{\rm i}$ is the ice abundance of each molecule after $10^6$ years, $t$ is the duration of the pre-stellar phase and $\alpha_{\rm i}$ is a positive unique constant for each molecule. 

We approximate the ice abundance of each molecule at 10$^6$\,yr, $A_{\rm i}$, using the ice abundance of each species at the end of the collapse stage of \cite{Garrod2022} models. These models are only used as an inspiration for the parametrization and are only an example of the chemical models available in the literature. The same can be done with other chemical models of the pre-stellar phase in the literature. Moreover, $\alpha_{\rm i}$ can be estimated using the ice abundances at $10^5$\,yr and $10^6$\,yr in Fig. 9 of \cite{Garrod2022} for each molecule. Therefore, Eq. \eqref{eq:dens_abund} gives the abundance of each species (in solids) between $10^5$\,yr and $10^6$\,yr into the collapse (typical age for the pre-stellar phase). The assumed $\alpha_{\rm i}$ are given in Table \ref{tab:alphas} and the resulting abundances are plotted in Fig. \ref{fig:abundances}. It is important to note that, as explained in Sect. \ref{sec:chem_models}, the \cite{Garrod2022} models likely underestimate the ice abundance of CH$_3$CN. Therefore, in our toy model we assume a larger $A_{\rm CH_3CN}$ to be consistent with observations. The same is done for CH$_3$NH$_2$, where the models likely overestimate the abundance of this molecule in ices (see Sect. \ref{sec:chem_models}).

Using Eq. \eqref{eq:dens_abund} we can find the abundance ratios (assumed to be the same as column density ratios) of two species 1 and 2 as

\begin{equation}
    X_{1,2} = A_{1,2} \left(\frac{t}{10^6}\right)^{\alpha_{1} - \alpha_{2}}.
    \label{eq:abund_ratio}
\end{equation}

\noindent Here $X_{1}/X_2$ and $A_{1}/A_2$ are written as $X_{1,2}$ and $A_{1,2}$ respectively. If the left hand side of Eq. \eqref{eq:abund_ratio} is the mean of the observed abundance ratios ($\overline{X_{1,2}}$), $t$ on the right hand side corresponds to the average time that various clouds spend in the pre-stellar phase ($\Bar{t}$); in other words:

\begin{equation}
    \overline{X_{1,2}} = A_{1,2} \left(\frac{\Bar{t}}{10^6}\right)^{\alpha_{1} - \alpha_{2}}.
    \label{eq:abund_ratio_mean}
\end{equation}

\noindent This equation can be used to replace $A_{1,2}$ in Eq. \eqref{eq:abund_ratio}, to give Eq. \eqref{eq:time_scatter} in the main text. If $\alpha_1 \simeq \alpha_2$, the correlation between $X_{1,2}$ and $t$ would be weak, making them degenerate. Hence, it will not be possible to infer $t$ from $X_{1,2}$ confidently. This is true for the ratios of C$_2$H$_3$CN/CH$_3$CN and HNCO/CH$_3$CN where $\alpha_1 - \alpha_2$ is 0.3 and 0.2 respectively (see Table \ref{tab:alphas}). However, since this is not the case for other ratios in Fig. \ref{fig:spread_factor} (i.e. $\alpha_1 \neq \alpha_2$), the distribution of the pre-stellar timescales can be inferred from the distributions of abundance ratios (see Fig. \ref{fig:time_spread}).

\section{Additional plots}
\label{sec:add_plots}

\begin{figure*}
    \centering
    \includegraphics[width=17cm]{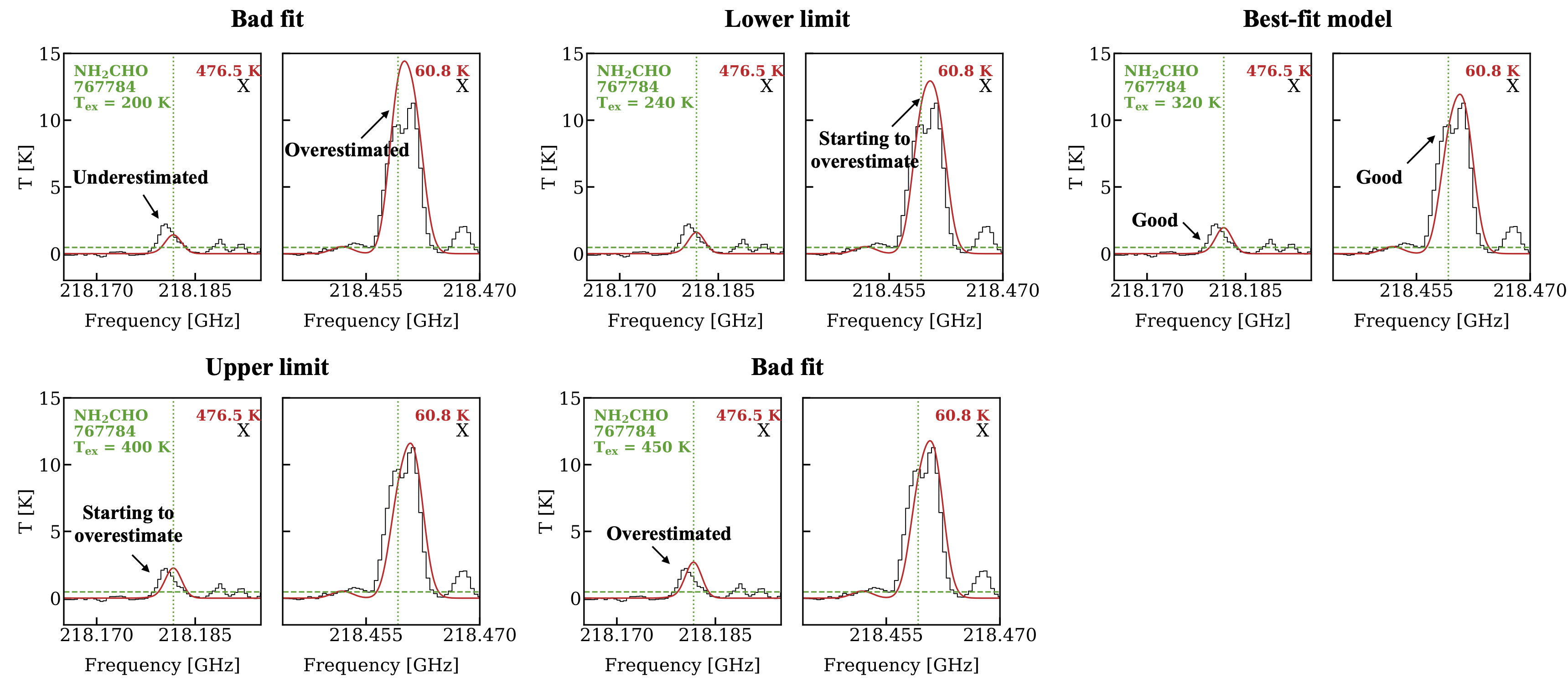}
    \caption{The demonstration of how excitation temperature upper and lower bounds are determined for NH$_2$CHO for source 767784. The various symbols are the same as Fig. \ref{fig:881427_CH3C-13-N}.}
    \label{fig:plots_NH2CHO}
\end{figure*}

\begin{figure*}
    \centering
    \includegraphics[width=17cm]{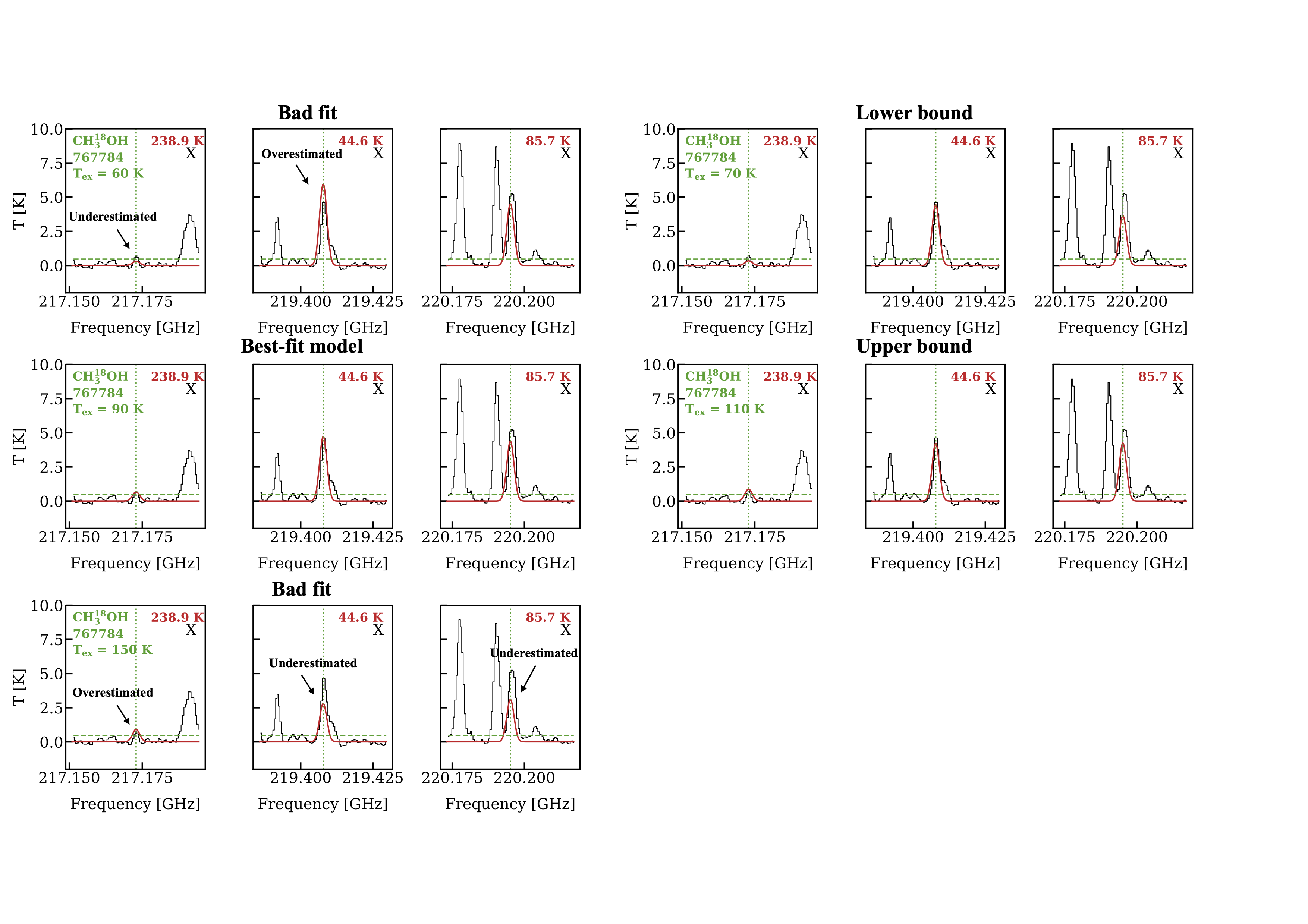}
    \caption{The demonstration of how excitation temperature upper and lower bounds are determined for CH$_3^{18}$OH for source 767784. The various symbols are the same as Fig. \ref{fig:881427_CH3C-13-N}.}
    \label{fig:plots_CH3O18H}
\end{figure*}

\begin{figure*}
    \centering
    \includegraphics[width=17cm]{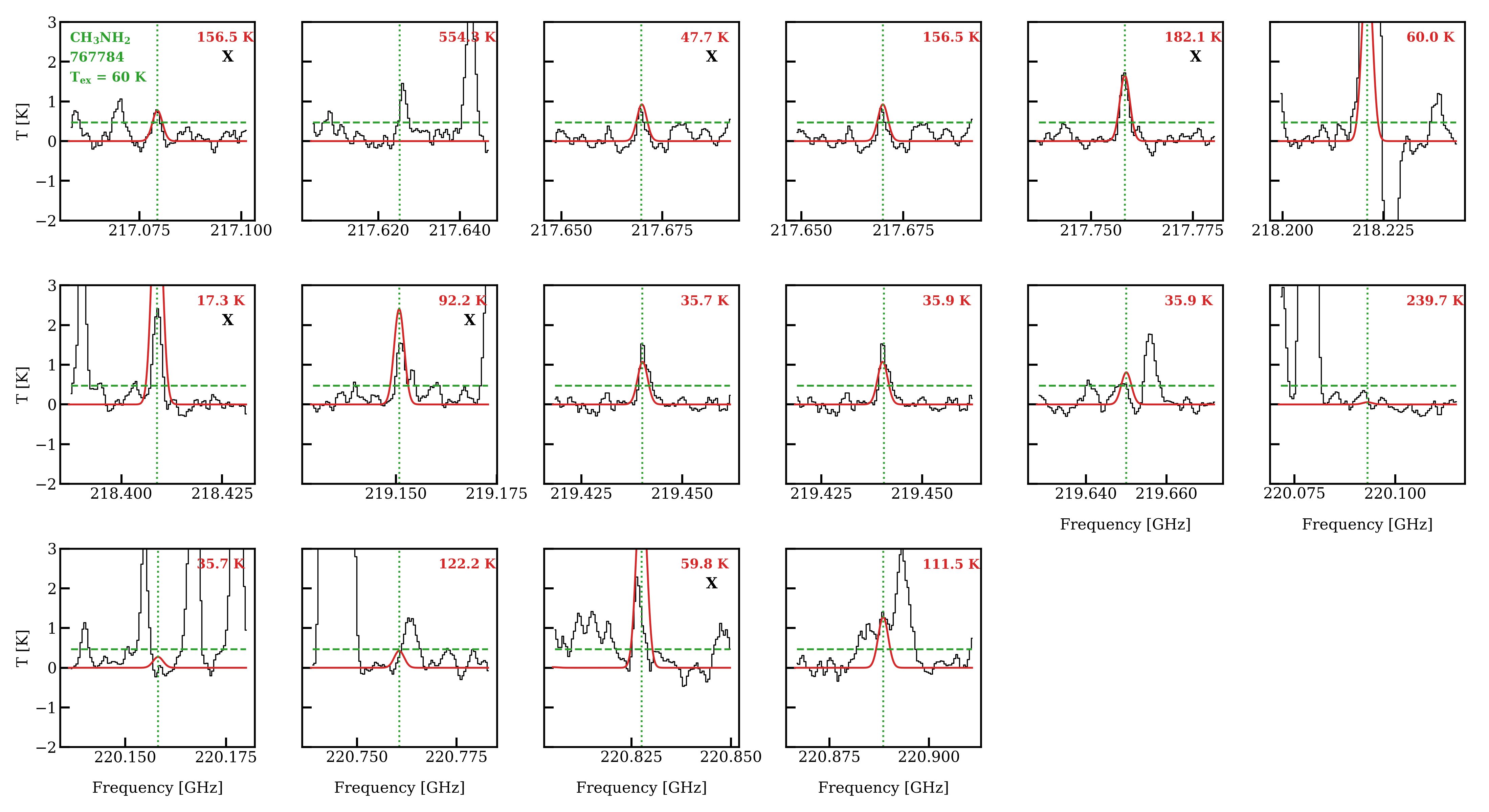}
    \caption{Showing that excitation temperature of 60\,K does not match the observations for CH$_3$NH$_2$ in source 767784. The various symbols are the same as Fig. \ref{fig:881427_CH3C-13-N}. Here a few lines are overestimated, e.g. the ones with $E_{\rm up}$ of 17.3\,K and 92.2\,K and if the column density is decreased lines such as the ones with $E_{\rm up}$ of 156.5\,K and 182.1\,K will be underestimated.}
    \label{fig:bad_CH3NH2_lower}
\end{figure*}

\begin{figure*}
    \centering
    \includegraphics[width=17cm]{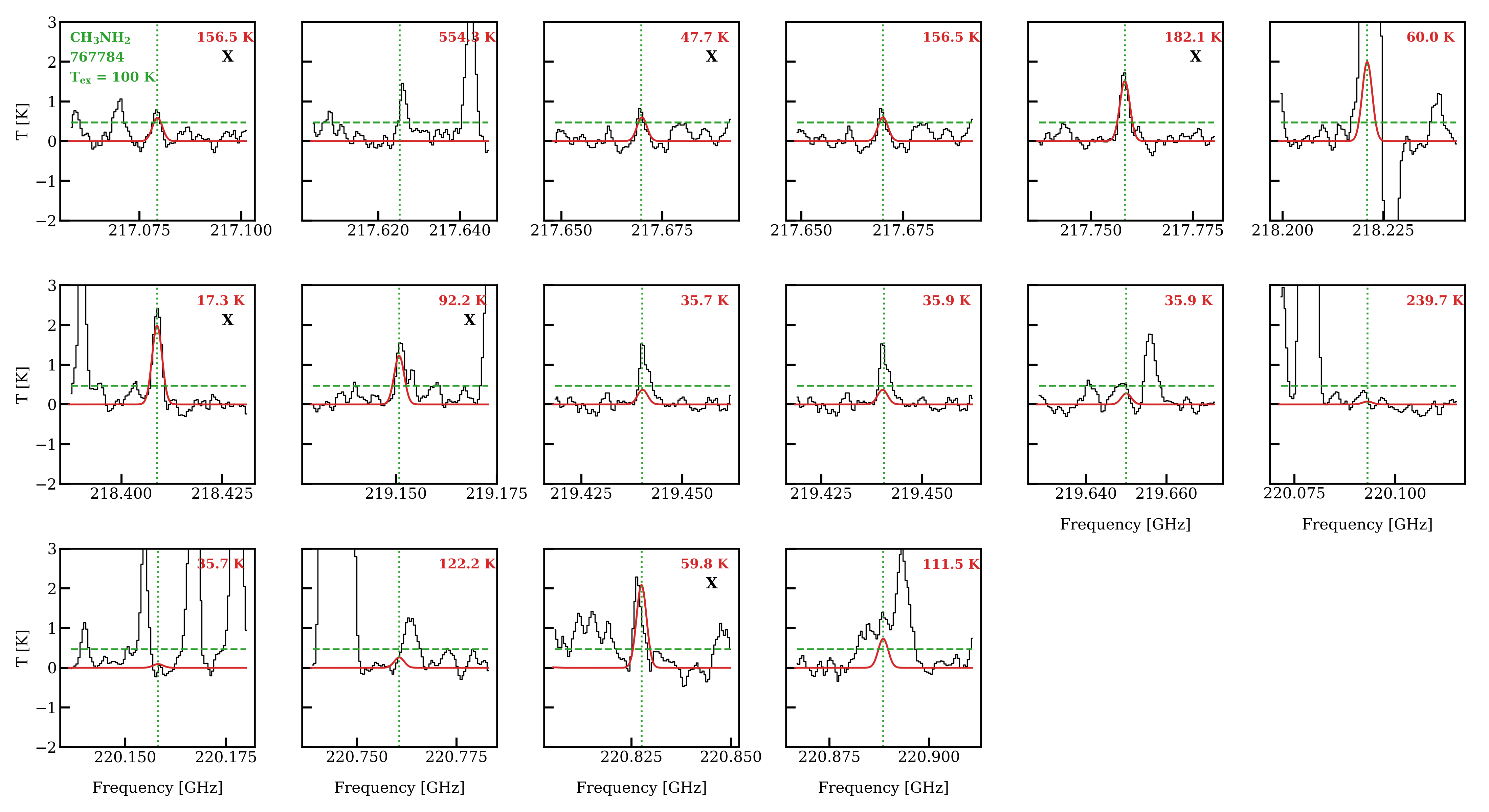}
    \caption{The best-fit model for CH$_3$NH$_2$ in source 767784. The various symbols are the same as Fig. \ref{fig:881427_CH3C-13-N}.}
    \label{fig:best_CH3NH2}
\end{figure*}

\begin{figure*}
    \centering
    \includegraphics[width=17cm]{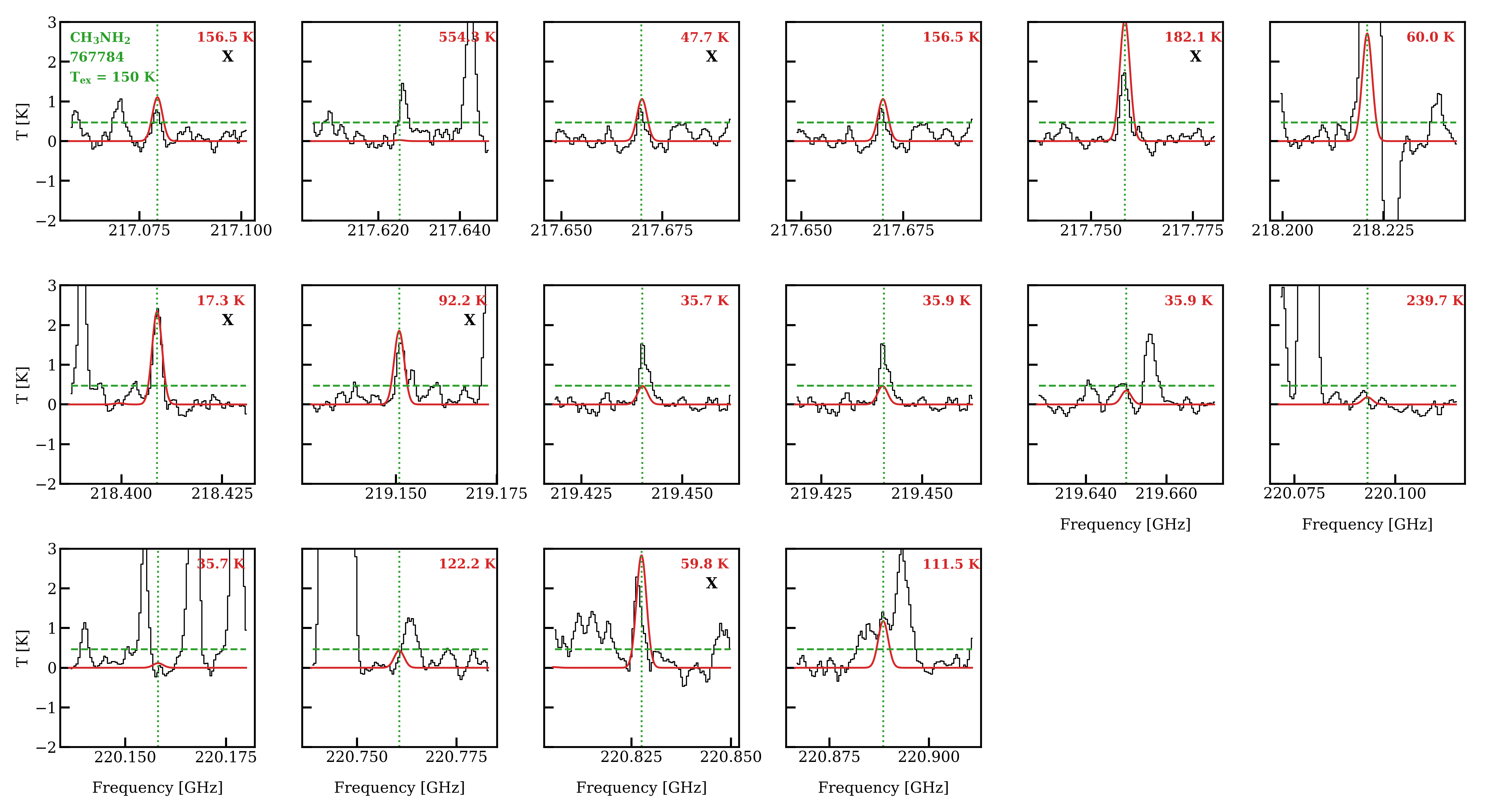}
    \caption{Showing that excitation temperature of 150\,K does not match the observations for CH$_3$NH$_2$ in source 767784. The various symbols are the same as Fig. \ref{fig:881427_CH3C-13-N}. Again here some lines are overestimated and if the column density is decreased the lines that are perfect fits now will be underestimated.}
    \label{fig:bad_CH3NH2_upper}
\end{figure*}

\begin{figure*}
    \centering
    \includegraphics[width=17cm]{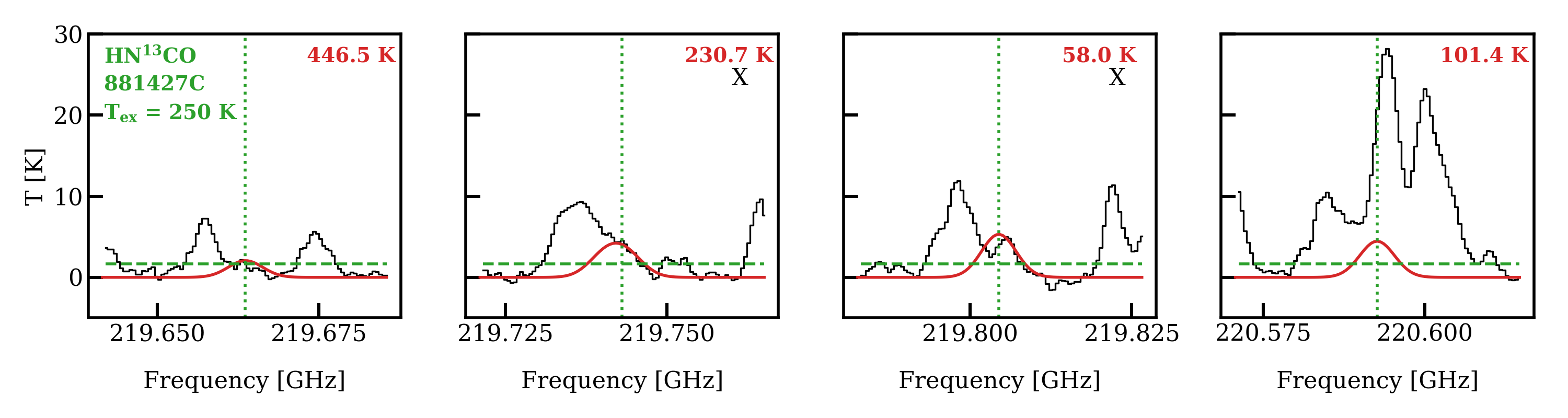}
    \caption{The best model for HN$^{13}$CO of source 881427 (red) on top of the data (black). In this figure only lines with $A_{\rm ij} > 10^{-6}$\,s$^{-1}$ and $E_{\rm up}< 700$\,K are shown. Moreover, only one representative plot of the hyper fine structure of this molecule is shown to avoid repetition. The detected lines that are mainly considered for
    the fits are marked with an ‘X’ on the top right. The dashed lines shows the 3$\sigma$ line.}
    \label{fig:881427_HNC-13-O}
\end{figure*}

\begin{figure*}
    \centering
    \includegraphics[width=17cm]{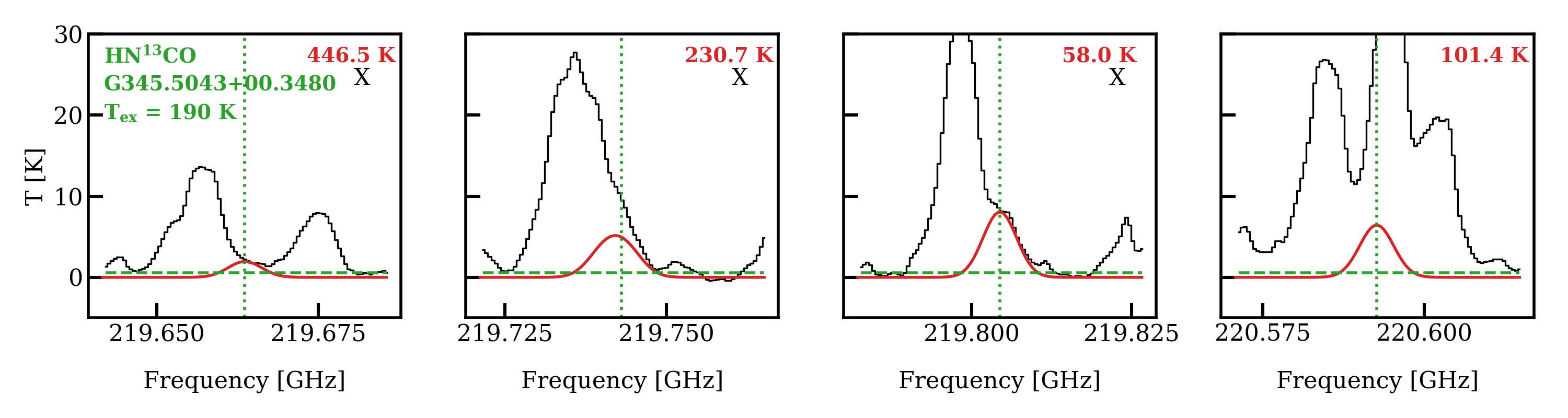}
    \caption{The same as Fig. \ref{fig:881427_HNC-13-O} but for G345.5043+00.3480.}
    \label{fig:G345_HNC-13-O}
\end{figure*}

\begin{figure*}
    \centering
    \includegraphics[width=17cm]{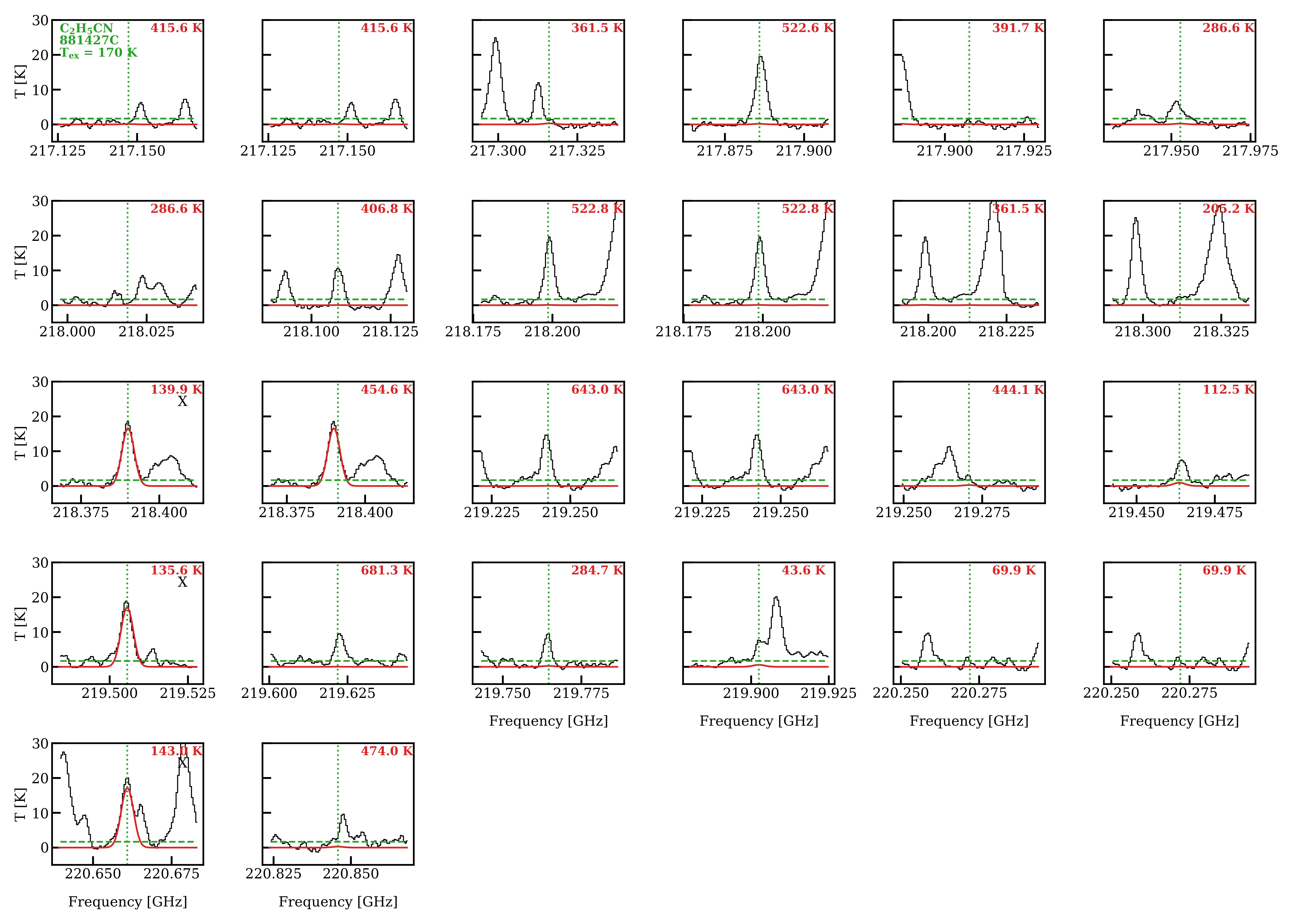}
    \caption{The same as Fig. \ref{fig:881427_HNC-13-O} but for C$_2$H$_5$CN.}
    \label{fig:881427_C2H5CN}
\end{figure*}

\begin{figure*}
    \centering
    \includegraphics[width=17cm]{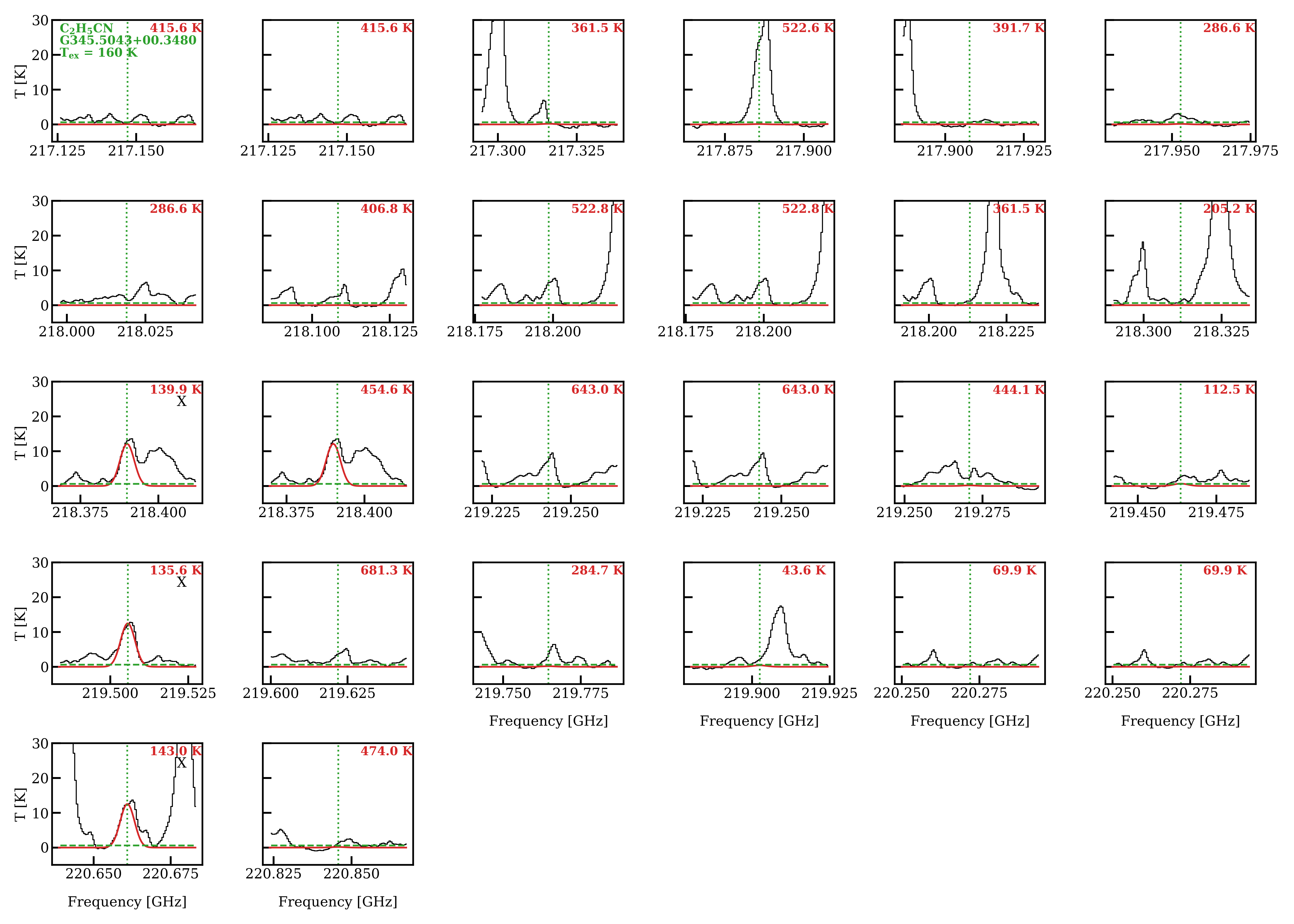}
    \caption{The same as Fig. \ref{fig:881427_HNC-13-O} but for G345.5043+00.3480 and C$_2$H$_5$CN.}
    \label{fig:G345_C2H5CN}
\end{figure*}

\begin{figure*}
    \centering
    \includegraphics[width=17cm]{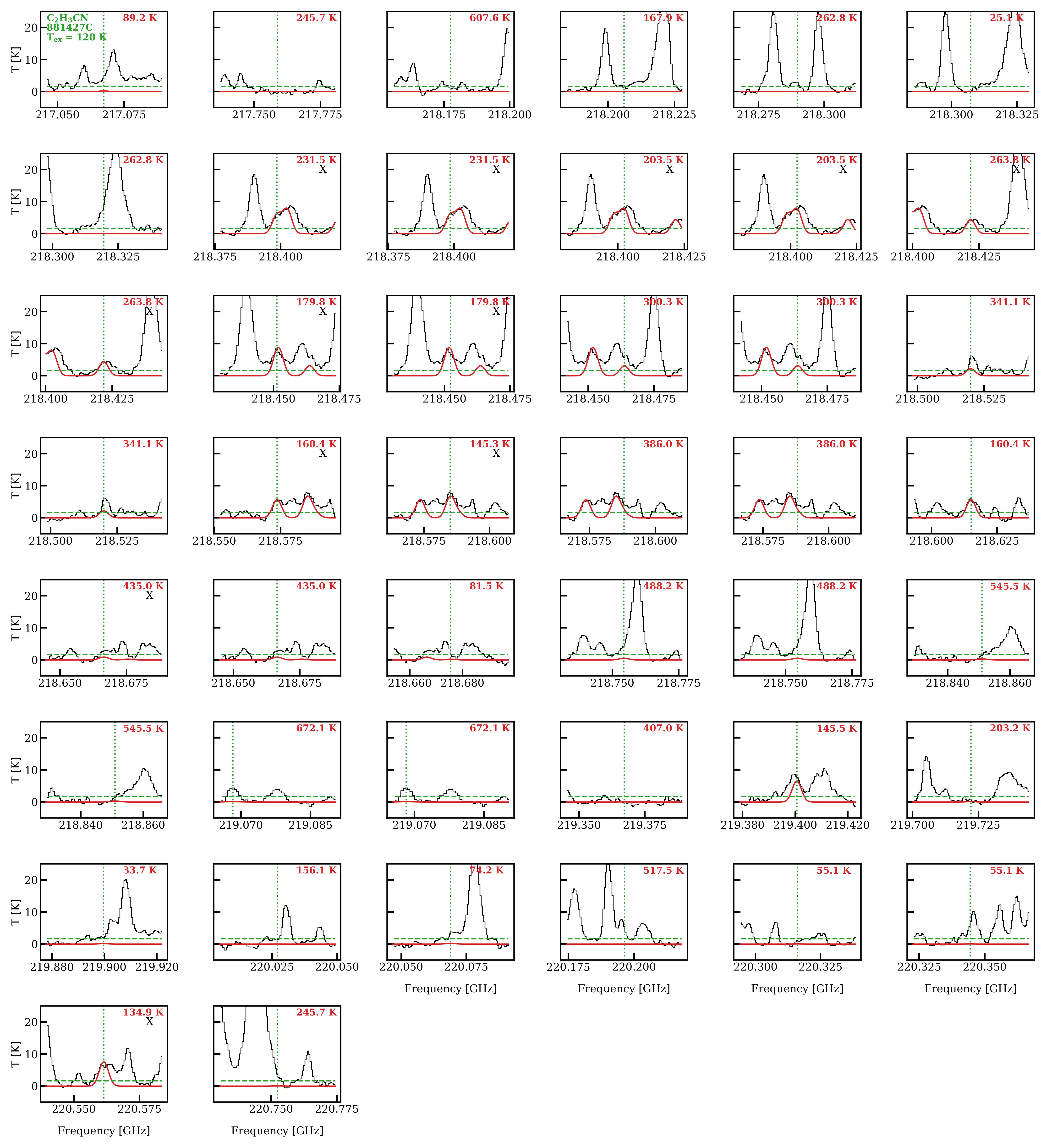}
    \caption{The same as Fig. \ref{fig:881427_HNC-13-O} but for C$_2$H$_3$CN.}
    \label{fig:881427_C2H3CN}
\end{figure*}

\begin{figure*}
    \centering
    \includegraphics[width=17cm]{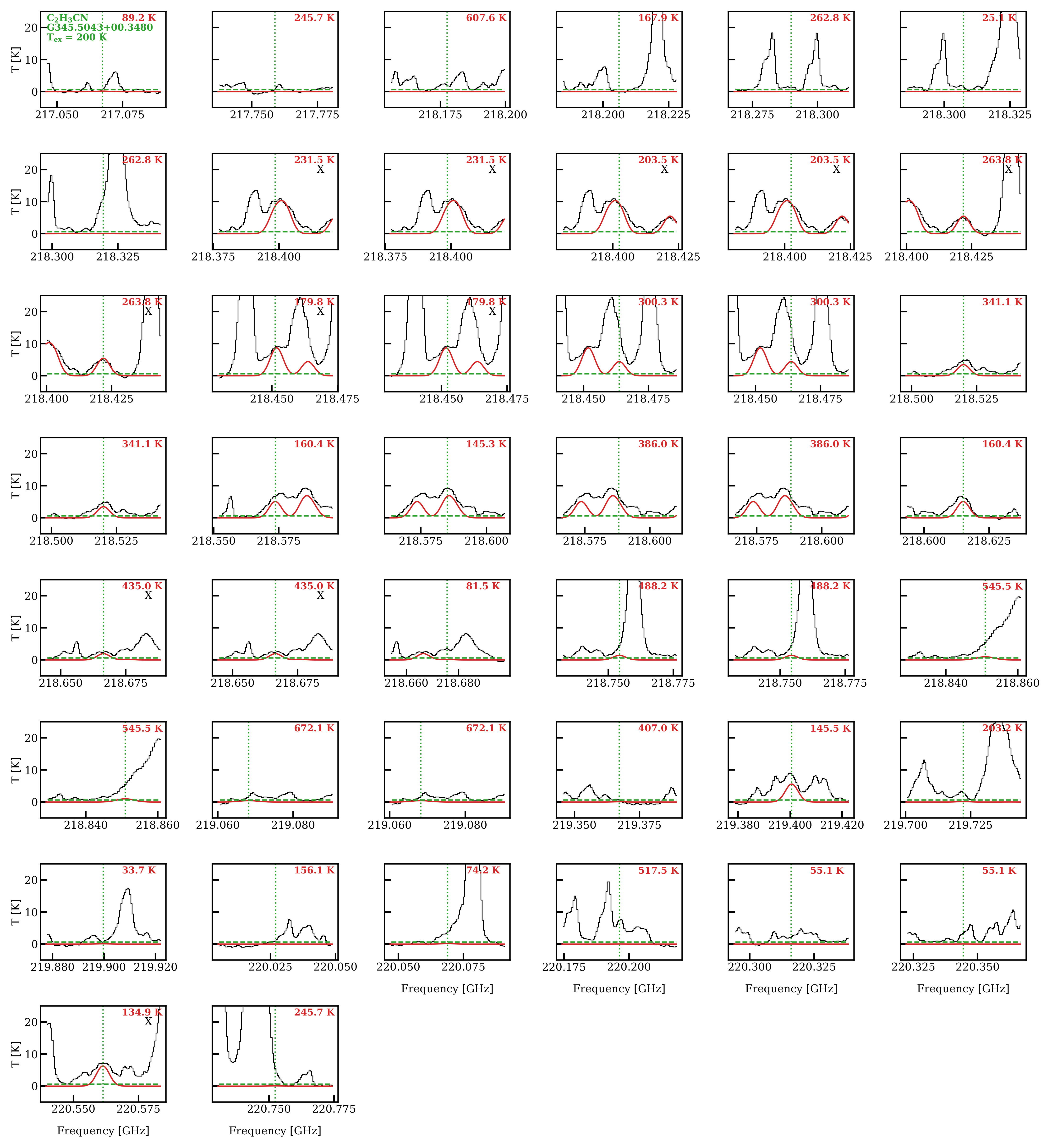}
    \caption{The same as Fig. \ref{fig:881427_HNC-13-O} but for G345.5043+00.3480 and C$_2$H$_3$CN.}
    \label{fig:G345_C2H3CN}
\end{figure*}

\begin{figure*}
    \centering
    \includegraphics[width=17cm]{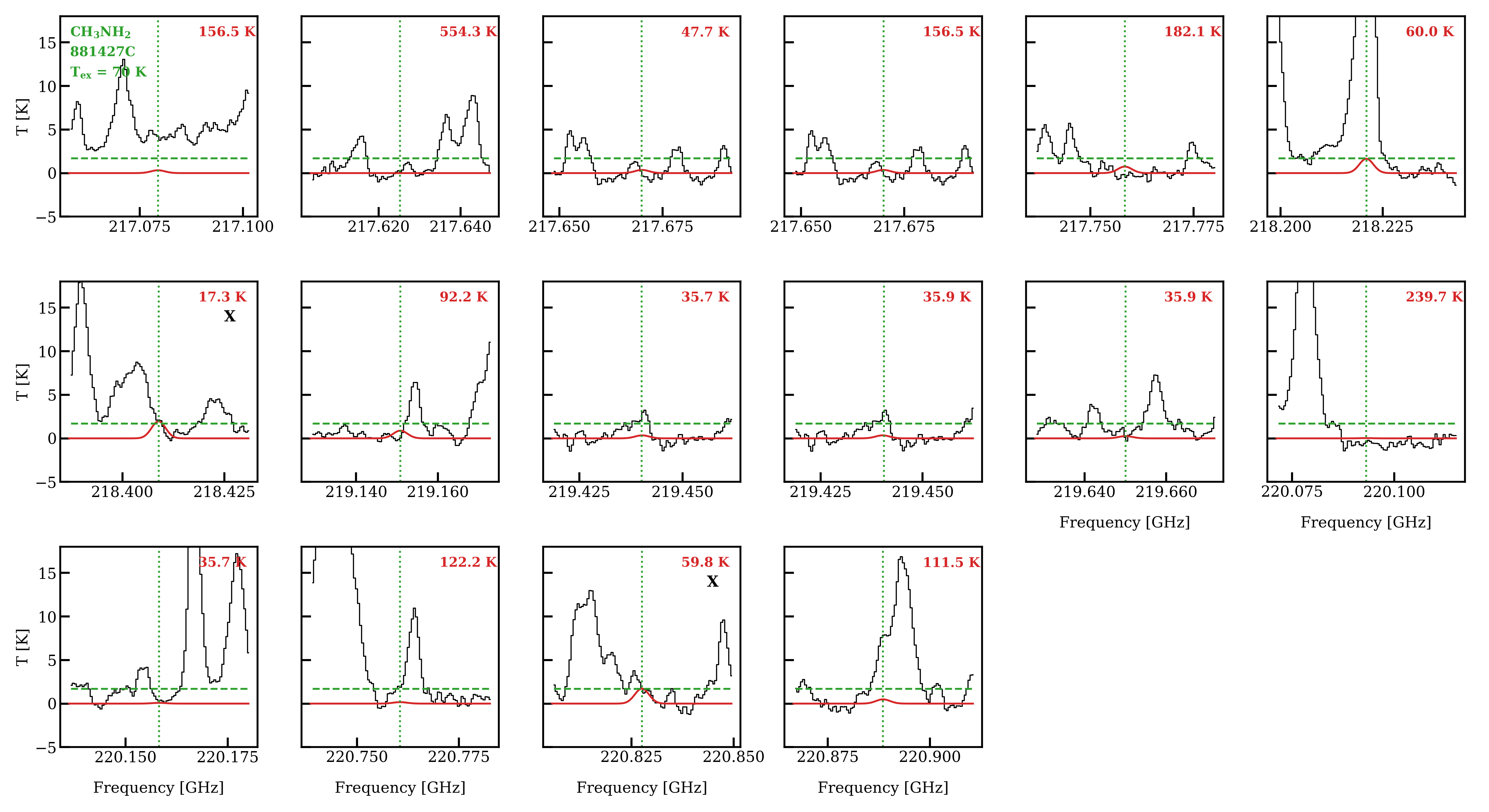}
    \caption{The same as Fig. \ref{fig:881427_HNC-13-O} but for CH$_3$NH$_2$.}
    \label{fig:881427_CH3NH2}
\end{figure*}

\begin{figure*}
    \centering
    \includegraphics[width=17cm]{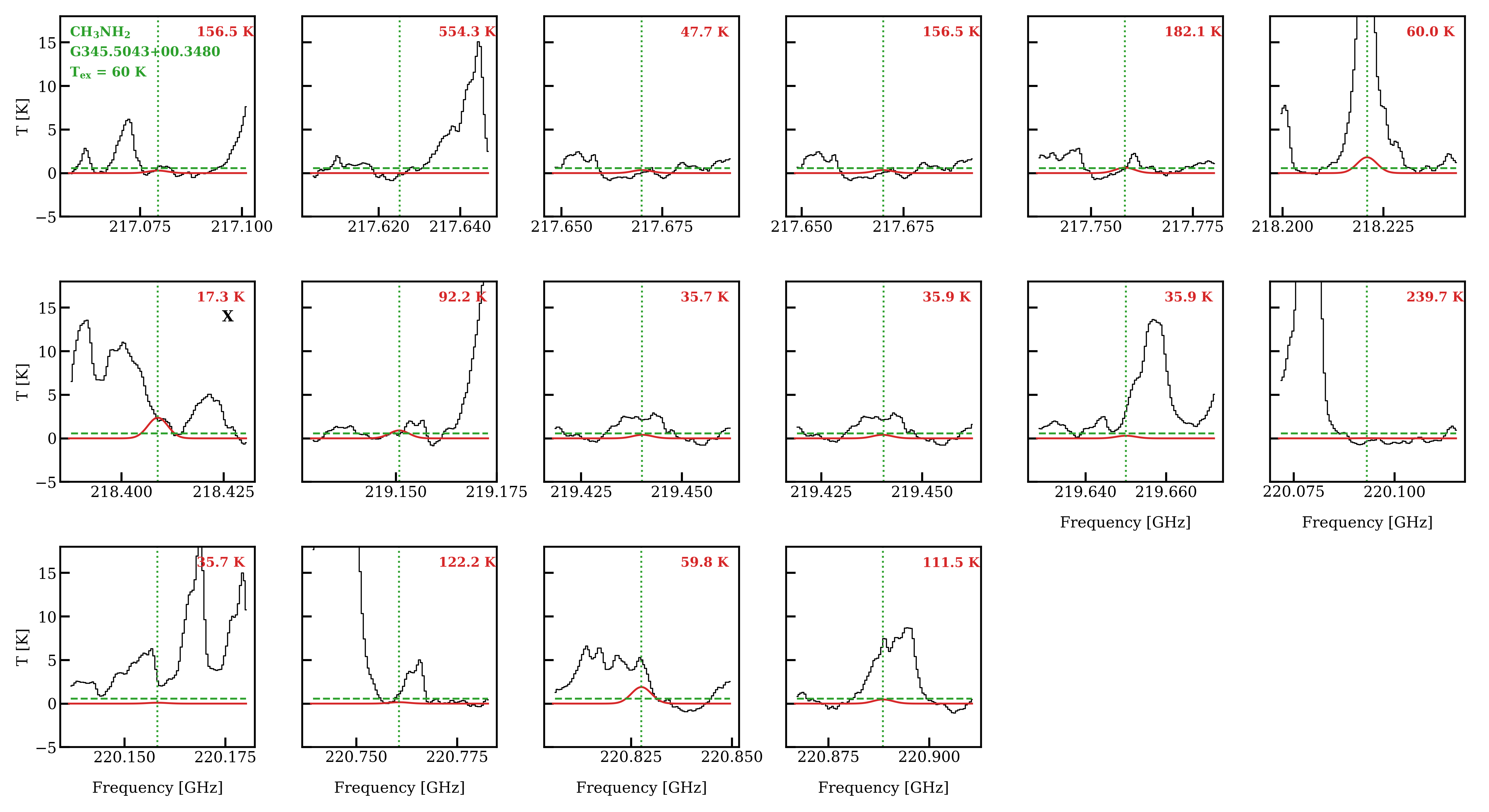}
    \caption{The same as Fig. \ref{fig:881427_HNC-13-O} but for G345.5043+00.3480 and CH$_3$NH$_2$.}
    \label{fig:G345_CH3NH2}
\end{figure*}

\begin{figure}
  \resizebox{\columnwidth}{!}{\includegraphics{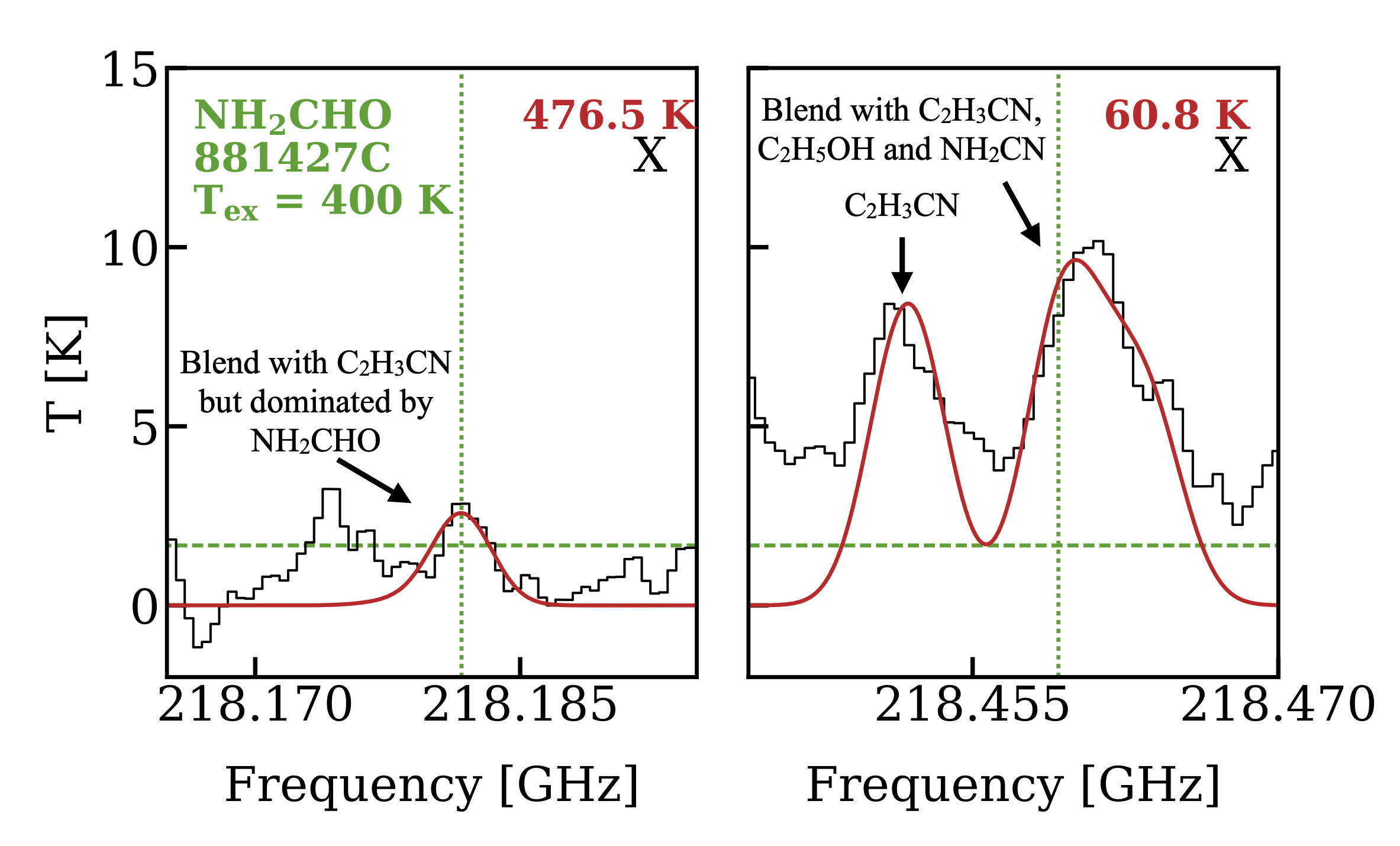}}
  \caption{The same as Fig. \ref{fig:881427_HNC-13-O} but for 881427C and NH$_2$CHO. Here only lines with $A_{\rm ij}>10^{-5}$\,s$^{-1}$ are shown.}
  \label{fig:881427_NH2CHO}
\end{figure}

\begin{figure}
  \resizebox{\columnwidth}{!}{\includegraphics{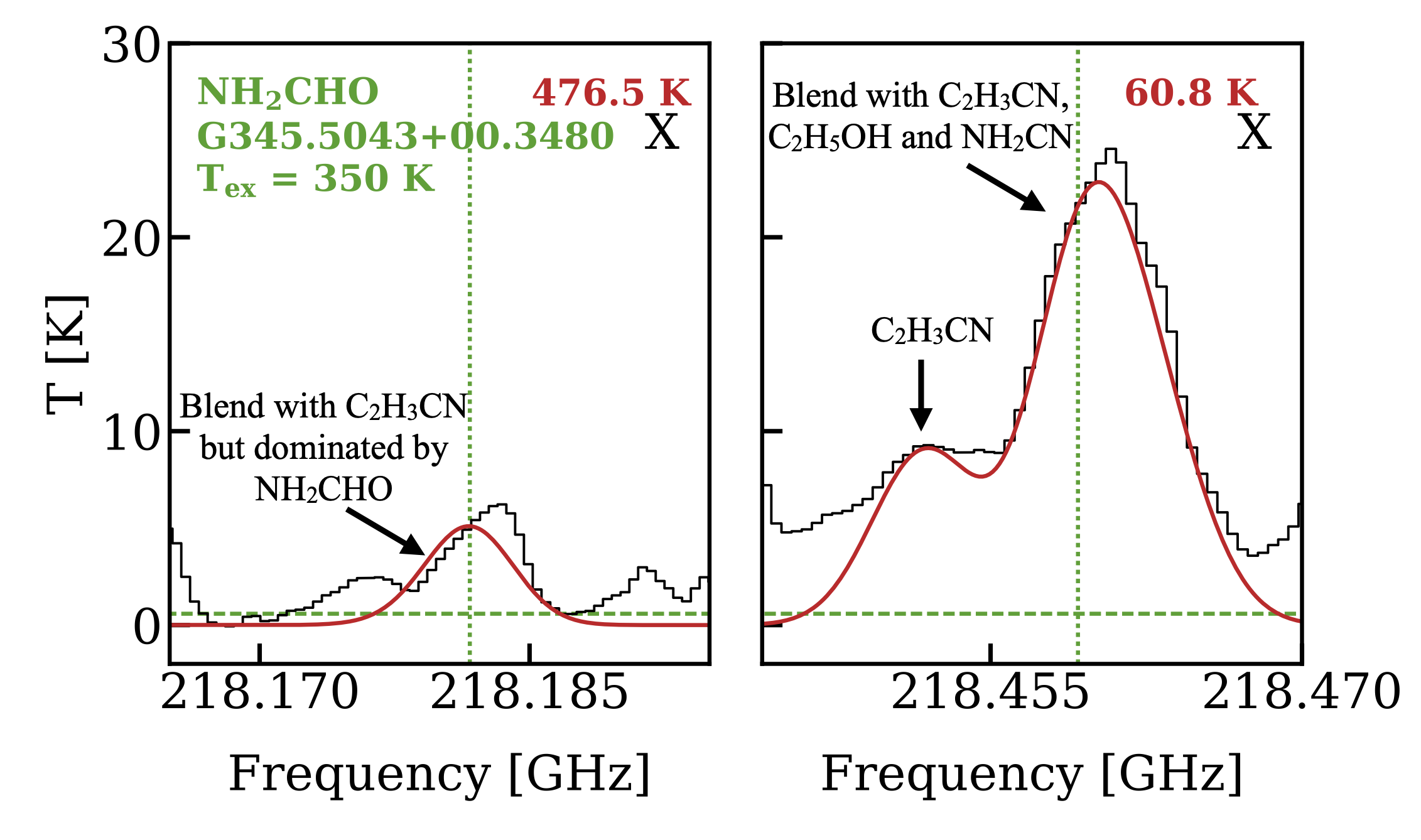}}
  \caption{The same as Fig. \ref{fig:881427_HNC-13-O} but for G345.5043+00.3480 and NH$_2$CHO. Here only lines with $A_{\rm ij}>10^{-5}$\,s$^{-1}$ are shown.}
  \label{fig:G345_NH2CHO}
\end{figure}

Figures \ref{fig:plots_NH2CHO}-\ref{fig:bad_CH3NH2_upper} show how the excitation temperature was measured for an example source 767784 and species NH$_2$CHO, CH$_3^{18}$OH and CH$_3$NH$_2$. Figures \ref{fig:881427_HNC-13-O}-\ref{fig:G345_NH2CHO} show the best-fit models of the N-bearing species considered in this work for two example sources 881427 (narrow lines) and G345.5043+00.3480 (broad lines). 

\begin{figure*}
    \centering
    \includegraphics[width=0.7\textwidth]{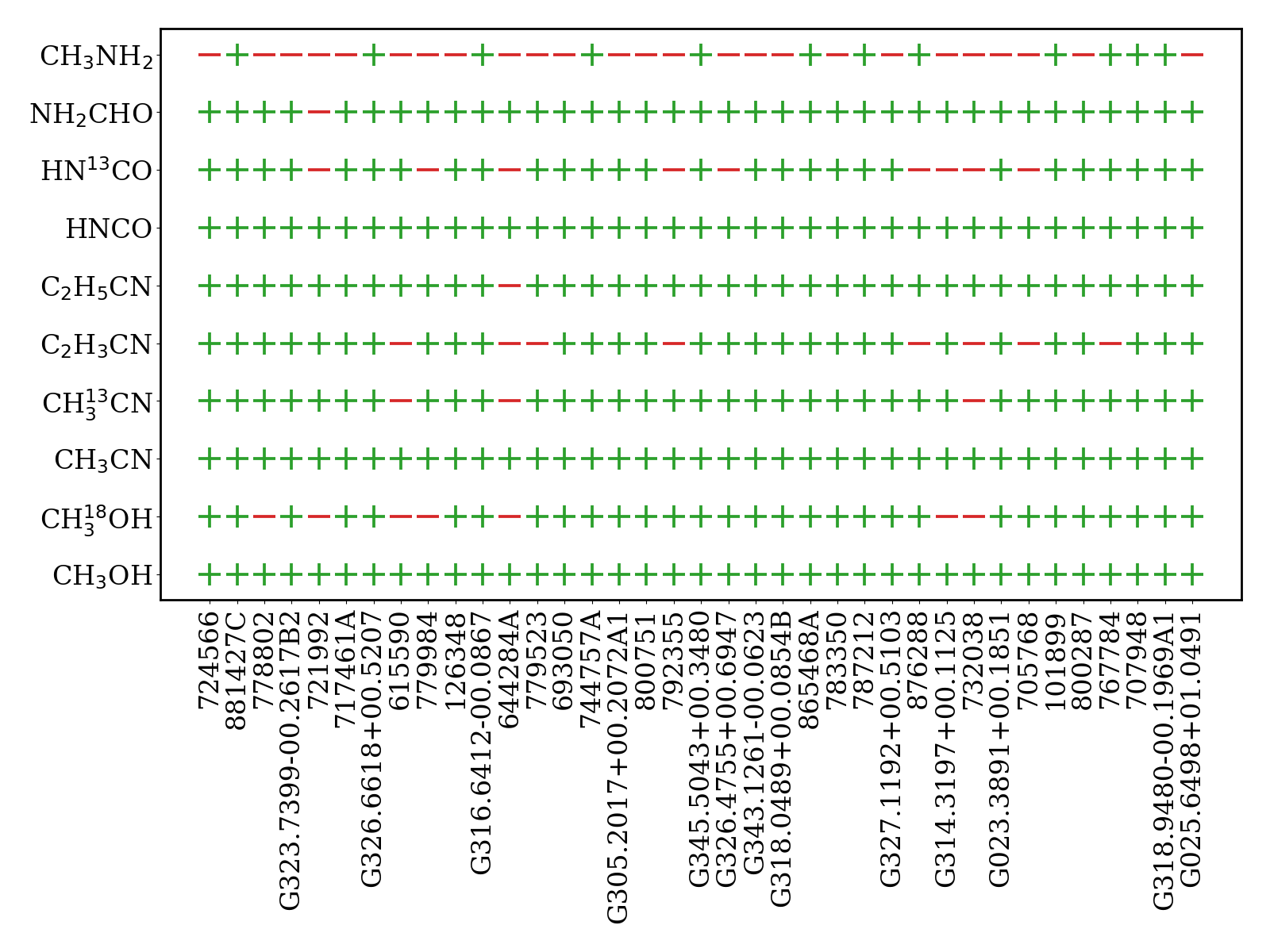}
    \caption{The same as Fig. \ref{fig:detections} but now the sources are ordered from left to right from the lowest to highest luminosity.} 
    \label{fig:detections_L}
\end{figure*}

\begin{figure*}
    \centering
    \includegraphics[width=1.05\textwidth]{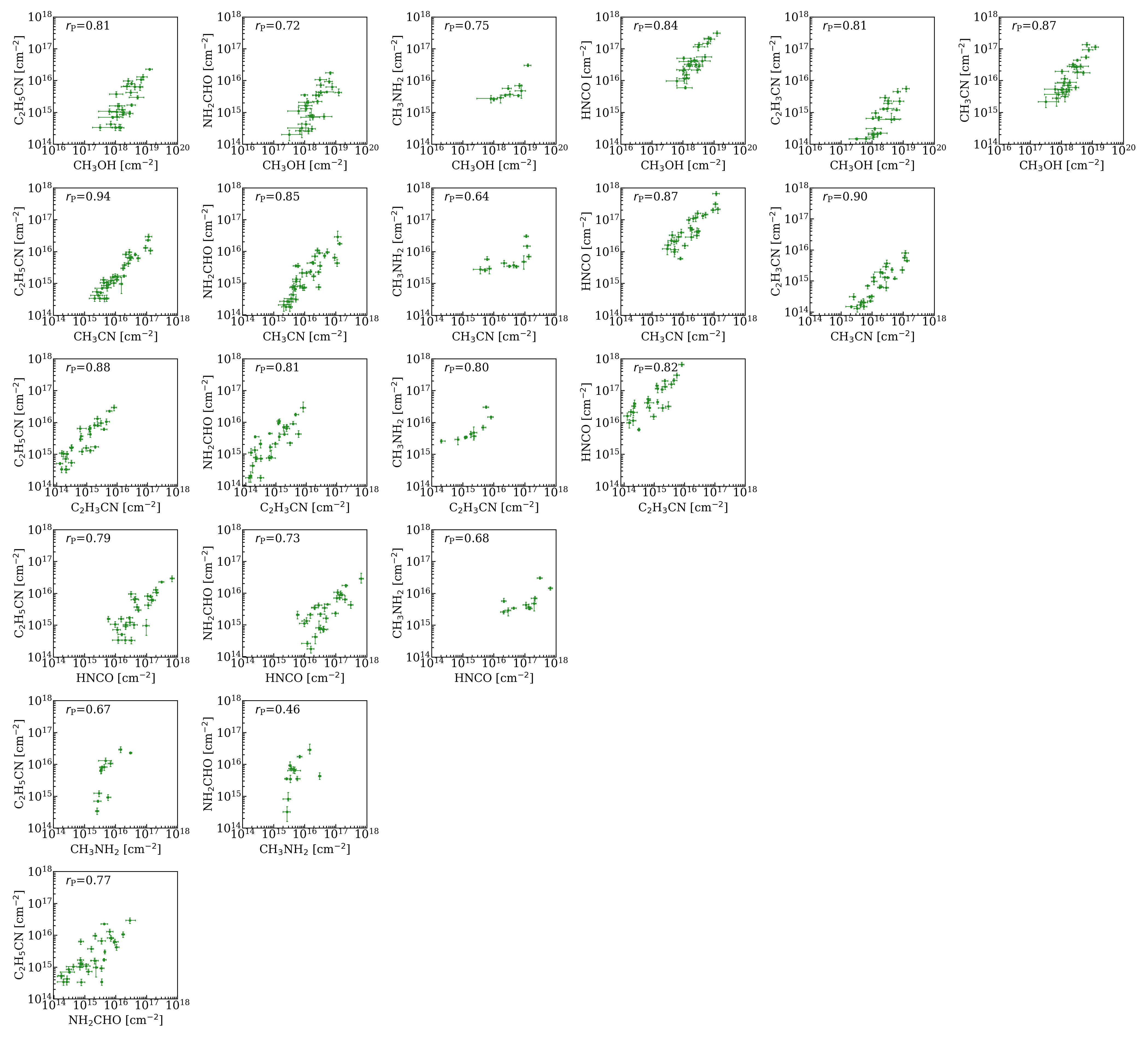}
    \caption{Correlations seen in column densities. Column densities of the molecules considered in this work plotted versus each other. Only detected molecules are plotted here. The Pearson's $r$ coefficient is given at the top left of each panel.} 
    \label{fig:column_column}
\end{figure*}

\begin{figure}
  \resizebox{\columnwidth}{!}{\includegraphics{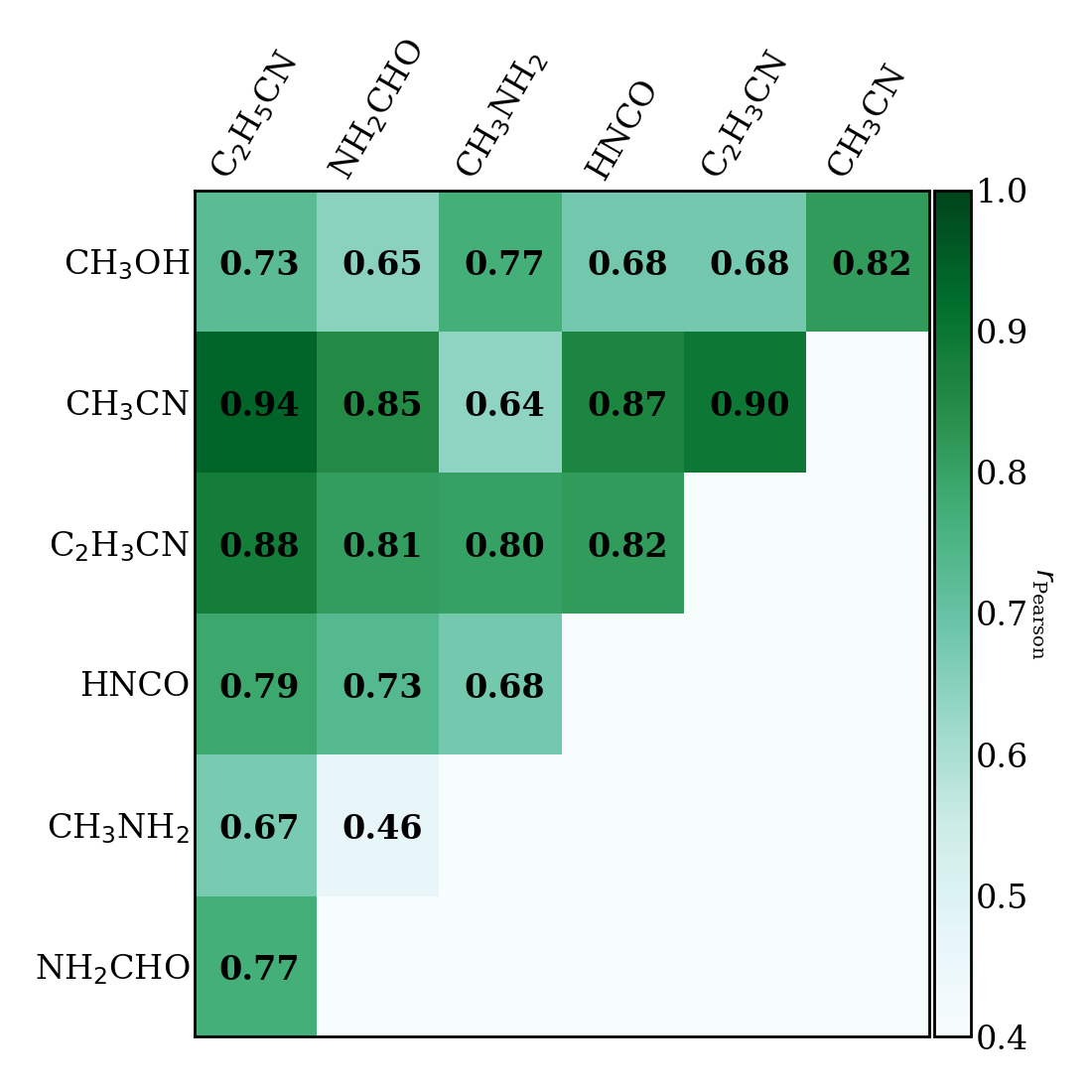}}
  \caption{The same as Fig. \ref{fig:matrix_cor} but column densities of methanol found from $^{13}$CH$_3$OH are also included.}
  \label{fig:matrix_cor_thick}
\end{figure}

\begin{figure*}
    \centering
    \includegraphics[width=15cm]{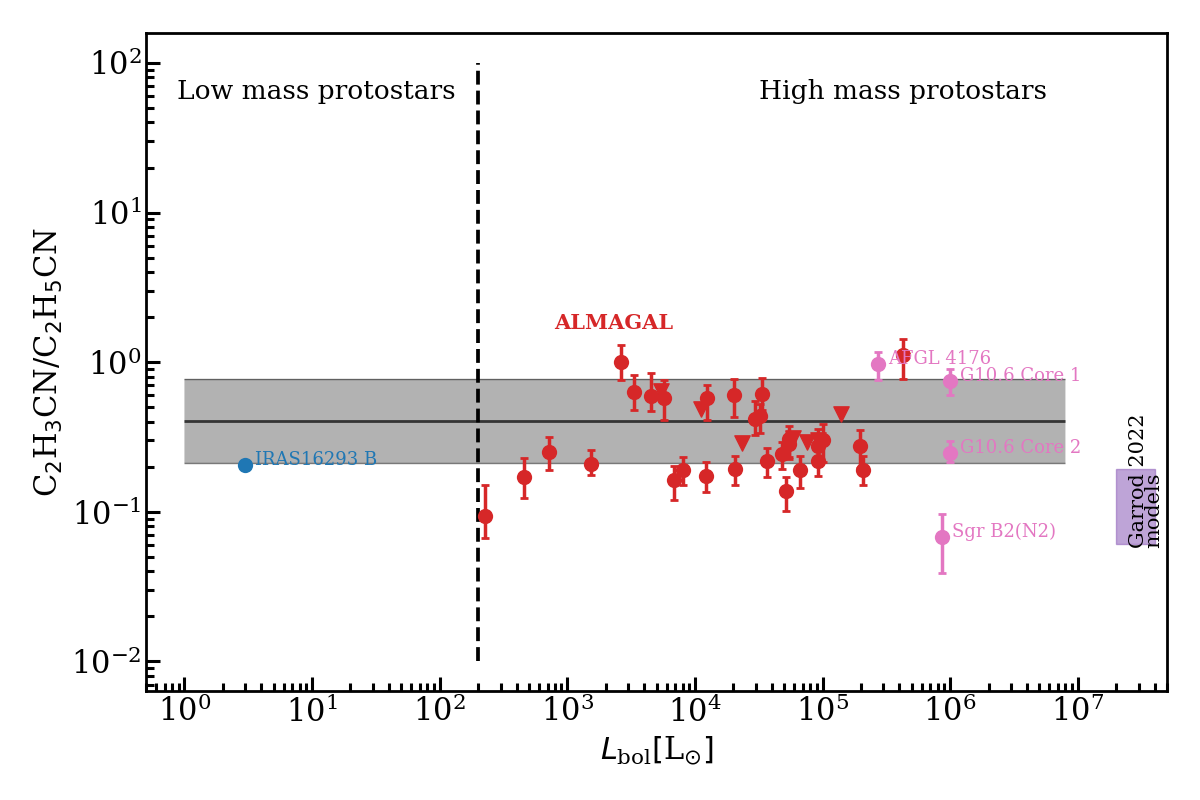}
    \caption{Same as Fig. \ref{fig:C2H5CN_CH3CN} but for C$_2$H$_3$CN/C$_2$H$_5$CN. The references are given in Table \ref{tab:refs}.} 
    \label{fig:C2H3CN_C2H5CN}
\end{figure*}

\begin{figure*}
    \centering
    \includegraphics[width=15cm]{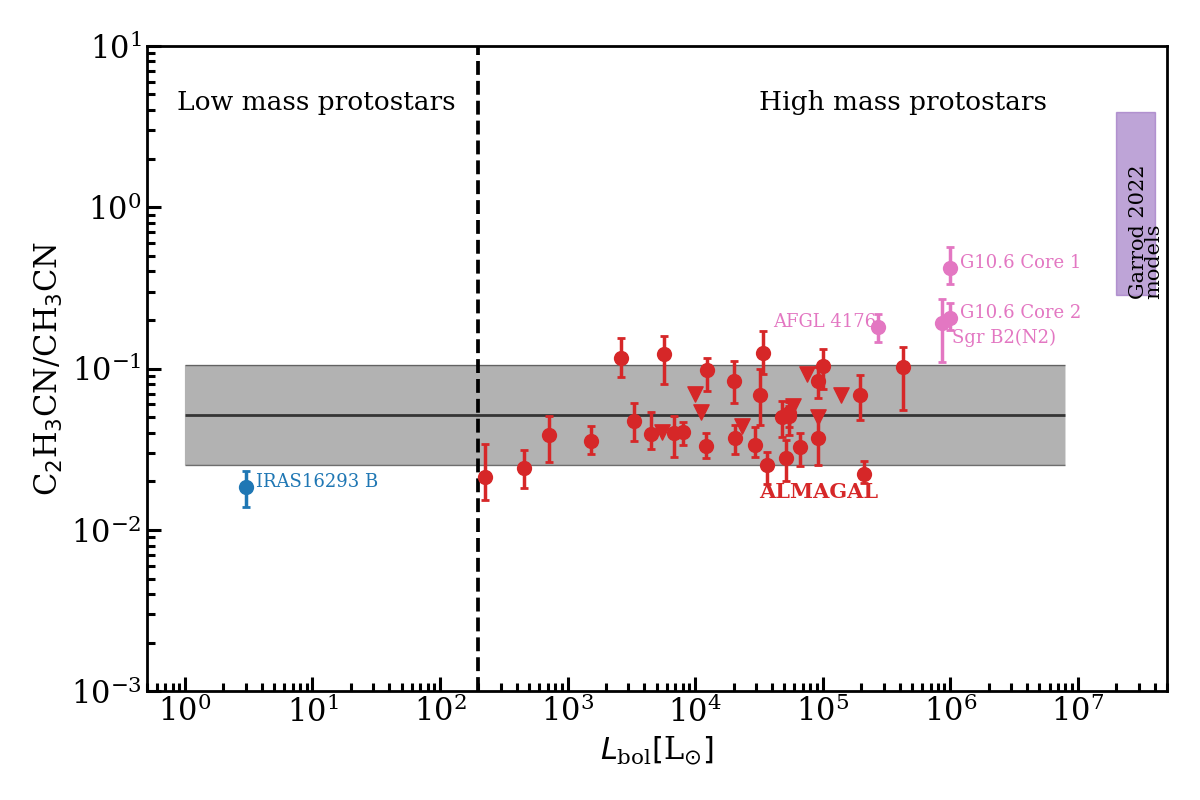}
    \caption{Same as Fig. \ref{fig:C2H5CN_CH3CN} but for C$_2$H$_3$CN/CH$_3$CN. The references are given in Table \ref{tab:refs}.} 
    \label{fig:C2H3CN_CH3CN}
\end{figure*}

\begin{figure*}
    \centering
    \includegraphics[width=14cm]{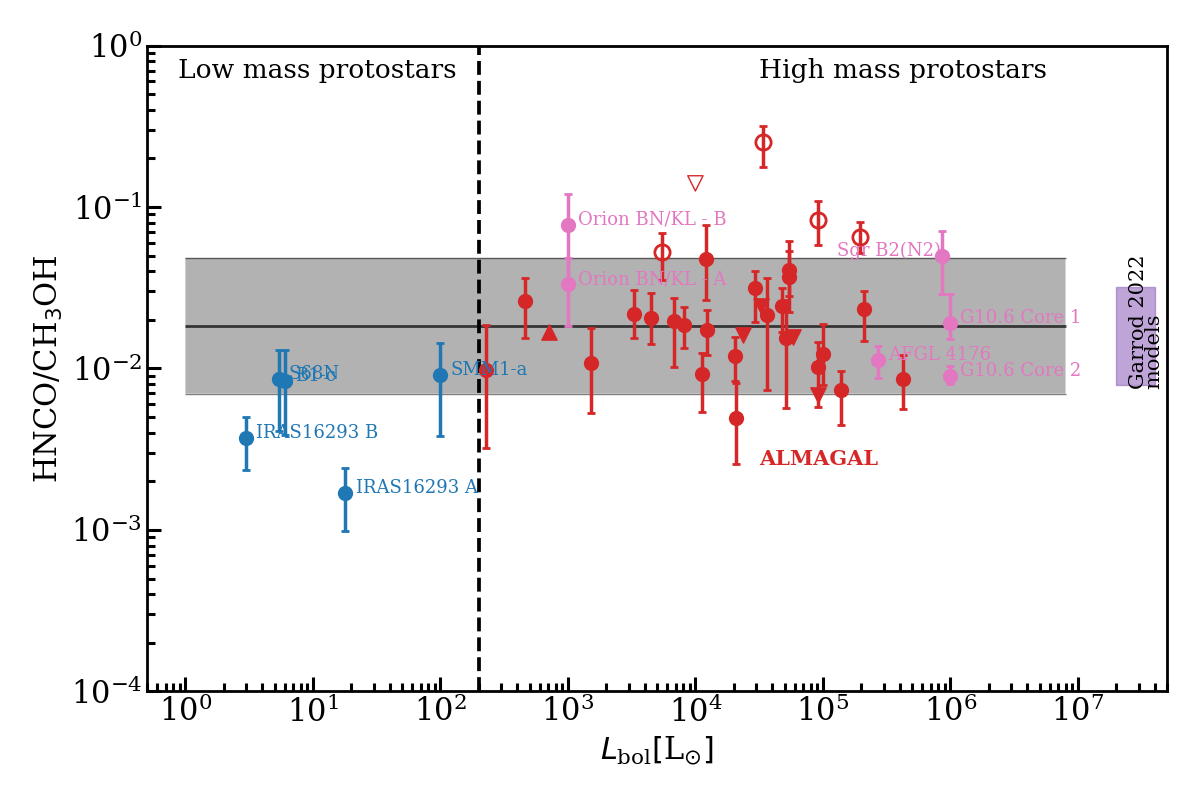}
    \caption{Same as Fig. \ref{fig:C2H5CN_CH3CN} but for HNCO/CH$_3$OH. In this plot SMM1-a is colored blue although it is an intermediate mass protostar. The references are given in Table \ref{tab:refs}. The red hollow circles indicate the sources for which $^{13}$CH$_3$OH was used to find the column density of CH$_3$OH.} 
    \label{fig:HNCO_CH3OH}
\end{figure*}

\begin{figure*}
    \centering
    \includegraphics[width=15cm]{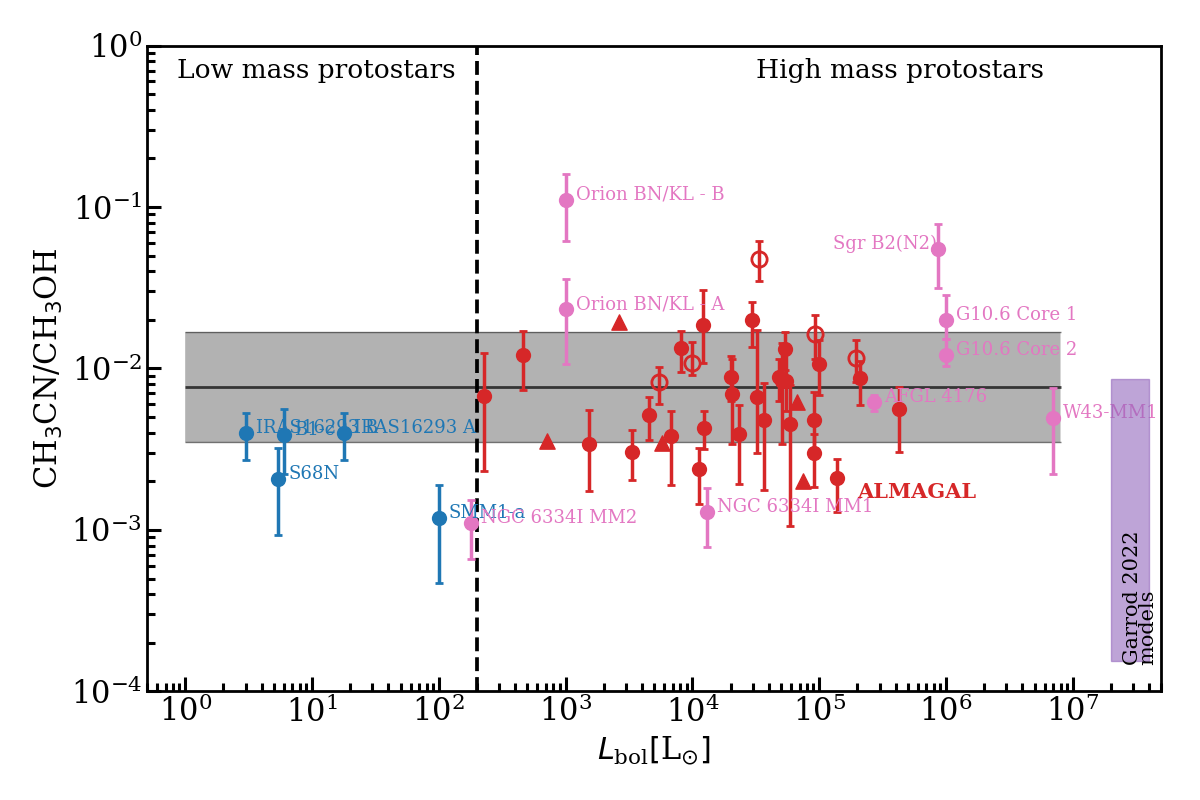}
    \caption{Same as Fig. \ref{fig:C2H5CN_CH3CN} but for CH$_3$CN/CH$_3$OH. In this plot SMM1-a is colored blue although it is an intermediate mass protostar. The references are given in Table \ref{tab:refs}. The red hollow circles indicate the sources for which $^{13}$CH$_3$OH was used to find the column density of CH$_3$OH.} 
    \label{fig:CH3CN_CH3OH}
\end{figure*}

\begin{figure*}
    \centering
    \includegraphics[width=15cm]{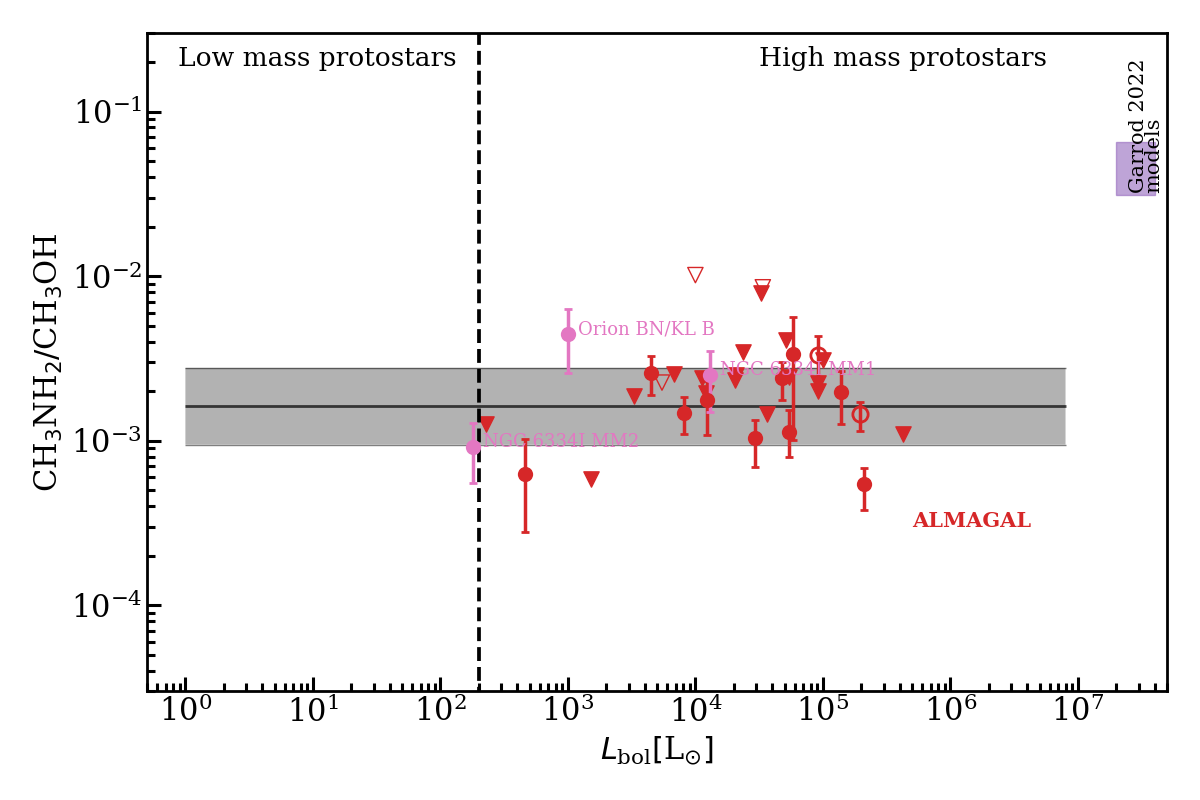}
    \caption{Same as Fig. \ref{fig:C2H5CN_CH3CN} but for CH$_3$NH$_2$/CH$_3$OH. The references are given in Table \ref{tab:refs}. The red hollow circles indicate the sources for which $^{13}$CH$_3$OH was used to find the column density of CH$_3$OH.} 
    \label{fig:CH3NH2_CH3OH}
\end{figure*}

\begin{figure*}
    \centering
    \includegraphics[width=15cm]{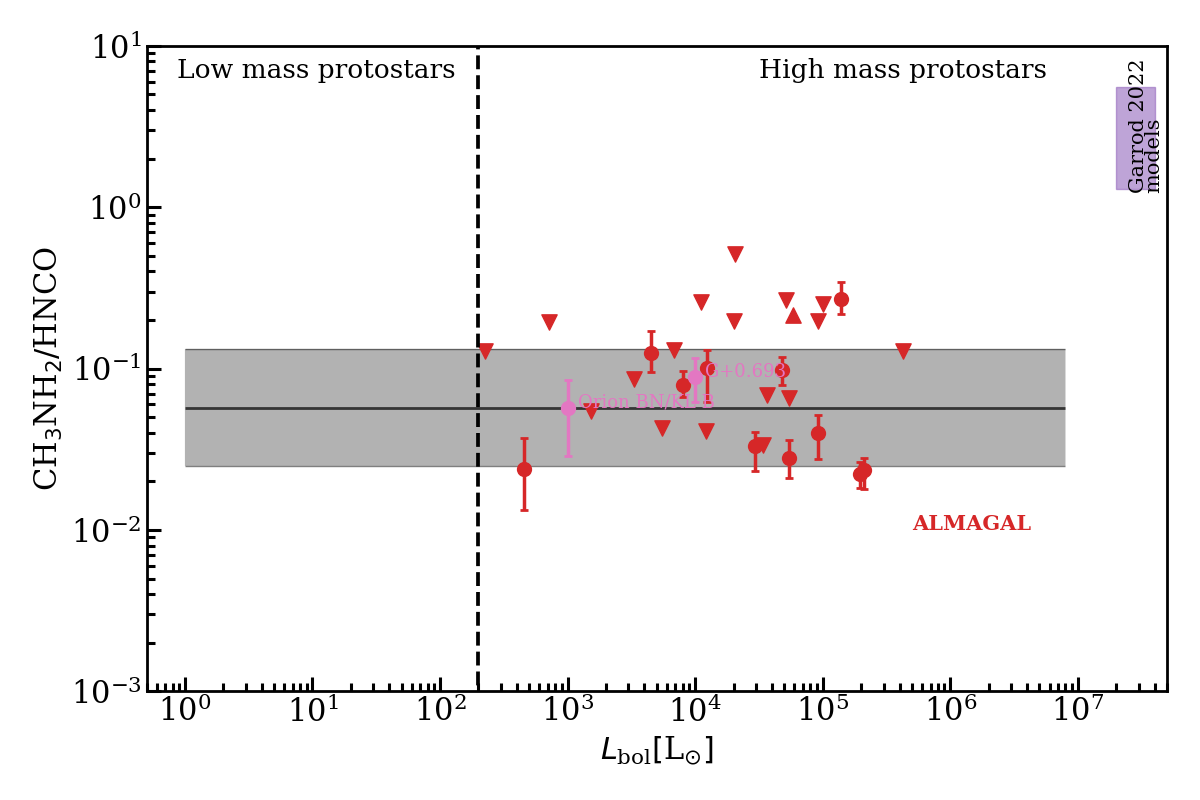}
    \caption{Same as Fig. \ref{fig:C2H5CN_CH3CN} but for CH$_3$NH$_2$/HNCO. The references are given in Table \ref{tab:refs}. The luminosity for source G+0.693 is assumed at 10$^4$ as no estimation of the luminosity was found for this source.} 
    \label{fig:CH3NH2_HNCO}
\end{figure*}

\begin{figure*}
    \centering
    \includegraphics[width=15cm]{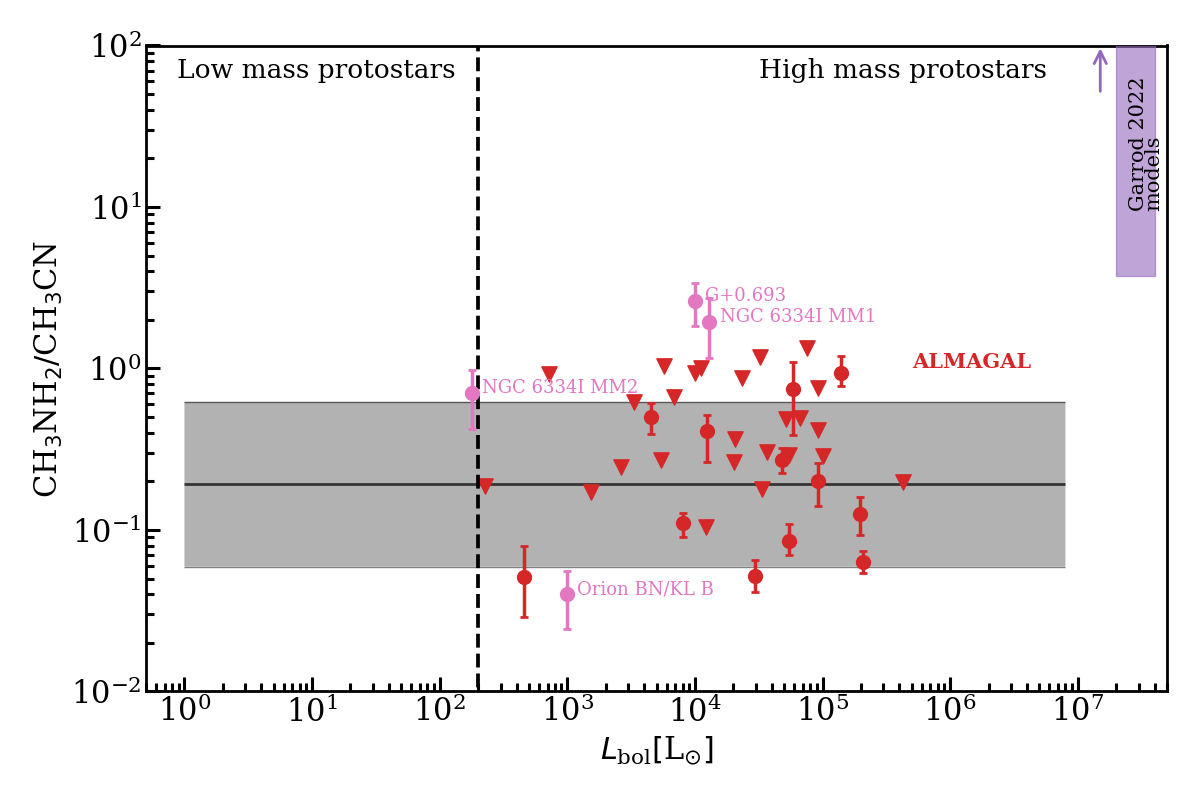}
    \caption{Same as Fig. \ref{fig:C2H5CN_CH3CN} but for CH$_3$NH$_2$/CH$_3$CN. The references are given in Table \ref{tab:refs}. The luminosity for source G+0.693 is assumed at 10$^4$ as no estimation of the luminosity was found for this source.} 
    \label{fig:CH3NH2_CH3CN}
\end{figure*}

\begin{figure*}
    \centering
    \includegraphics[width=15cm]{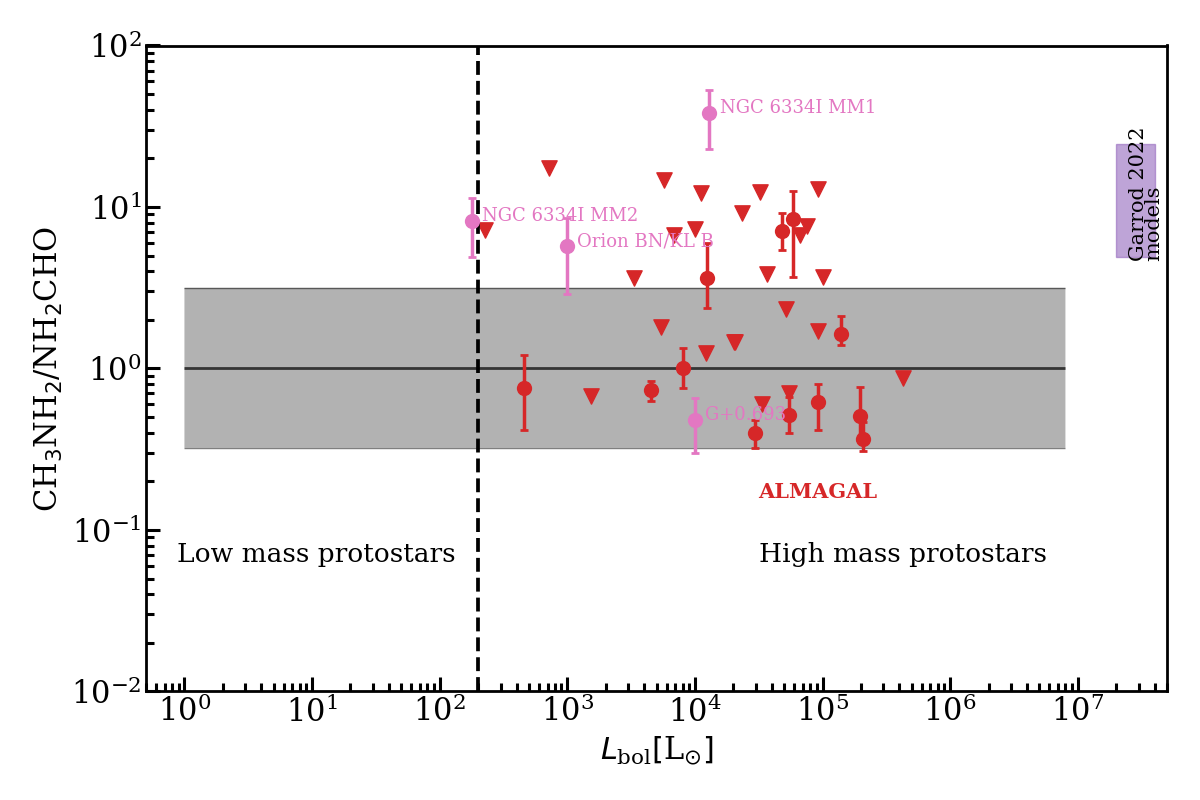}
    \caption{Same as Fig. \ref{fig:C2H5CN_CH3CN} but for CH$_3$NH$_2$/NH$_2$CHO. The references are given in Table \ref{tab:refs}. The luminosity for source G+0.693 is assumed at 10$^4$ as no estimation of the luminosity was found for this source.} 
    \label{fig:CH3NH2_NH2CHO}
\end{figure*}

\begin{figure*}
    \centering
    \includegraphics[width=15cm]{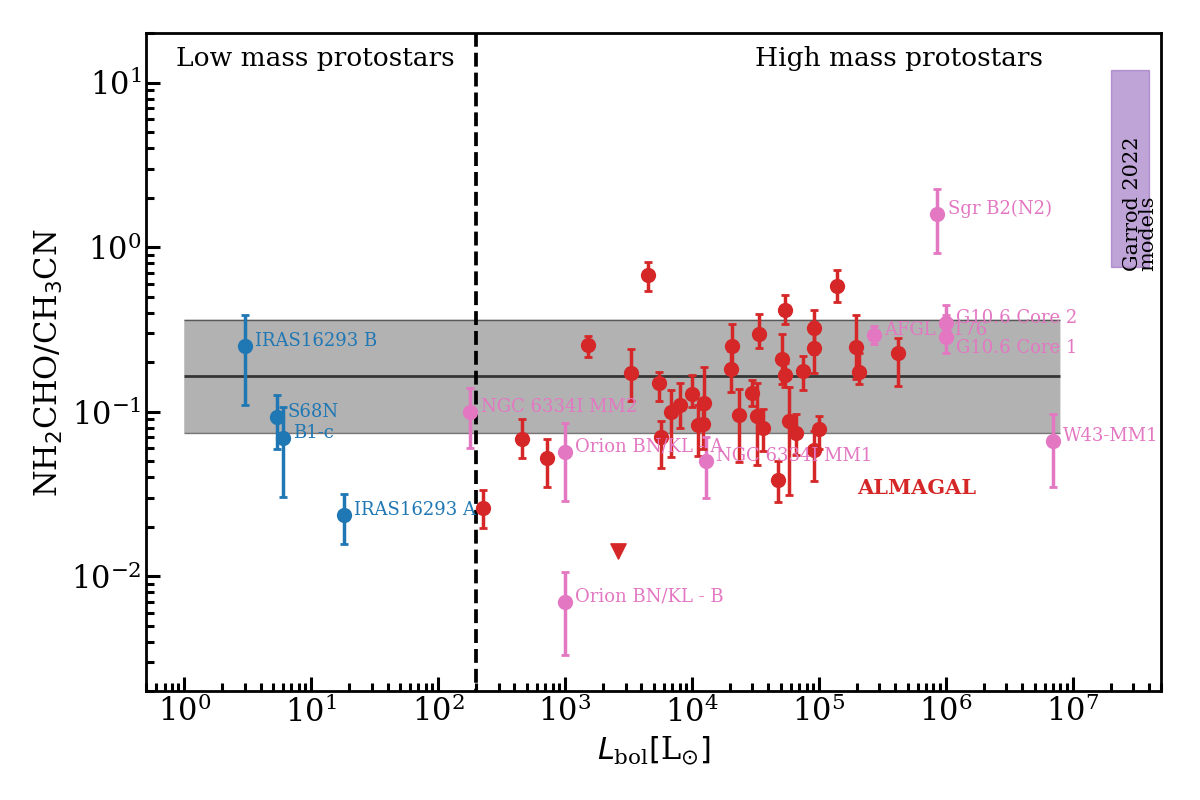}
    \caption{Same as Fig. \ref{fig:C2H5CN_CH3CN} but for NH$_2$CHO/CH$_3$CN. The references are given in Table \ref{tab:refs}.}
    \label{fig:NH2CHO_CH3CN}
\end{figure*}

\begin{figure*}
    \centering
    \includegraphics[width=15cm]{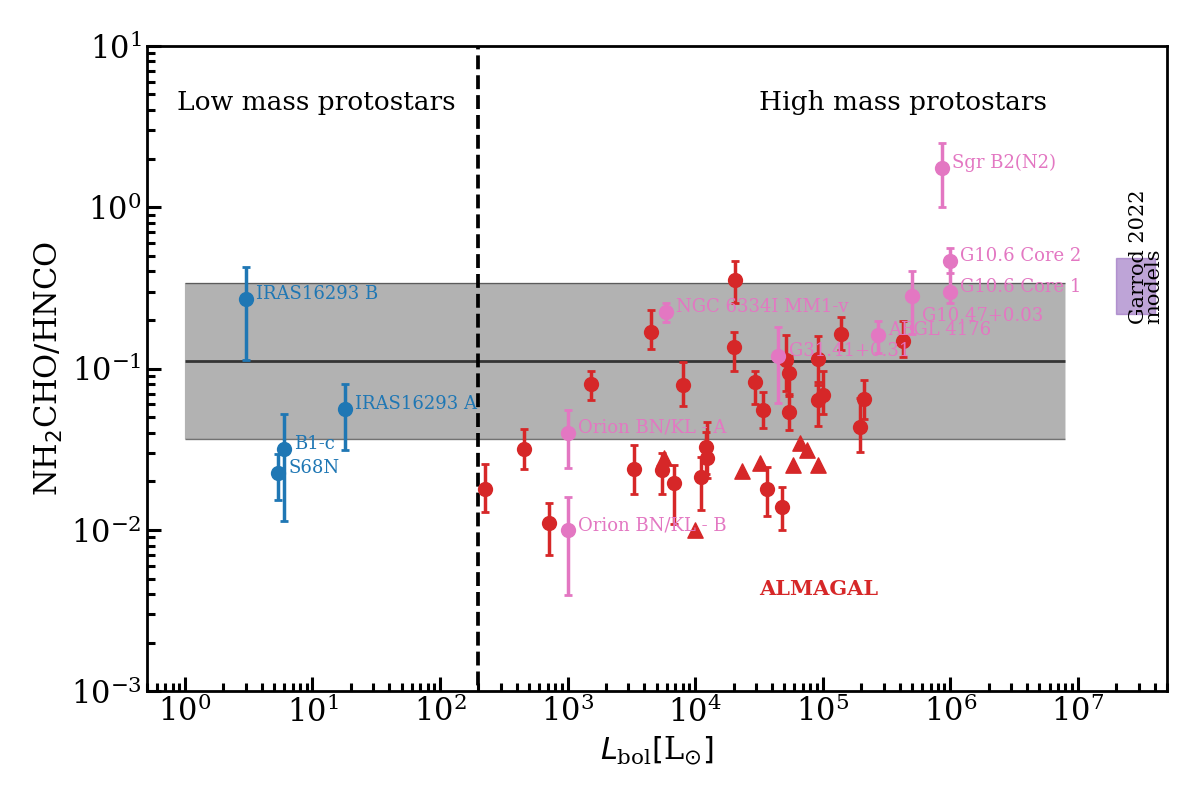}
    \caption{Same as Fig. \ref{fig:C2H5CN_CH3CN} but for NH$_2$CHO/HNCO. The references are given in Table \ref{tab:refs}.}
    \label{fig:NH2CHO_HNCO}
\end{figure*}

Figure \ref{fig:detections_L} presents what molecule is (tentatively) detected (plus signs) and what molecule is not detected (minus signs) in each source where the sources are ordered by increasing luminosity from left to right. Figure \ref{fig:column_column} presents the column density of the species considered here versus each other with the Pearson's $r$ printed on the top left of each plot. Figure \ref{fig:matrix_cor_thick} is the same as Fig. \ref{fig:matrix_cor} but including the sources where methanol column density was measured from its (potentially optically thick) $^{13}$C isotopologue. Figures \ref{fig:C2H3CN_C2H5CN}-\ref{fig:NH2CHO_HNCO} show column density ratios of various species with respect to each other versus luminosity for the ALMAGAL sources in addition to the literature sources. The values for literature sources are taken from the references given in Table \ref{tab:refs}. 

\begin{figure}
  \resizebox{\columnwidth}{!}{\includegraphics{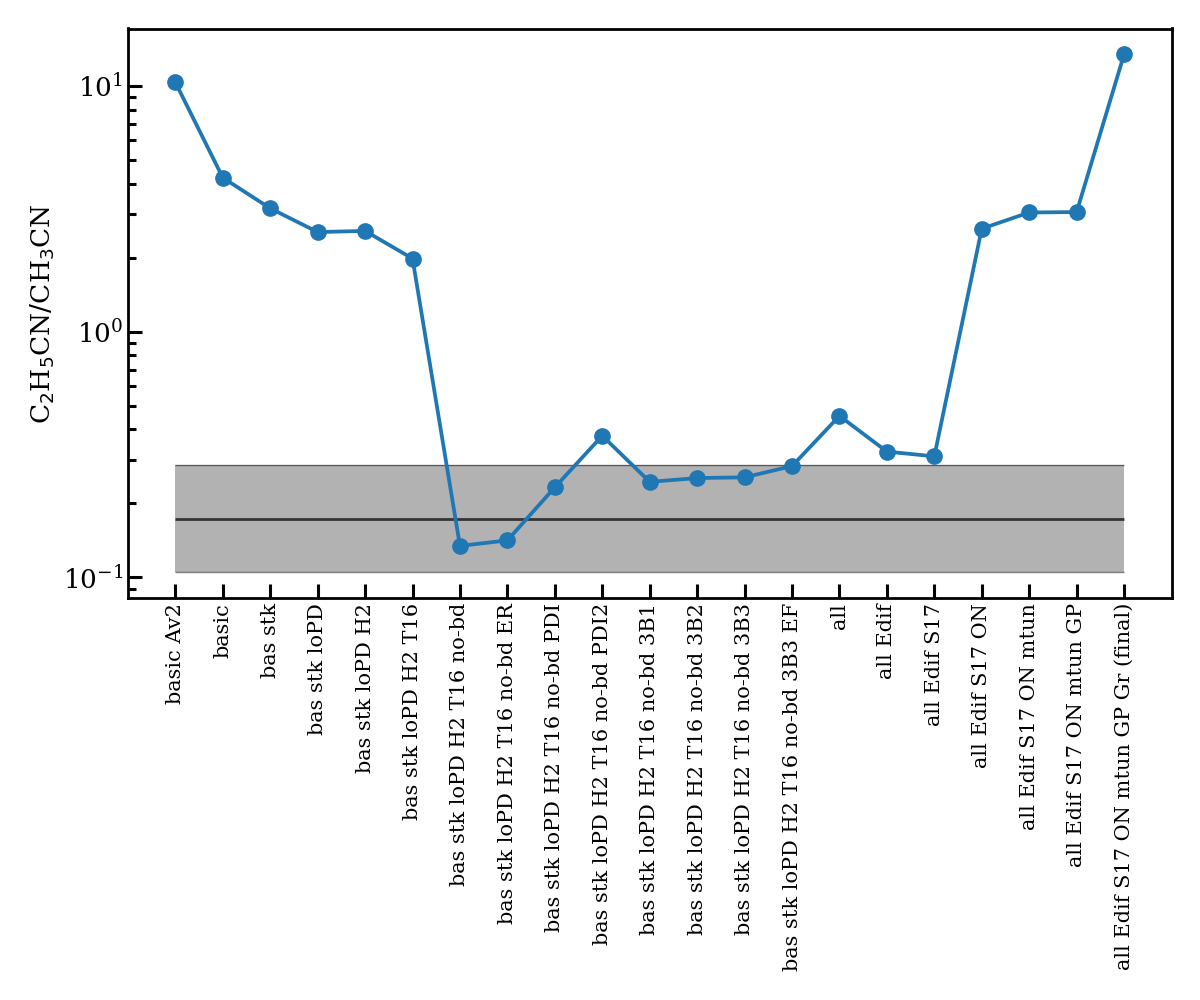}}
  \caption{The spread highlighted in Fig. \ref{fig:C2H5CN_CH3CN} is over plotted (gray) with values from Table 15 of \cite{Garrod2022} that show the peak gas phase abundances of a few N-bearing species during their warm up phase (blue).}
  \label{fig:garrod_C2H5CN_CH3CN}
\end{figure}

\begin{figure}
  \resizebox{\columnwidth}{!}{\includegraphics{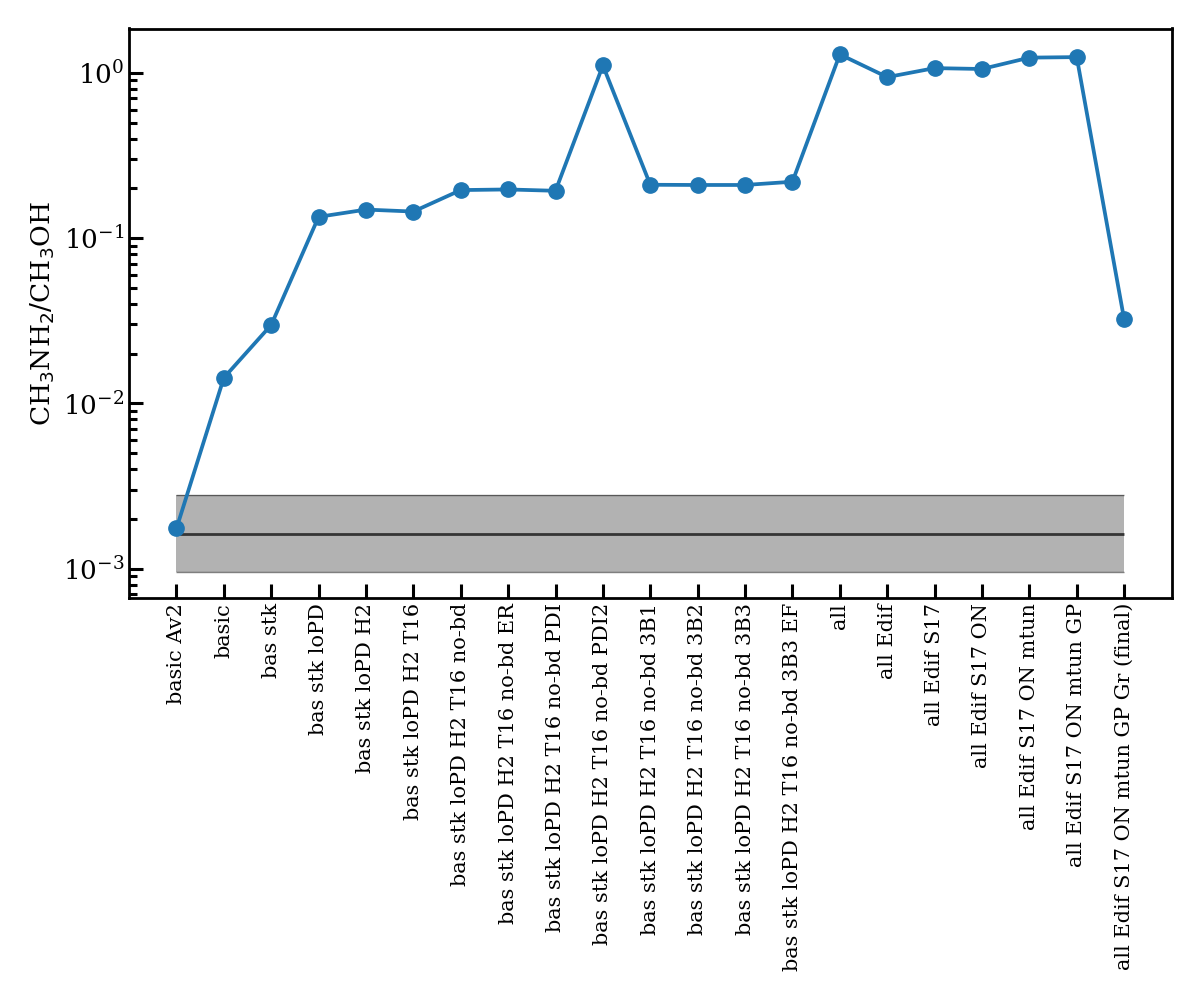}}
  \caption{The same as Fig. \ref{fig:garrod_C2H5CN_CH3CN} but for CH$_3$NH$_2$/CH$_3$OH.}
  \label{fig:garrod_CH3NH2_CH3OH}
\end{figure}

\begin{figure}
  \resizebox{\columnwidth}{!}{\includegraphics{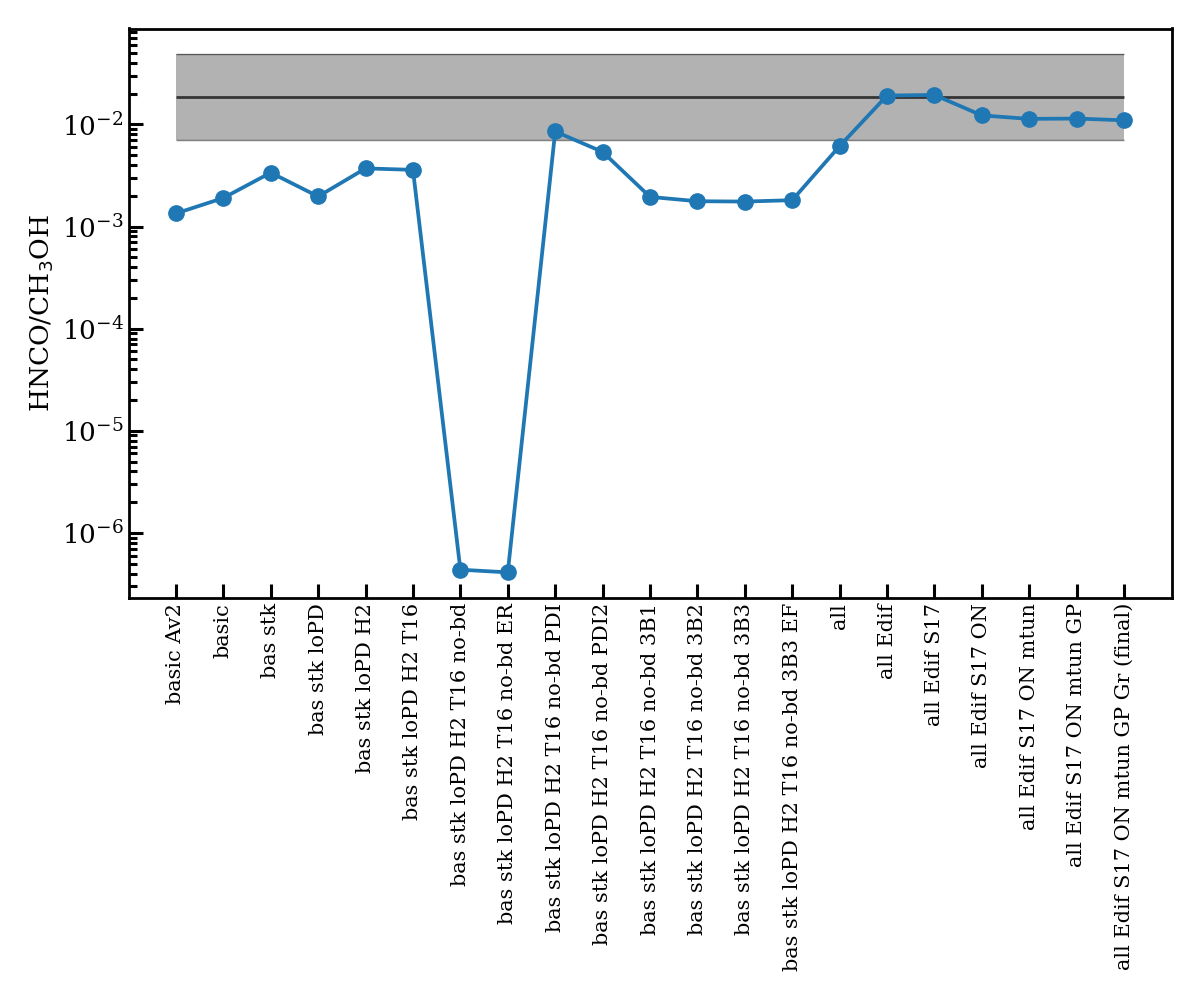}}
  \caption{The same as Fig. \ref{fig:garrod_C2H5CN_CH3CN} but for HNCO/CH$_3$OH.}
  \label{fig:garrod_HNCO_CH3OH}
\end{figure}

\begin{figure}
  \resizebox{\columnwidth}{!}{\includegraphics{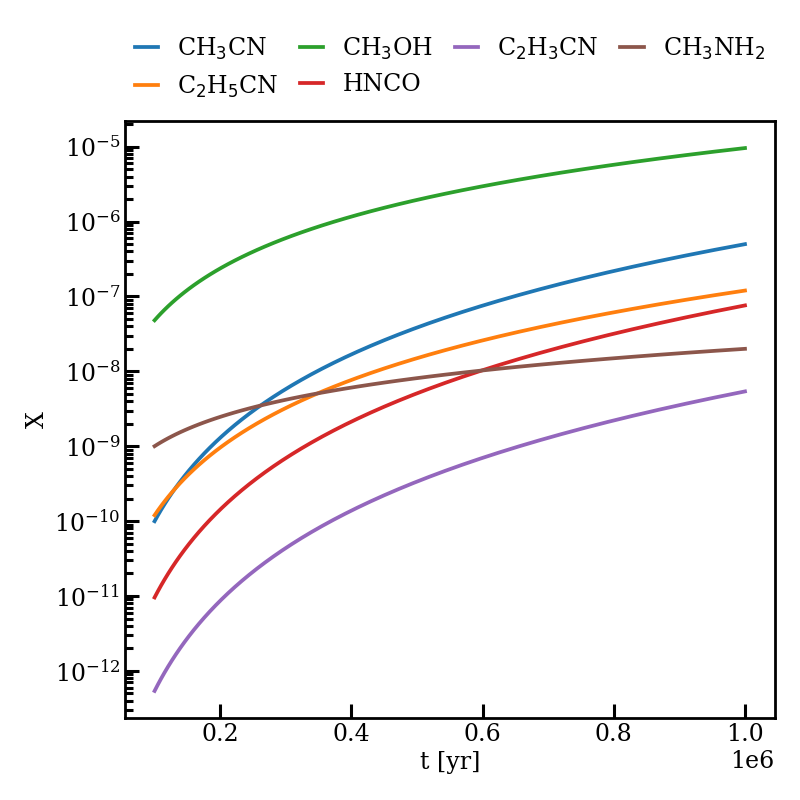}}
  \caption{The abundances assumed for the toy model.}
  \label{fig:abundances}
\end{figure}

Figures \ref{fig:garrod_C2H5CN_CH3CN}-\ref{fig:garrod_HNCO_CH3OH} present the column density ratios calculated from Table 15 of \cite{Garrod2022} that show the peak gas phase abundances of a few N-bearing species during the warm-up phase of their chemical models with medium pace warm-up. These plots also show the scatter seen in the data for both low- and high-mass protostars found in this work. Figure \ref{fig:abundances} shows the abundances of the species considered here calculated with the toy model as a function of time spent in the pre-stellar phase.

\section{Additional tables}


\begin{table*}
\tiny
\renewcommand{\arraystretch}{1.3}
\centering
    \caption{Fitted parameters for methanol.}
    \label{tab:results_methanol}
    \resizebox{0.8\textwidth}{!}{\begin{tabular}{@{\extracolsep{1mm}}*{5}{l}}
          \toprule
          \toprule      
          & \multicolumn{1}{c}{CH$_3$OH}& \multicolumn{3}{c}{CH$_3^{18}$OH}\\
          \cmidrule{2-2} \cmidrule{3-5}
        Source & $N (\rm cm^{-2})$ & $N (\rm cm^{-2})$ &  $T_{\rm ex} (\rm K)$ & FWHM (km s$^{-1}$) \\
        \midrule     

101899 & $^*$1.3$^{+0.3}_{-0.3}$ $\times 10^{18}$ & -- & -- & -- \\ 
126348 & 1.1$^{+0.4}_{-0.5}$ $\times 10^{18}$ & 3.6$^{+0.9}_{-1.3}$ $\times 10^{15}$ & 110$^{+30}_{-40}$ & [7.5] \\ 
615590 & $^*$1.9$^{+0.4}_{-0.4}$ $\times 10^{18}$ & <\,8.4 $\times 10^{15}$ & [150] & [4.5] \\ 
644284A & $^*$4.6$^{+1.5}_{-0.6}$ $\times 10^{17}$ & <\,4.7 $\times 10^{15}$ & [150] & [6.7] \\ 
693050 & 1.0$^{+0.7}_{-0.4}$ $\times 10^{18}$ & 2.4$^{+1.4}_{-0.8}$ $\times 10^{15}$ & 70$^{+50}_{-20}$ & [4.7] \\ 
705768 & 1.7$^{+0.5}_{-0.4}$ $\times 10^{18}$ & 3.9$^{+0.7}_{-0.3}$ $\times 10^{15}$ & [140] & [6.7] \\ 
707948 & $^*$1.0$^{+0.2}_{-0.2}$ $\times 10^{19}$ & -- & -- & -- \\ 
717461A & 1.5$^{+0.4}_{-0.4}$ $\times 10^{18}$ & 3.6$^{+0.6}_{-0.6}$ $\times 10^{15}$ & [150] & [6.3] \\ 
721992 & <\,5.7 $\times 10^{17}$ & <\,1.4 $\times 10^{15}$ & [90] & [3.5] \\ 
724566 & 4.2$^{+3.4}_{-2.6}$ $\times 10^{18}$ & 1.1$^{+0.8}_{-0.6}$ $\times 10^{16}$ & 150$^{+70}_{-80}$ & [6.0] \\ 
732038 & <\,2.0 $\times 10^{18}$ & <\,5.3 $\times 10^{15}$ & [150] & [6.0] \\ 
744757A & 1.7$^{+0.4}_{-0.4}$ $\times 10^{18}$ & 4.0$^{+0.5}_{-0.4}$ $\times 10^{15}$ & [130] & [5.7] \\ 
767784 & 2.9$^{+0.7}_{-1.0}$ $\times 10^{18}$ & 8.2$^{+1.2}_{-2.3}$ $\times 10^{15}$ & 90$^{+20}_{-20}$ & [4.0] \\ 
778802 & <\,9.5 $\times 10^{17}$ & <\,2.3 $\times 10^{15}$ & [150] & [7.0] \\ 
779523 & 1.3$^{+0.3}_{-0.3}$ $\times 10^{18}$ & 3.3$^{+0.4}_{-0.4}$ $\times 10^{15}$ & [150] & [8.4] \\ 
779984 & <\,7.3 $\times 10^{17}$ & <\,2.1 $\times 10^{15}$ & [150] & [7.0] \\ 
783350 & 6.4$^{+4.2}_{-3.5}$ $\times 10^{17}$ & 2.0$^{+1.2}_{-1.0}$ $\times 10^{15}$ & 110$^{+50}_{-70}$ & [6.5] \\ 
787212 & 3.3$^{+0.8}_{-0.8}$ $\times 10^{18}$ & 8.7$^{+1.2}_{-1.2}$ $\times 10^{15}$ & 90$^{+10}_{-10}$ & [7.5] \\ 
792355 & 7.0$^{+2.1}_{-1.9}$ $\times 10^{17}$ & 2.2$^{+0.4}_{-0.3}$ $\times 10^{15}$ & [150] & [6.7] \\ 
800287 & 2.6$^{+0.9}_{-0.8}$ $\times 10^{18}$ & 8.8$^{+2.2}_{-1.5}$ $\times 10^{15}$ & [140] & 5.0 \\ 
800751 & 1.2$^{+0.8}_{-0.6}$ $\times 10^{18}$ & 3.6$^{+2.2}_{-1.5}$ $\times 10^{15}$ & 100$^{+40}_{-50}$ & [5.5] \\ 
865468A & 1.3$^{+0.3}_{-0.3}$ $\times 10^{19}$ & 3.7$^{+0.2}_{-0.4}$ $\times 10^{16}$ & 80$^{+10}_{-20}$ & [8.0] \\ 
876288 & 8.1$^{+5.2}_{-5.3}$ $\times 10^{17}$ & 4.3$^{+0.7}_{-0.9}$ $\times 10^{15}$ & [150] & [4.5] \\ 
881427C & 7.6$^{+2.8}_{-2.8}$ $\times 10^{18}$ & 1.8$^{+0.6}_{-0.6}$ $\times 10^{16}$ & 80$^{+30}_{-40}$ & [5.5] \\ 
G023.3891+00.1851 & 1.1$^{+0.4}_{-0.4}$ $\times 10^{18}$ & 3.6$^{+0.6}_{-0.4}$ $\times 10^{15}$ & [200] & [4.0] \\ 
G025.6498+01.0491 & 3.3$^{+1.1}_{-0.9}$ $\times 10^{18}$ & 8.4$^{+1.7}_{-1.1}$ $\times 10^{15}$ & [150] & [8.3] \\ 
G305.2017+00.2072A1 & 1.3$^{+0.3}_{-0.3}$ $\times 10^{18}$ & 3.0$^{+0.4}_{-0.4}$ $\times 10^{15}$ & [130] & [5.5] \\ 
G314.3197+00.1125 & <\,1.6 $\times 10^{18}$ & <\,3.8 $\times 10^{15}$ & [150] & [8.5] \\ 
G316.6412-00.0867 & 2.3$^{+0.5}_{-0.6}$ $\times 10^{18}$ & 5.7$^{+0.6}_{-0.8}$ $\times 10^{15}$ & 90$^{+20}_{-20}$ & [6.0] \\ 
G318.0489+00.0854B & 1.8$^{+1.2}_{-1.1}$ $\times 10^{18}$ & 4.7$^{+2.9}_{-2.6}$ $\times 10^{15}$ & 150$^{+60}_{-80}$ & [7.3] \\ 
G318.9480-00.1969A1 & 6.2$^{+1.4}_{-1.8}$ $\times 10^{18}$ & 1.4$^{+0.1}_{-0.3}$ $\times 10^{16}$ & 90$^{+20}_{-20}$ & [6.3] \\ 
G323.7399-00.2617B2 & 5.1$^{+3.2}_{-2.4}$ $\times 10^{18}$ & 1.4$^{+0.8}_{-0.6}$ $\times 10^{16}$ & 90$^{+40}_{-40}$ & [5.5] \\ 
G326.4755+00.6947 & 3.2$^{+4.9}_{-1.4}$ $\times 10^{17}$ & 7.7$^{+11.7}_{-2.9}$ $\times 10^{14}$ & 70$^{+100}_{-30}$ & [6.7] \\ 
G326.6618+00.5207 & 1.0$^{+0.2}_{-0.2}$ $\times 10^{18}$ & 2.5$^{+0.3}_{-0.3}$ $\times 10^{15}$ & [150] & [5.5] \\ 
G327.1192+00.5103 & 3.1$^{+2.0}_{-1.0}$ $\times 10^{18}$ & 9.7$^{+5.7}_{-2.0}$ $\times 10^{15}$ & 80$^{+40}_{-20}$ & [8.0] \\ 
G343.1261-00.0623 & $^*$6.2$^{+1.0}_{-1.3}$ $\times 10^{17}$ & -- & -- & -- \\ 
G345.5043+00.3480 & 6.7$^{+1.5}_{-2.0}$ $\times 10^{18}$ & 1.7$^{+0.2}_{-0.3}$ $\times 10^{16}$ & 70$^{+10}_{-20}$ & 6.5 \\

\bottomrule
        \end{tabular}}
        \tablefoot{The same as Table \ref{tab:results} but for CH$_3^{18}$OH and CH$_3$OH. The stars indicate the values found for CH$_3$OH from $^{13}$C isotopologue of this molecule (instead of its $^{18}$O) using a FWHM specifically fitted to the single lines of $^{13}$CH$_3$OH and CH$_3$OH and $T_{\rm ex}$ fixed to that of CH$_3^{13}$CN. The reason that for some sources no values of CH$_3^{18}$OH column density are reported is that it is not possible to measure this value due to line blending. The FWHM of 800287 and G345.5043+00.3480 are fixed to a smaller value than those of CH$_3$CN lines to match the observations.}
\end{table*}

\begin{table*}
    \caption{Observational parameters for each source.}
    \label{tab:obs_params}
    \centering
    \begin{tabular}{l l l l l l} 
    \toprule
    \toprule    
Source & R.A. ($^{\rm{h\, m\, s}}$) & Dec. ($^{\circ}\, \arcmin \, \arcsec$) & Beam ($\arcsec$) & Continuum & Line \\
&&&&rms (mJy beam$^{-1}$) & rms (K)\\
\midrule     

101899 & 18:34:40.29 & -09:00:38.44 & 1.45 $\times$ 1.07 & 1.4 & 0.24\\
126348 & 18:42:51.98 & -03:59:54.37 & 1.26 $\times$ 1.08 & 1.5 & 0.16\\
615590 & 09:24:41.96 & -52:02:8.04 & 0.68 $\times$ 0.59 & 0.5 & 0.50\\
644284A & 10:31:29.78 & -58:02:19.27 & 0.96 $\times$ 0.80 & 1.7 & 0.32\\
693050 & 12:35:35.05 & -63:02:31.19 & 1.03 $\times$ 0.94 & 13.0 & 0.19\\
705768 & 13:12:36.17 & -62:33:34.43 & 0.93 $\times$ 0.81 & 1.5 & 0.24\\
707948 & 13:16:43.19 & -62:58:32.83 & 0.94 $\times$ 0.80 & 3.2 & 0.26\\
717461A & 13:43:1.68 & -62:08:51.42 & 1.34 $\times$ 1.21 & 1.0 & 0.22\\
721992 & 13:51:58.23 & -61:15:40.88 & 0.90 $\times$ 0.79 & 0.9 & 0.37\\
724566 & 13:59:30.92 & -61:48:38.27 & 0.87 $\times$ 0.78 & 1.0 & 0.34\\
732038 & 14:13:15.05 & -61:16:53.19 & 0.87 $\times$ 0.78 & 1.3 & 0.40\\
744757A & 14:45:26.35 & -59:49:15.55 & 1.31 $\times$ 1.27 & 1.5 & 0.15\\
767784 & 15:29:19.31 & -56:31:22.02 & 1.31 $\times$ 1.24 & 6.5 & 0.16\\
778802 & 15:44:32.97 & -54:05:28.31 & 1.39 $\times$ 1.19 & 1.2 & 0.15\\
779523 & 15:44:59.61 & -54:02:22.44 & 1.39 $\times$ 1.19 & 1.5 & 0.13\\
779984 & 15:48:55.17 & -54:40:38.13 & 1.39 $\times$ 1.19 & 0.6 & 0.11\\
783350 & 15:49:19.46 & -53:45:14.17 & 1.39 $\times$ 1.19 & 2.2 & 0.14\\
787212 & 15:57:59.74 & -53:58:0.92 & 1.39 $\times$ 1.19 & 2.6 & 0.15\\
792355 & 16:03:32.13 & -53:09:30.1 & 1.38 $\times$ 1.17 & 1.2 & 0.16\\
800287 & 16:11:26.57 & -51:41:57.14 & 0.84 $\times$ 0.72 & 0.6 & 0.45\\
800751 & 16:12:26.43 & -51:46:16.28 & 1.38 $\times$ 1.19 & 0.7 & 0.15\\
865468A & 17:05:10.9 & -41:29:6.99 & 1.32 $\times$ 1.14 & 7.6 & 0.22\\
876288 & 17:11:51.02 & -39:09:29.18 & 0.92 $\times$ 0.69 & 1.4 & 0.29\\
881427C & 17:20:6.12 & -38:57:15.84 & 1.31 $\times$ 1.13 & 3.4 & 0.56\\
G023.3891+00.1851 & 18:33:14.32 & -08:23:57.61 & 1.43 $\times$ 1.07 & 0.8 & 0.18\\
G025.6498+01.0491 & 18:34:20.92 & -05:59:42.08 & 1.24 $\times$ 1.07 & 2.7 & 0.15\\
G305.2017+00.2072A1 & 13:11:10.45 & -62:34:38.6 & 1.38 $\times$ 1.21 & 2.6 & 0.21\\
G314.3197+00.1125 & 14:26:26.25 & -60:38:31.26 & 1.36 $\times$ 1.23 & 1.5 & 0.21\\
G316.6412-00.0867 & 14:44:18.35 & -59:55:11.28 & 1.32 $\times$ 1.25 & 1.1 & 0.15\\
G318.0489+00.0854B & 14:53:42.64 & -59:08:53.02 & 1.31 $\times$ 1.27 & 1.6 & 0.15\\
G318.9480-00.1969A1 & 15:00:55.28 & -58:58:52.6 & 1.30 $\times$ 1.27 & 2.0 & 0.15\\
G323.7399-00.2617B2 & 15:31:45.45 & -56:30:49.84 & 1.29 $\times$ 1.25 & 1.9 & 0.15\\
G326.4755+00.6947 & 15:43:18.91 & -54:07:35.4 & 1.38 $\times$ 1.23 & 0.9 & 0.12\\
G326.6618+00.5207 & 15:45:2.87 & -54:09:3.0 & 1.38 $\times$ 1.23 & 0.9 & 0.12\\
G327.1192+00.5103 & 15:47:32.72 & -53:52:38.6 & 0.90 $\times$ 0.72 & 0.9 & 0.34\\
G343.1261-00.0623 & 16:58:17.22 & -42:52:7.54 & 1.33 $\times$ 1.15 & 5.3 & 0.21\\
G345.5043+00.3480 & 17:04:22.89 & -40:44:23.06 & 1.34 $\times$ 1.15 & 4.1 & 0.20\\

\bottomrule
\end{tabular}
\tablefoot{R.A. and Dec. refer to the coordinates of the maximum pixel in the moment zero maps of K=4 CH$_3$CN. 787212 is an exception where the spectrum is extracted ${\sim} 1$ beam off-source to decrease line blending and infall-related signatures.}
\end{table*}

\begin{table*}
    \caption{Source properties from the Hi-GAL survey.}
    \label{tab:source_features}
    \centering
    \begin{tabular}{l l l l l l l l} 
    \toprule
    \toprule    
Source & $d^{a}$ & $R_{\rm GC}^{b}$  & $v_{\rm LSR} ^{c}$  & $L^{d}$  & $^{16}$O/$^{18}$O & $^{12}$C/$^{13}$C \\
&(kpc) & (kpc) & (km s$^{-1}$) & (L$_{\odot}$) & & \\
\midrule     
101899 & 4.60 & 4.22 & 77.8 & 92000 & 285 & 44 \\
126348 & 4.41 & 4.67 & 75.8 & 6798 & 311 & 47 \\
615590 & 2.70 & 8.31 & 40.1 & 5470 & 525 & 70 \\
644284A & 4.75 & 8.20 & 1.8 & -- & 519 & 69 \\
693050 & 4.31 & 6.89 & -41.2 & 12200 & 442 & 61 \\
705768 & 6.88 & 6.93 & -34.1 & 91728 & 444 & 61 \\
707948 & 7.14 & 6.97 & -33.3 & 196800 & 446 & 61 \\
717461A & 4.29 & 6.31 & -52.0 & 3323 & 408 & 57 \\
721992 & 5.38 & 6.16 & -57.1 & 2630 & 399 & 56 \\
724566 & 4.93 & 6.10 & -56.2 & 226 & 395 & 56 \\
732038 & 5.64 & 5.93 & -63.1 & 74930 & 385 & 55 \\
744757A & 2.51 & 6.45 & -40.4 & 12381 & 416 & 58 \\
767784 & 4.04 & 5.37 & -67.8 & 139400 & 353 & 52 \\
778802 & 2.35 & 6.22 & -39.5 & 715 & 403 & 57 \\
779523 & 2.57 & 6.07 & -41.3 & 11178 & 393 & 56 \\
779984 & 3.83 & 5.28 & -63.3 & 5738 & 347 & 51 \\
783350 & 4.62 & 4.85 & -79.8 & 51260 & 322 & 48 \\
787212 & 2.90 & 5.79 & -41.2 & 54225 & 377 & 54 \\
792355 & 4.35 & 4.84 & -70.6 & 23455 & 321 & 48 \\
800287 & 4.89 & 4.44 & -87.0 & 100000 & 297 & 46 \\
800751 & 3.88 & 5.00 & -66.3 & 20659 & 331 & 49 \\
865468A & 3.04 & 5.17 & -27.3 & 47824 & 341 & 50 \\
876288 & 5.95 & 2.58 & -95.5 & 58410 & 188 & 34 \\
881427C & 1.50 & 6.59 & -10.9 & 458 & 424 & 59 \\
G023.3891+00.1851 & 10.86 & 4.72 & 75.7 & 91560 & 314 & 48 \\
G025.6498+01.0491 & 12.20 & 6.05 & 41.4 & 424000 & 392 & 56 \\
G305.2017+00.2072A1 & 4.00 & 6.61 & -40.9 & 20301 & 425 & 59 \\
G314.3197+00.1125 & 8.25 & 6.33 & -48.5 & 66440 & 409 & 58 \\
G316.6412-00.0867 & 2.73 & 6.35 & -19.1 & 8080 & 410 & 58 \\
G318.0489+00.0854B & 3.18 & 6.07 & -50.4 & 36480 & 393 & 56 \\
G318.9480-00.1969A1 & 10.40 & 6.83 & -33.8 & 209000 & 438 & 61 \\
G323.7399-00.2617B2 & 3.20 & 5.79 & -51.6 & 1529 & 377 & 54 \\
G326.4755+00.6947 & 11.34 & 6.42 & -41.4 & 32440 & 414 & 58 \\
G326.6618+00.5207 & 2.48 & 6.13 & -41.0 & 4523 & 397 & 56 \\
G327.1192+00.5103 & 4.74 & 4.81 & -84.5 & 54270 & 320 & 48 \\
G343.1261-00.0623 & 2.00 & 6.16 & -32.4 & 33800 & 399 & 56 \\
G345.5043+00.3480 & 2.00 & 6.13 & -17.4 & 29498 & 397 & 56 \\

\bottomrule
\end{tabular}
\tablefoot{$^{\rm (a)}$ Distance to the source. All from \cite{Mege2021}, except for sources G305.2017+00.2072A1 and G323.7399-00.2617B2 that are from \cite{Lumsden2013}. The typical uncertainties on the source distances are ${\sim}0.5$\,kpc and that has been assumed to calculate the uncertainty on the isotopologue ratios. $^{\rm (b)}$ Distances to the galactic center calculated using Sun's distance to the galactic center as 8.05\,kpc from \cite{Honma2015}. $^{\rm (c)}$ Source velocity obtained using the CH$_3$CN ladder mainly $K=0$ to $K=5$ lines in each source. $^{\rm (d)}$ Bolometric luminosities from \cite{Elia2017} for all sources except 61559, G305.2017+00.2072A1, G323.7399-00.2617B2 and G343.1261-00.0623 that are from \cite{Lumsden2013}. These values are corrected based on the distances presented in the second column. There are no estimates of the bolometric luminosity for source 644284A and hence it is assumed as a generic 10$^4$\,L$_{\odot}$ in the various plots. If multiple sources are present, the luminosities are divided between the sources with each source weighted by its peak continuum flux following \cite{vanGelder2022}. The isotopologue ratios are calculated using the equations in \cite{Wilson1994} ($^{16}$O/$^{18}$O) and \cite{Milam2005} ($^{12}$C/$^{13}$C) and the distances to the galactic center presented in the third column. The uncertainties on $^{12}$C/$^{13}$C are between 6 and 12. The uncertainties on $^{16}$O/$^{18}$O are between 75 and 116. These are calculated assuming a typical 0.5\,kpc error on the distances of these sources.}
\end{table*}

\begin{table*}
\small
    \caption{The references for column densities used in the column density ratio plots and the calculation of the scatter.}
    \label{tab:refs}
    \centering
    \begin{tabular}{l l l l l} 
    \toprule
    \toprule 
Source & CH$_3$CN & C$_2$H$_3$CN & C$_2$H$_5$CN & HNCO \\
\midrule  
IRAS16293 A & \cite{Calcutt2018} & -- & \cite{Calcutt2018} & \cite{Ligterink2017}\\
IRAS16293 B & \cite{Calcutt2018} & \cite{Calcutt2018} & \cite{Calcutt2018} & \cite{Ligterink2017} \\
S68N & \cite{Nazari2021} & -- & \cite{Nazari2021} & \cite{Nazari2021} \\
B1-c & \cite{Nazari2021} & -- & \cite{Nazari2021} & \cite{Nazari2021} \\
NGC 1333 IRAS 4A2 & \cite{Lopez2017} & -- &\cite{Lopez2017} & -- \\
HH212 & -- & -- & -- & -- \\
L483 & -- & -- & -- & -- \\
SMM1-a & \cite{Ligterink2021} & -- & -- & \cite{Ligterink2021} \\
NGC 6334I MM1 & \cite{Bogelund2019} & -- & -- & -- \\
NGC 6334I MM2 & \cite{Bogelund2019} & -- & -- & -- \\
Orion BN/KL A & \cite{Cernicharo2016} & -- & \cite{Cernicharo2016} & \cite{Cernicharo2016}\\
Orion BN/KL B & \cite{Cernicharo2016} & -- & \cite{Cernicharo2016} & \cite{Cernicharo2016}\\
AFGL 4176 & \cite{Bogelund2019AFGL} & \cite{Bogelund2019AFGL} & \cite{Bogelund2019AFGL} & \cite{Bogelund2019AFGL}\\
G10.6 Core 1 & \cite{Law2021}& \cite{Law2021} & \cite{Law2021}&\cite{Law2021}\\
G10.6 Core 2 & \cite{Law2021}& \cite{Law2021} & \cite{Law2021}&\cite{Law2021}\\
W43-MM1 & \cite{Molet2019} & --&  \cite{Molet2019} & --\\
G10.47+0.03 & -- &-- &-- &\cite{Gorai2020}\\
G31.41+0.31 & --&-- &-- &\cite{Colzi2021}\\
NGC 6334I MM1-v & -- &-- &-- &\cite{Ligterink2020}\\
Sgr B2(N2) & \cite{Belloche2016}& \cite{Belloche2016}&\cite{Belloche2016} &\cite{Belloche2017}\\
G+0.693 & \cite{Zeng2018} & --& --& \cite{Zeng2018}\\

    \toprule
    \toprule 
Source & NH$_2$CHO & CH$_3$NH$_2$ &  CH$_3$OH &\\
\midrule  
IRAS16293 A & \cite{Manigand2020} & -- & \cite{Ligterink2017} &\\
IRAS16293 B & \cite{Coutens2016} & -- & \cite{Jorgensen2018} &\\
S68N & \cite{Nazari2021} & -- & \cite{vanGelder2020} &\\
B1-c & \cite{Nazari2021} & -- & \cite{vanGelder2020} &\\
NGC 1333 IRAS 4A2 & \cite{Lopez2017} & -- & -- &\\
HH212 & \cite{Lee2019} & -- & \cite{Lee2019}  &\\
L483 & \cite{Jacobsen2019} & -- & \cite{Jacobsen2019}  &\\
SMM1-a & -- & -- & \cite{Ligterink2021} &\\
NGC 6334I MM1 & \cite{Bogelund2019} & \cite{Bogelund2019} & \cite{Bogelund2018} &\\
NGC 6334I MM2 & \cite{Bogelund2019} & \cite{Bogelund2019} & \cite{Bogelund2018} &\\
Orion BN/KL A & \cite{Cernicharo2016} &-- & \cite{Cernicharo2016}&\\
Orion BN/KL B & \cite{Cernicharo2016} & \cite{Cernicharo2016} & \cite{Cernicharo2016}&\\
AFGL 4176 & \cite{Bogelund2019AFGL}& --& \cite{Bogelund2019AFGL}&\\
G10.6 Core 1 &\cite{Law2021} &-- &\cite{Law2021} &\\
G10.6 Core 2 &\cite{Law2021} &-- &\cite{Law2021} &\\
W43-MM1 &  \cite{Molet2019} &-- &  \cite{Molet2019}&\\
G10.47+0.03 & \cite{Gorai2020} &-- & --&\\
G31.41+0.31 & \cite{Colzi2021} &-- &-- &\\
NGC 6334I MM1-v & \cite{Ligterink2020} & -- &-- &\\
Sgr B2(N2) & \cite{Belloche2017}&-- &\cite{Belloche2016} &\\
G+0.693 &\cite{Zeng2018} & \cite{Zeng2018}& --&\\
\bottomrule
\end{tabular}
\tablefoot{If no reference is presented here, it indicates that either no value was found or the value given in a study was not used in this work for various reasons such as the column density found from optically thick lines. The values for the last source are only used when considering ratios including CH$_3$NH$_2$.}
\end{table*}

\begin{table}
    \caption{Weighted averages}
    \label{tab:means}
    \centering
    \begin{tabular}{l l l l l} 
    \toprule
    \toprule    
Ratio & ALMAGAL & Low- & High- & All\\
&&mass&mass&\\
\midrule     

C$_2$H$_3$CN/CH$_3$CN    &0.0447         &--         &0.0528         &0.0514 \\
C$_2$H$_3$CN/C$_2$H$_5$CN        &0.2905         &--         &0.4774         &0.4036 \\
C$_2$H$_5$CN/CH$_3$CN    &0.1672         &0.1071         &0.1859         &0.1736 \\
CH$_3$NH$_2$/CH$_3$CN    &0.1624         &--         &0.1920         &0.1920 \\
CH$_3$NH$_2$/CH$_3$OH    &0.0016         &--         &0.0016         &0.0016 \\
CH$_3$NH$_2$/HNCO        &0.0560         &--         &0.0574         &0.0574 \\
CH$_3$NH$_2$/NH$_2$CHO   &0.8489         &--         &1.0093         &1.0093 \\
NH$_2$CHO/CH$_3$CN       &0.1584         &0.0587         &0.1694         &0.1640 \\
NH$_2$CHO/CH$_3$OH       &0.0014         &0.0003         &0.0017         &0.0016 \\
NH$_2$CHO/HNCO   &0.0607         &0.0405         &0.1150         &0.1112 \\
HNCO/CH$_3$CN    &2.7236         &2.0746         &2.4525         &2.4217 \\
HNCO/CH$_3$OH    &0.0274         &0.0043         &0.0204         &0.0184 \\
CH$_3$CN/CH$_3$OH        &0.0085         &0.0034         &0.0081         &0.0076 \\

\bottomrule
\end{tabular}
\tablefoot{The same as Table \ref{tab:spread_factor} but for 10 to the power of weighted mean of $\log_{10}$ of the column density ratios.}
\end{table}

\begin{table}
    \caption{Assumed values for $\alpha$.}
    \label{tab:alphas}
    \centering
    \begin{tabular}{l l} 
    \toprule
    \toprule    
Speacies & $\alpha$\\
\midrule     
CH$_3$CN	&3.7\\
C$_2$H$_5$CN	&3.0\\
CH$_3$OH	&2.3\\
HNCO	&3.9\\
C$_2$H$_3$CN	&4.0\\
CH$_3$NH$_2$	&1.3\\

\bottomrule
\end{tabular}
\end{table}

\onecolumn
\begin{longtable}{lllll}
\caption{All transitions of the N-bearing species studied here in the data that have $E_{\rm up} < 1000$\,K.} \\
\hline \hline
Species & Transition & Frequency & $A_{\rm ij}$ & $E_\mathrm{up}$ \\
 & J K L M & (MHz) & (s$^{-1}$) & (K) \\
\hline
\endfirsthead
\caption{continued.}\\
\hline\hline
Species & Transition & Frequency & $A_{\rm ij}$ & $E_\mathrm{up}$ \\
 & J K L M & (MHz) & (s$^{-1}$) & (K) \\
\hline
\endhead
\hline
\endfoot

CH$_3$CN & 12 11 - 11 11 & 220\,235.031 & $1.47 \times 10^{-4}$ & 931.4\\ 
(JPL) & 12 10 - 11 10 & 220\,323.631 & $2.81 \times 10^{-4}$ & 782.0\\ 
  & 12 9 - 11 9 & 220\,403.9 & $4.03 \times 10^{-4}$ & 646.7\\ 
  & 12 8 - 11 8 & 220\,475.807 & $5.12 \times 10^{-4}$ & 525.6\\ 
  & 12 7 - 11 7 & 220\,539.323 & $6.08 \times 10^{-4}$ & 418.6\\ 
  & 12 6 - 11 6 & 220\,594.423 & $6.92 \times 10^{-4}$ & 325.9\\ 
  & 12 5 - 11 5 & 220\,641.084 & $7.63 \times 10^{-4}$ & 247.4\\ 
  & 12 4 - 11 4 & 220\,679.287 & $8.21 \times 10^{-4}$ & 183.1\\ 
  & 12 3 - 11 3 & 220\,709.016 & $8.66 \times 10^{-4}$ & 133.2\\ 
  & 12 2 - 11 2 & 220\,730.261 & $8.98 \times 10^{-4}$ & 97.4\\ 
  & 12 1 - 11 1 & 220\,743.011 & $9.18 \times 10^{-4}$ & 76.0\\ 
  & 12 0 - 11 0 & 220\,747.261 & $9.24 \times 10^{-4}$ & 68.9\\ 
\hline 
CH$_{3}^{13}$CN & 12 11 - 11 11 & 220\,127.94 & $1.46 \times 10^{-4}$ & 931.4\\ 
(JPL) & 12 10 - 11 10 & 220\,216.178 & $2.80 \times 10^{-4}$ & 782.0\\ 
  & 12 9 - 11 9 & 220\,296.118 & $4.02 \times 10^{-4}$ & 646.7\\ 
  & 12 8 - 11 8 & 220\,367.731 & $5.11 \times 10^{-4}$ & 525.6\\ 
  & 12 7 - 11 7 & 220\,430.986 & $6.07 \times 10^{-4}$ & 418.6\\ 
  & 12 6 - 11 6 & 220\,485.859 & $6.91 \times 10^{-4}$ & 325.9\\ 
  & 12 5 - 11 5 & 220\,532.327 & $7.62 \times 10^{-4}$ & 247.4\\ 
  & 12 4 - 11 4 & 220\,570.373 & $8.20 \times 10^{-4}$ & 183.1\\ 
  & 12 3 - 11 3 & 220\,599.98 & $8.65 \times 10^{-4}$ & 133.1\\ 
  & 12 2 - 11 2 & 220\,621.136 & $8.97 \times 10^{-4}$ & 97.4\\ 
  & 12 1 - 11 1 & 220\,633.834 & $9.17 \times 10^{-4}$ & 76.0\\ 
  & 12 0 - 11 0 & 220\,638.066 & $9.23 \times 10^{-4}$ & 68.8\\ 
\hline 
C$_2$H$_3$CN & 17 3 14 - 17 2 15 & 217\,067.326 & $2.53 \times 10^{-5}$ & 89.2\\ 
(CDMS) & 29 3 27 - 30 0 30 & 217\,434.016 & $3.74 \times 10^{-7}$ & 217.8\\ 
  & 28 3 26 - 29 0 29 & 217\,489.31 & $3.69 \times 10^{-7}$ & 204.6\\ 
  & 32 3 30 - 33 1 33 & 217\,588.868 & $7.32 \times 10^{-8}$ & 260.2\\ 
  & 32 1 31 - 32 1 32 & 217\,758.802 & $2.28 \times 10^{-6}$ & 245.7\\ 
  & 30 3 28 - 31 0 31 & 217\,870.372 & $3.78 \times 10^{-7}$ & 231.5\\ 
  & 27 3 25 - 28 0 28 & 218\,047.067 & $3.64 \times 10^{-7}$ & 191.8\\ 
  & 50 3 47 - 50 2 48 & 218\,177.547 & $3.37 \times 10^{-5}$ & 607.6\\ 
  & 54 3 51 - 54 3 52 & 218\,199.996 & $3.61 \times 10^{-6}$ & 704.9\\ 
  & 26 2 25 - 26 1 26 & 218\,206.058 & $1.64 \times 10^{-5}$ & 167.9\\ 
  & 28 6 23 - 29 5 24 & 218\,289.844 & $7.93 \times 10^{-6}$ & 262.8\\ 
  & 8 2 6 - 7 1 7 & 218\,307.361 & $1.41 \times 10^{-5}$ & 25.1\\ 
  & 28 6 22 - 29 5 25 & 218\,319.509 & $7.93 \times 10^{-6}$ & 262.8\\ 
  & 23 7 17 - 22 7 16 & 218\,398.555 & $7.84 \times 10^{-4}$ & 231.5\\ 
  & 23 7 16 - 22 7 15 & 218\,398.555 & $7.84 \times 10^{-4}$ & 231.5\\ 
  & 23 6 18 - 22 6 17 & 218\,402.435 & $8.05 \times 10^{-4}$ & 203.5\\ 
  & 23 6 17 - 22 6 16 & 218\,402.451 & $8.05 \times 10^{-4}$ & 203.5\\ 
  & 22 3 20 - 23 1 23 & 218\,406.015 & $7.34 \times 10^{-8}$ & 134.8\\ 
  & 23 8 15 - 22 8 14 & 218\,421.801 & $7.59 \times 10^{-4}$ & 263.8\\ 
  & 23 8 16 - 22 8 15 & 218\,421.801 & $7.59 \times 10^{-4}$ & 263.8\\ 
  & 23 5 19 - 22 5 18 & 218\,451.297 & $8.23 \times 10^{-4}$ & 179.8\\ 
  & 23 5 18 - 22 5 17 & 218\,452.357 & $8.23 \times 10^{-4}$ & 179.8\\ 
  & 23 9 14 - 22 9 13 & 218\,463.739 & $7.32 \times 10^{-4}$ & 300.3\\ 
  & 23 9 15 - 22 9 14 & 218\,463.739 & $7.32 \times 10^{-4}$ & 300.3\\ 
  & 23 10 13 - 22 10 12 & 218\,519.997 & $7.02 \times 10^{-4}$ & 341.1\\ 
  & 23 10 14 - 22 10 13 & 218\,519.997 & $7.02 \times 10^{-4}$ & 341.1\\ 
  & 56 9 48 - 57 8 49 & 218\,548.431 & $8.55 \times 10^{-6}$ & 901.9\\ 
  & 56 9 47 - 57 8 50 & 218\,552.66 & $8.55 \times 10^{-6}$ & 901.9\\ 
  & 23 4 20 - 22 4 19 & 218\,573.646 & $8.39 \times 10^{-4}$ & 160.4\\ 
  & 23 3 21 - 22 3 20 & 218\,585.072 & $8.51 \times 10^{-4}$ & 145.3\\ 
  & 23 11 12 - 22 11 11 & 218\,588.106 & $6.68 \times 10^{-4}$ & 386.0\\ 
  & 23 11 13 - 22 11 12 & 218\,588.106 & $6.68 \times 10^{-4}$ & 386.0\\ 
  & 23 4 19 - 22 4 18 & 218\,615.092 & $8.40 \times 10^{-4}$ & 160.4\\ 
  & 23 12 11 - 22 12 10 & 218\,666.47 & $6.31 \times 10^{-4}$ & 435.0\\ 
  & 23 12 12 - 22 12 11 & 218\,666.47 & $6.31 \times 10^{-4}$ & 435.0\\ 
  & 16 3 13 - 16 2 14 & 218\,675.574 & $2.55 \times 10^{-5}$ & 81.5\\ 
  & 23 13 10 - 22 13 9 & 218\,754.33 & $5.91 \times 10^{-4}$ & 488.2\\ 
  & 23 13 11 - 22 13 10 & 218\,754.33 & $5.91 \times 10^{-4}$ & 488.2\\ 
  & 31 3 29 - 32 0 32 & 218\,786.607 & $3.83 \times 10^{-7}$ & 245.6\\ 
  & 23 14 9 - 22 14 8 & 218\,850.96 & $5.47 \times 10^{-4}$ & 545.5\\ 
  & 23 14 10 - 22 14 9 & 218\,850.96 & $5.47 \times 10^{-4}$ & 545.5\\ 
  & 23 16 7 - 22 16 6 & 219\,068.22 & $4.50 \times 10^{-4}$ & 672.1\\ 
  & 23 16 8 - 22 16 7 & 219\,068.22 & $4.50 \times 10^{-4}$ & 672.1\\ 
  & 26 3 24 - 27 0 27 & 219\,116.926 & $3.59 \times 10^{-7}$ & 179.5\\ 
  & 23 17 6 - 22 17 5 & 219\,188.47 & $3.96 \times 10^{-4}$ & 741.4\\ 
  & 23 17 7 - 22 17 6 & 219\,188.47 & $3.96 \times 10^{-4}$ & 741.4\\ 
  & 23 18 5 - 22 18 4 & 219\,315.93 & $3.39 \times 10^{-4}$ & 814.7\\ 
  & 23 18 6 - 22 18 5 & 219\,315.93 & $3.39 \times 10^{-4}$ & 814.7\\ 
  & 41 2 39 - 41 1 40 & 219\,367.176 & $2.59 \times 10^{-5}$ & 407.0\\ 
  & 23 3 20 - 22 3 19 & 219\,400.585 & $8.61 \times 10^{-4}$ & 145.5\\ 
  & 23 19 4 - 22 19 3 & 219\,450.43 & $2.78 \times 10^{-4}$ & 891.8\\ 
  & 23 19 5 - 22 19 4 & 219\,450.43 & $2.78 \times 10^{-4}$ & 891.8\\ 
  & 23 20 3 - 22 20 2 & 219\,592.0 & $2.14 \times 10^{-4}$ & 972.7\\ 
  & 23 20 4 - 22 20 3 & 219\,592.0 & $2.14 \times 10^{-4}$ & 972.7\\ 
  & 29 1 28 - 28 2 27 & 219\,722.157 & $2.00 \times 10^{-5}$ & 203.2\\ 
  & 33 3 31 - 34 1 34 & 219\,785.766 & $7.38 \times 10^{-8}$ & 275.2\\ 
  & 10 2 9 - 9 1 8 & 219\,899.652 & $1.61 \times 10^{-5}$ & 33.7\\ 
  & 25 2 24 - 25 0 25 & 220\,027.012 & $2.07 \times 10^{-6}$ & 156.1\\ 
  & 15 3 12 - 15 2 13 & 220\,068.962 & $2.57 \times 10^{-5}$ & 74.2\\ 
  & 32 3 30 - 33 0 33 & 220\,170.167 & $3.89 \times 10^{-7}$ & 260.2\\ 
  & 46 3 43 - 45 4 42 & 220\,196.36 & $1.02 \times 10^{-5}$ & 517.5\\ 
  & 9 4 6 - 10 3 7 & 220\,316.125 & $5.50 \times 10^{-6}$ & 55.1\\ 
  & 9 4 5 - 10 3 8 & 220\,344.454 & $5.50 \times 10^{-6}$ & 55.1\\ 
  & 24 1 24 - 23 1 23 & 220\,561.394 & $8.88 \times 10^{-4}$ & 134.9\\ 
  & 25 3 23 - 26 0 26 & 220\,707.088 & $3.52 \times 10^{-7}$ & 167.6\\ 
  & 32 1 31 - 32 0 32 & 220\,752.367 & $1.56 \times 10^{-5}$ & 245.7\\ 
  & 21 3 19 - 22 1 22 & 220\,797.187 & $7.34 \times 10^{-8}$ & 124.8\\ 
  & 63 3 60 - 64 2 63 & 220\,880.958 & $5.37 \times 10^{-7}$ & 949.8\\ 
\hline 
C$_2$H$_5$CN & 55 9 47 - 56 7 50 & 217\,130.895 & $1.98 \times 10^{-7}$ & 752.4\\ 
(CDMS) & 34 12 22 - 35 11 25 & 217\,147.226 & $1.03 \times 10^{-5}$ & 415.6\\ 
  & 34 12 23 - 35 11 24 & 217\,147.226 & $1.03 \times 10^{-5}$ & 415.6\\ 
  & 60 5 56 - 59 6 53 & 217\,168.729 & $7.18 \times 10^{-6}$ & 816.0\\ 
  & 40 3 38 - 40 2 39 & 217\,316.137 & $3.11 \times 10^{-5}$ & 361.5\\ 
  & 16 5 12 - 17 3 15 & 217\,403.316 & $5.99 \times 10^{-8}$ & 86.3\\ 
  & 65 5 60 - 65 5 61 & 217\,406.574 & $6.04 \times 10^{-6}$ & 961.2\\ 
  & 62 2 60 - 63 1 63 & 217\,474.283 & $1.19 \times 10^{-7}$ & 838.5\\ 
  & 62 3 60 - 63 0 63 & 217\,648.198 & $1.19 \times 10^{-7}$ & 838.5\\ 
  & 60 7 53 - 59 8 52 & 217\,780.866 & $1.50 \times 10^{-5}$ & 843.5\\ 
  & 65 7 58 - 65 6 59 & 217\,826.656 & $6.34 \times 10^{-5}$ & 980.2\\ 
  & 48 3 45 - 48 2 46 & 217\,885.862 & $4.04 \times 10^{-5}$ & 522.6\\ 
  & 41 4 37 - 40 5 36 & 217\,907.697 & $1.62 \times 10^{-5}$ & 391.7\\ 
  & 36 2 35 - 36 1 36 & 217\,952.765 & $1.74 \times 10^{-5}$ & 286.6\\ 
  & 57 5 53 - 56 6 50 & 217\,995.808 & $9.09 \times 10^{-6}$ & 740.1\\ 
  & 36 2 35 - 36 0 36 & 218\,019.016 & $3.09 \times 10^{-6}$ & 286.6\\ 
  & 42 4 39 - 42 2 40 & 218\,108.303 & $5.07 \times 10^{-6}$ & 406.8\\ 
  & 39 13 26 - 40 12 29 & 218\,198.608 & $1.09 \times 10^{-5}$ & 522.8\\ 
  & 39 13 27 - 40 12 28 & 218\,198.608 & $1.09 \times 10^{-5}$ & 522.8\\ 
  & 40 3 38 - 40 1 39 & 218\,213.228 & $4.84 \times 10^{-6}$ & 361.5\\ 
  & 29 4 26 - 29 2 27 & 218\,311.807 & $2.09 \times 10^{-6}$ & 205.2\\ 
  & 59 5 55 - 58 6 52 & 218\,361.75 & $7.90 \times 10^{-6}$ & 790.3\\ 
  & 24 3 21 - 23 3 20 & 218\,389.97 & $8.67 \times 10^{-4}$ & 139.9\\ 
  & 44 5 40 - 44 4 41 & 218\,391.341 & $5.11 \times 10^{-5}$ & 454.6\\ 
  & 58 5 54 - 57 6 51 & 218\,633.246 & $8.55 \times 10^{-6}$ & 765.0\\ 
  & 44 14 30 - 45 13 33 & 219\,242.932 & $1.15 \times 10^{-5}$ & 643.0\\ 
  & 44 14 31 - 45 13 32 & 219\,242.932 & $1.15 \times 10^{-5}$ & 643.0\\ 
  & 44 4 41 - 44 3 42 & 219\,270.802 & $4.30 \times 10^{-5}$ & 444.1\\ 
  & 22 2 21 - 21 1 20 & 219\,463.64 & $4.29 \times 10^{-5}$ & 112.5\\ 
  & 24 2 22 - 23 2 21 & 219\,505.59 & $8.88 \times 10^{-4}$ & 135.6\\ 
  & 54 6 48 - 53 7 47 & 219\,621.822 & $1.53 \times 10^{-5}$ & 681.3\\ 
  & 13 3 10 - 13 0 13 & 219\,699.089 & $8.22 \times 10^{-7}$ & 49.2\\ 
  & 35 3 32 - 34 4 31 & 219\,764.572 & $2.01 \times 10^{-5}$ & 284.7\\ 
  & 12 3 10 - 11 2 9 & 219\,902.495 & $3.04 \times 10^{-5}$ & 43.6\\ 
  & 57 6 52 - 56 7 49 & 220\,013.47 & $1.43 \times 10^{-5}$ & 752.7\\ 
  & 35 7 28 - 36 5 31 & 220\,204.183 & $1.49 \times 10^{-7}$ & 325.5\\ 
  & 8 7 1 - 9 6 4 & 220\,272.041 & $1.86 \times 10^{-6}$ & 69.9\\ 
  & 8 7 2 - 9 6 3 & 220\,272.041 & $1.86 \times 10^{-6}$ & 69.9\\ 
  & 49 15 34 - 50 14 37 & 220\,291.895 & $1.19 \times 10^{-5}$ & 776.0\\ 
  & 49 15 35 - 50 14 36 & 220\,291.895 & $1.19 \times 10^{-5}$ & 776.0\\ 
  & 25 2 24 - 24 2 23 & 220\,660.916 & $9.01 \times 10^{-4}$ & 143.0\\ 
  & 45 8 38 - 46 6 41 & 220\,816.769 & $1.83 \times 10^{-7}$ & 516.4\\ 
  & 45 5 41 - 45 4 42 & 220\,845.741 & $5.26 \times 10^{-5}$ & 474.0\\ 
\hline 
HNCO & 10 4 6 9 - 9 4 5 9 & 219\,546.723 & $1.40 \times 10^{-6}$ & 750.4\\ 
(JPL) & 10 4 7 9 - 9 4 6 9 & 219\,546.723 & $1.40 \times 10^{-6}$ & 750.4\\ 
  & 10 4 6 9 - 9 4 5 10 & 219\,547.054 & $3.50 \times 10^{-9}$ & 750.4\\ 
  & 10 4 7 9 - 9 4 6 10 & 219\,547.054 & $3.50 \times 10^{-9}$ & 750.4\\ 
  & 10 4 6 9 - 9 4 5 8 & 219\,547.091 & $1.25 \times 10^{-4}$ & 750.4\\ 
  & 10 4 7 9 - 9 4 6 8 & 219\,547.082 & $1.25 \times 10^{-4}$ & 750.4\\ 
  & 10 4 6 11 - 9 4 5 10 & 219\,547.095 & $1.26 \times 10^{-4}$ & 750.4\\ 
  & 10 4 7 11 - 9 4 6 10 & 219\,547.082 & $1.26 \times 10^{-4}$ & 750.4\\ 
  & 10 4 6 10 - 9 4 5 9 & 219\,547.165 & $1.25 \times 10^{-4}$ & 750.4\\ 
  & 10 4 7 10 - 9 4 6 9 & 219\,547.082 & $1.25 \times 10^{-4}$ & 750.4\\ 
  & 10 4 7 10 - 9 4 6 10 & 219\,547.497 & $1.26 \times 10^{-6}$ & 750.4\\ 
  & 10 4 6 10 - 9 4 5 10 & 219\,547.497 & $1.26 \times 10^{-6}$ & 750.4\\ 
  & 10 3 8 9 - 9 3 7 9 & 219\,656.239 & $1.52 \times 10^{-6}$ & 447.5\\ 
  & 10 3 7 9 - 9 3 6 9 & 219\,656.24 & $1.52 \times 10^{-6}$ & 447.5\\ 
  & 10 3 8 9 - 9 3 7 10 & 219\,656.736 & $3.80 \times 10^{-9}$ & 447.5\\ 
  & 10 3 7 9 - 9 3 6 10 & 219\,656.738 & $3.80 \times 10^{-9}$ & 447.5\\ 
  & 10 3 8 11 - 9 3 7 10 & 219\,656.71 & $1.37 \times 10^{-4}$ & 447.5\\ 
  & 10 3 7 11 - 9 3 6 10 & 219\,656.71 & $1.37 \times 10^{-4}$ & 447.5\\ 
  & 10 3 8 9 - 9 3 7 8 & 219\,656.71 & $1.36 \times 10^{-4}$ & 447.5\\ 
  & 10 3 7 9 - 9 3 6 8 & 219\,656.71 & $1.36 \times 10^{-4}$ & 447.5\\ 
  & 10 3 8 10 - 9 3 7 9 & 219\,656.71 & $1.36 \times 10^{-4}$ & 447.5\\ 
  & 10 3 7 10 - 9 3 6 9 & 219\,656.71 & $1.36 \times 10^{-4}$ & 447.5\\ 
  & 10 3 8 10 - 9 3 7 10 & 219\,657.329 & $1.37 \times 10^{-6}$ & 447.5\\ 
  & 10 3 7 10 - 9 3 6 10 & 219\,657.33 & $1.37 \times 10^{-6}$ & 447.5\\ 
  & 10 2 9 9 - 9 2 8 9 & 219\,733.115 & $1.60 \times 10^{-6}$ & 231.1\\ 
  & 10 2 9 9 - 9 2 8 10 & 219\,733.731 & $4.01 \times 10^{-9}$ & 231.1\\ 
  & 10 2 9 11 - 9 2 8 10 & 219\,733.85 & $1.45 \times 10^{-4}$ & 231.1\\ 
  & 10 2 9 9 - 9 2 8 8 & 219\,733.85 & $1.43 \times 10^{-4}$ & 231.1\\ 
  & 10 2 9 10 - 9 2 8 9 & 219\,733.85 & $1.43 \times 10^{-4}$ & 231.1\\ 
  & 10 2 9 10 - 9 2 8 10 & 219\,734.431 & $1.45 \times 10^{-6}$ & 231.1\\ 
  & 10 2 8 9 - 9 2 7 9 & 219\,736.543 & $1.60 \times 10^{-6}$ & 231.1\\ 
  & 10 2 8 9 - 9 2 7 10 & 219\,737.159 & $4.01 \times 10^{-9}$ & 231.1\\ 
  & 10 2 8 11 - 9 2 7 10 & 219\,737.193 & $1.45 \times 10^{-4}$ & 231.1\\ 
  & 10 2 8 9 - 9 2 7 8 & 219\,737.193 & $1.43 \times 10^{-4}$ & 231.1\\ 
  & 10 2 8 10 - 9 2 7 9 & 219\,737.193 & $1.43 \times 10^{-4}$ & 231.1\\ 
  & 10 2 8 10 - 9 2 7 10 & 219\,737.859 & $1.45 \times 10^{-6}$ & 231.1\\ 
  & 10 0 10 9 - 9 0 9 9 & 219\,797.536 & $1.67 \times 10^{-6}$ & 58.0\\ 
  & 10 0 10 9 - 9 0 9 10 & 219\,798.247 & $4.18 \times 10^{-9}$ & 58.0\\ 
  & 10 0 10 11 - 9 0 9 10 & 219\,798.32 & $1.51 \times 10^{-4}$ & 58.0\\ 
  & 10 0 10 10 - 9 0 9 9 & 219\,798.32 & $1.50 \times 10^{-4}$ & 58.0\\ 
  & 10 0 10 9 - 9 0 9 8 & 219\,798.32 & $1.49 \times 10^{-4}$ & 58.0\\ 
  & 10 0 10 10 - 9 0 9 10 & 219\,799.033 & $1.51 \times 10^{-6}$ & 58.0\\ 
  & 10 1 9 9 - 9 1 8 9 & 220\,584.205 & $1.67 \times 10^{-6}$ & 101.5\\ 
  & 10 1 9 9 - 9 1 8 10 & 220\,584.797 & $4.19 \times 10^{-9}$ & 101.5\\ 
  & 10 1 9 11 - 9 1 8 10 & 220\,585.2 & $1.51 \times 10^{-4}$ & 101.5\\ 
  & 10 1 9 9 - 9 1 8 8 & 220\,585.2 & $1.49 \times 10^{-4}$ & 101.5\\ 
  & 10 1 9 10 - 9 1 8 9 & 220\,585.2 & $1.50 \times 10^{-4}$ & 101.5\\ 
  & 10 1 9 10 - 9 1 8 10 & 220\,585.457 & $1.51 \times 10^{-6}$ & 101.5\\ 
\hline 
HN$^{13}$CO & 10 4 6 9 - 9 4 5 9 & 219\,552.523 & $1.40 \times 10^{-6}$ & 748.7\\ 
(JPL) & 10 4 7 9 - 9 4 6 9 & 219\,552.523 & $1.40 \times 10^{-6}$ & 748.7\\ 
  & 10 4 7 9 - 9 4 6 8 & 219\,552.897 & $1.25 \times 10^{-4}$ & 748.7\\ 
  & 10 4 6 9 - 9 4 5 8 & 219\,552.897 & $1.25 \times 10^{-4}$ & 748.7\\ 
  & 10 4 6 11 - 9 4 5 10 & 219\,552.9 & $1.26 \times 10^{-4}$ & 748.7\\ 
  & 10 4 7 11 - 9 4 6 10 & 219\,552.9 & $1.26 \times 10^{-4}$ & 748.7\\ 
  & 10 4 6 10 - 9 4 5 9 & 219\,552.972 & $1.25 \times 10^{-4}$ & 748.7\\ 
  & 10 4 7 10 - 9 4 6 9 & 219\,552.972 & $1.25 \times 10^{-4}$ & 748.7\\ 
  & 10 4 7 10 - 9 4 6 10 & 219\,553.307 & $1.26 \times 10^{-6}$ & 748.7\\ 
  & 10 4 6 10 - 9 4 5 10 & 219\,553.307 & $1.26 \times 10^{-6}$ & 748.7\\ 
  & 10 3 8 9 - 9 3 7 9 & 219\,663.066 & $1.52 \times 10^{-6}$ & 446.5\\ 
  & 10 3 7 9 - 9 3 6 9 & 219\,663.067 & $1.52 \times 10^{-6}$ & 446.5\\ 
  & 10 3 8 11 - 9 3 7 10 & 219\,663.625 & $1.37 \times 10^{-4}$ & 446.5\\ 
  & 10 3 7 11 - 9 3 6 10 & 219\,663.625 & $1.37 \times 10^{-4}$ & 446.5\\ 
  & 10 3 8 9 - 9 3 7 8 & 219\,663.626 & $1.36 \times 10^{-4}$ & 446.5\\ 
  & 10 3 7 9 - 9 3 6 8 & 219\,663.627 & $1.36 \times 10^{-4}$ & 446.5\\ 
  & 10 3 8 10 - 9 3 7 9 & 219\,663.666 & $1.36 \times 10^{-4}$ & 446.5\\ 
  & 10 3 7 10 - 9 3 6 9 & 219\,663.667 & $1.36 \times 10^{-4}$ & 446.5\\ 
  & 10 3 8 10 - 9 3 7 10 & 219\,664.17 & $1.37 \times 10^{-6}$ & 446.5\\ 
  & 10 3 7 10 - 9 3 6 10 & 219\,664.171 & $1.37 \times 10^{-6}$ & 446.5\\ 
  & 10 2 9 9 - 9 2 8 9 & 219\,739.762 & $1.60 \times 10^{-6}$ & 230.7\\ 
  & 10 2 9 11 - 9 2 8 10 & 219\,740.451 & $1.45 \times 10^{-4}$ & 230.7\\ 
  & 10 2 9 9 - 9 2 8 8 & 219\,740.456 & $1.43 \times 10^{-4}$ & 230.7\\ 
  & 10 2 9 10 - 9 2 8 9 & 219\,740.471 & $1.43 \times 10^{-4}$ & 230.7\\ 
  & 10 2 9 10 - 9 2 8 10 & 219\,741.095 & $1.45 \times 10^{-6}$ & 230.7\\ 
  & 10 2 8 9 - 9 2 7 9 & 219\,743.054 & $1.60 \times 10^{-6}$ & 230.7\\ 
  & 10 2 8 11 - 9 2 7 10 & 219\,743.742 & $1.45 \times 10^{-4}$ & 230.7\\ 
  & 10 2 8 9 - 9 2 7 8 & 219\,743.747 & $1.43 \times 10^{-4}$ & 230.7\\ 
  & 10 2 8 10 - 9 2 7 9 & 219\,743.762 & $1.43 \times 10^{-4}$ & 230.7\\ 
  & 10 2 8 10 - 9 2 7 10 & 219\,744.386 & $1.45 \times 10^{-6}$ & 230.7\\ 
  & 10 0 10 9 - 9 0 9 9 & 219\,803.645 & $1.67 \times 10^{-6}$ & 58.0\\ 
  & 10 0 10 9 - 9 0 9 10 & 219\,804.366 & $4.18 \times 10^{-9}$ & 58.0\\ 
  & 10 0 10 11 - 9 0 9 10 & 219\,804.439 & $1.51 \times 10^{-4}$ & 58.0\\ 
  & 10 0 10 10 - 9 0 9 9 & 219\,804.442 & $1.50 \times 10^{-4}$ & 58.0\\ 
  & 10 0 10 9 - 9 0 9 8 & 219\,804.446 & $1.49 \times 10^{-4}$ & 58.0\\ 
  & 10 0 10 10 - 9 0 9 10 & 219\,805.163 & $1.51 \times 10^{-6}$ & 58.0\\ 
  & 10 1 9 9 - 9 1 8 9 & 220\,592.054 & $1.67 \times 10^{-6}$ & 101.4\\ 
  & 10 1 9 9 - 9 1 8 10 & 220\,592.555 & $4.19 \times 10^{-9}$ & 101.4\\ 
  & 10 1 9 11 - 9 1 8 10 & 220\,592.606 & $1.51 \times 10^{-4}$ & 101.4\\ 
  & 10 1 9 9 - 9 1 8 8 & 220\,592.611 & $1.49 \times 10^{-4}$ & 101.4\\ 
  & 10 1 9 10 - 9 1 8 9 & 220\,592.613 & $1.50 \times 10^{-4}$ & 101.4\\ 
  & 10 1 9 10 - 9 1 8 10 & 220\,593.114 & $1.51 \times 10^{-6}$ & 101.4\\ 
\hline 
H$^{15}$NCO & 29 1 29 - 30 0 30 & 219\,775.87 & $5.65 \times 10^{-5}$ & 486.2\\ 
(JPL) & & & & \\ 
\hline 
NH$_2$CHO v=0 & 39 4 36 - 38 5 33 & 218\,177.973 & $5.20 \times 10^{-6}$ & 845.0\\ 
(CDMS) & 10 1 9 - 9 1 8 & 218\,459.213 & $7.48 \times 10^{-4}$ & 60.8\\ 
  & 39 5 34 - 38 6 33 & 219\,389.054 & $7.20 \times 10^{-6}$ & 875.0\\ 
  & 27 2 25 - 28 1 28 & 220\,058.164 & $3.11 \times 10^{-7}$ & 406.3\\ 
  & 10 3 7 - 11 1 10 & 220\,538.374 & $2.27 \times 10^{-7}$ & 82.9\\ 
  & 27 2 25 - 28 0 28 & 220\,668.548 & $1.37 \times 10^{-7}$ & 406.3\\ 
\hline 
NH$_2$CHO v$_{12}$=1 & 9 2 8 - 9 1 9 & 218\,179.121 & $1.89 \times 10^{-5}$ & 473.2\\ 
(CDMS) & 10 1 9 - 9 1 8 & 218\,181.68 & $7.45 \times 10^{-4}$ & 476.5\\ 
  & 19 4 15 - 20 2 18 & 218\,692.835 & $3.23 \times 10^{-7}$ & 656.5\\ 
  & 17 1 16 - 17 1 17 & 219\,417.827 & $6.31 \times 10^{-6}$ & 578.3\\ 
  & 3 3 1 - 4 2 2 & 220\,490.762 & $1.64 \times 10^{-6}$ & 448.3\\ 
  & 3 3 0 - 4 2 3 & 220\,911.024 & $1.64 \times 10^{-6}$ & 448.3\\ 
\hline 
CH$_3$NH$_2$ & 11 2 0 - 11 -1 1 & 217\,079.4 & $5.06 \times 10^{-5}$ & 156.5\\ 
(JPL) & 25 5 2 - 24 -6 3 & 217\,134.012 & $1.42 \times 10^{-5}$ & 788.4\\ 
  & 25 -5 3 - 24 6 2 & 217\,152.204 & $1.42 \times 10^{-5}$ & 788.4\\ 
  & 18 7 7 - 19 6 7 & 217\,625.208 & $1.05 \times 10^{-5}$ & 554.3\\ 
  & 3 3 5 - 4 2 5 & 217\,669.865 & $3.38 \times 10^{-6}$ & 47.7\\ 
  & 11 2 5 - 11 1 5 & 217\,670.017 & $1.41 \times 10^{-5}$ & 156.5\\ 
  & 11 2 7 - 11 1 6 & 217\,754.62 & $2.97 \times 10^{-7}$ & 156.5\\ 
  & 12 -2 3 - 12 1 2 & 217\,758.328 & $5.20 \times 10^{-5}$ & 182.1\\ 
  & 7 0 5 - 6 1 5 & 218\,220.975 & $3.39 \times 10^{-5}$ & 60.0\\ 
  & 25 5 4 - 24 6 4 & 218\,399.414 & $1.44 \times 10^{-5}$ & 788.3\\ 
  & 3 1 5 - 2 0 5 & 218\,408.868 & $4.71 \times 10^{-5}$ & 17.3\\ 
  & 25 5 6 - 24 6 6 & 218\,833.011 & $1.45 \times 10^{-5}$ & 788.7\\ 
  & 8 2 4 - 8 1 5 & 219\,150.831 & $2.54 \times 10^{-5}$ & 92.2\\ 
  & 5 1 2 - 4 -1 3 & 219\,440.099 & $5.06 \times 10^{-6}$ & 35.7\\ 
  & 5 1 0 - 4 -1 1 & 219\,440.487 & $5.06 \times 10^{-6}$ & 35.9\\ 
  & 30 1 5 - 29 4 5 & 219\,499.757 & $1.26 \times 10^{-7}$ & 997.9\\ 
  & 5 1 4 - 4 1 4 & 219\,650.046 & $5.04 \times 10^{-6}$ & 35.9\\ 
  & 18 3 5 - 17 4 4 & 220\,082.666 & $1.08 \times 10^{-7}$ & 398.8\\ 
  & 14 2 5 - 13 3 4 & 220\,093.113 & $3.88 \times 10^{-6}$ & 239.7\\ 
  & 5 1 7 - 4 1 7 & 220\,158.161 & $5.05 \times 10^{-6}$ & 35.7\\ 
  & 30 1 0 - 29 -4 1 & 220\,417.41 & $3.98 \times 10^{-7}$ & 998.0\\ 
  & 25 5 5 - 24 6 5 & 220\,449.349 & $1.48 \times 10^{-5}$ & 788.6\\ 
  & 7 4 5 - 8 3 5 & 220\,760.6 & $8.23 \times 10^{-6}$ & 122.2\\ 
  & 10 2 7 - 10 1 6 & 220\,805.517 & $7.45 \times 10^{-7}$ & 133.0\\ 
  & 7 0 3 - 6 1 2 & 220\,827.559 & $3.59 \times 10^{-5}$ & 59.8\\ 
  & 9 2 6 - 9 1 6 & 220\,888.511 & $4.95 \times 10^{-5}$ & 111.5\\ 
\hline

\label{tab:lines}
\end{longtable}

\end{appendix}

\end{document}